\documentclass[reqno,11pt]{article}
\pdfoutput=1
\usepackage{jheppub}
\usepackage{epsfig}
\usepackage{amssymb}
\usepackage{verbatim}
\usepackage{amsmath}
\usepackage{dsfont}
\usepackage{slashed}
\usepackage[dvipsnames]{xcolor}
\usepackage{epstopdf}
\usepackage{natbib}
\usepackage{graphicx}
\def\M{\mathfrak{M}}
\def\ph{\Phi}

\def\l{\lambda}

\def\L{\Lambda}

\def\s{\sigma}
\def\half{{1 \over 2}}
\def\a{\alpha}
\def\b{\beta}

\newcommand{\be}{\begin{eqnarray}}
\newcommand{\ee}{\end{eqnarray}}

\newcommand{\bc}{\begin{center}}
\newcommand{\ec}{\end{center}}

\title{CHY-Graphs on a Torus}
\author[a]{Carlos Cardona}
\author[b,c]{and Humberto Gomez}
\affiliation[a]{Physics Division, National Center for Theoretical Sciences, National Tsing-Hua
University,\\ Hsinchu, Taiwan 30013, Republic of China.}
\affiliation[b]{Instituto de Fisica -- Universidade de S\~ao Paulo,\\
Caixa Postal  66318, 05315-970 S\~ao Paulo, SP, Brazil.}
\affiliation[c]{Facultad de Ciencias Basicas,  Universidad Santiago de Cali,\\
Calle 5 $N^\circ$  62-00 Barrio Pampalinda, Cali, Valle, Colombia.}

\emailAdd{carlosandres@mx.nthu.edu.tw}
\emailAdd{humgomzu@gmail.com}

\abstract{Recently, we proposed a new approach using a punctured Elliptic curve in
the CHY framework in order to compute one-loop scattering amplitudes. In this note, we further 
develop this approach by introducing a set of connectors, which become the main ingredient to build integrands on $\mathfrak{M}_{1,n}$, the moduli space of n-punctured Elliptic curves. As a particular application, we study the $\Phi^3$ bi-adjoint scalar theory. We propose a set of rules to construct integrands on $\mathfrak{M}_{1,n}$  from $\Phi^ 3$ integrands on $\mathfrak{M}_{0,n}$, the moduli space of n-punctured spheres. We illustrate these rules by computing a variety of $\Phi^3$ one-loop Feynman diagrams. Conversely, we also provide another set of rules to compute the corresponding CHY-integrand on $\mathfrak{M}_{1,n}$  by
starting instead from a given $\Phi^ 3$ one-loop Feynman diagram. In addition, our results can easily be extended to higher loops.}

\begin{document}

\begin{flushright}
\vspace{10pt} \hfill{NCTS-TH/1604} \vspace{20mm}
\end{flushright}

\maketitle


\section{Introduction}\label{sect:intro}

During the last decade we have witnessed substantial progress in the developing of on-shell methods for the computation of S-matrix elements, following after the breakthrough paper of E. Witten \cite{Witten:2003nn} on scattering amplitudes of pure YM theory in four dimensions. A particular promising development for a wider class of theories and in arbitrary dimension has been done in recent years by F. Cachazo, S. He and E. Yuan (CHY), who have proposed a compact form for the tree-level massless S-matrix in terms of integrals on the moduli space of  $n-$punctured spheres  \cite{Cachazo:2013hca, Cachazo:2013iea, Cachazo:2013gna}. Those integrals are localized over solutions of the so-called scattering equations, which can be written as,
\begin{equation}\label{CHYSE}
E_a=\sum_{b=1\atop b\neq a}^n \frac{k_a\cdot k_b}{\sigma_a-\sigma_b} =  0, \qquad a\in \{1,2,\ldots ,n\},
\end{equation}
where $k_a^\mu$ denotes the momentum of the $a^{\rm th}$ external particle, associated to the location $\sigma_a$ on the sphere.

CHY have extended their approach to the study of scattering of scalars, gauge bosons, gravitons and mixing interactions among them \cite{Cachazo:2013hca,Cachazo:2013iea,Cachazo:2013iaa,Cachazo:2014xea,Cachazo:2014nsa,Berkovits:2013xba,Gomez:2013wza}. More recently, interesting interaction mixings have been found in the single-soft limit of some of those theories \cite{Cachazo:2016njl}. Althought the formalism remains conjectural, it have already passed several non-trivial checks. It has been proved to reproduce the expected soft-limits \cite{Cachazo:2013hca} in the theories where it can be applied. It also has been proved to reproduce the correct BCFW  \cite{Britto:2005fq} recurrence relations in Yang-Mills and Bi-adjoint $\Phi^3$ theories \cite{Dolan:2013isa}.

 So far many methods have been developed to deal with the integration over the punctured sphere at the solutions of the scattering equations. Early attempts considered solutions of \eqref{CHYSE} at particular kinematics \cite{Cachazo:2013iea, Kalousios:2013eca, Lam:2014tga} as well as at particular dimensions \cite{Weinzierl:2014vwa, Cachazo:2013iaa, Cachazo:2016sdc, He:2016vfi}. Later, methods which avoid solving the equations were developed \cite{Kalousios:2015fya,Dolan:2014ega, Huang:2015yka, Cardona:2015ouc, Cardona:2015eba, Dolan:2015iln, Sogaard:2015dba, Cachazo:2015nwa,Mafra:2016ltu} and some mathematical new structures have been found recently in \cite{Bosma:2016ttj, Zlotnikov:2016wtk}. Generalized Feynman rules for single poles diagrams were developed in \cite{Baadsgaard:2015ifa,Baadsgaard:2015voa} and a generalization to second order poles was done in \cite{Huang:2016zzb}. More recently, by using cross-ratio identities coming from the scattering equations, an algorithm have been proposed to reduce the order of any higher-order integrand to simple poles \cite{Cardona:2016gon}.

A natural task in order to move forward is upgrading the CHY formalism to loop-level. Some progress in this direction has been done recently from sligtly different approaches.  By using ambitwistor string \cite{Mason:2013sva}, a proposal was made for the integration measure as well as the corresponding scattering equations at one-loop has been obtained in \cite{Geyer:2015bja, Geyer:2015jch}. By performing a forward limit on the scattering equations for massive particles formulated previously in \cite{Naculich:2014naa, Dolan:2013isa}, the scattering equations at one-loop have also been obtained and used in \cite{Cachazo:2015aol,He:2015yua}. A generalization of this approach to higher loops has been considered in \cite{Feng:2016nrf}.

On a parallel approach, we have generalized the double-cover formulation made at tree level in \cite{Gomez:2016bmv} to the one-loop case by embedding the Torus in a $\mathbb{CP}^2$ throught a Elliptic curve and we have used it to reproduce the Feynman $n-$gon diagram \cite{Cardona:2016bpi}.

In this paper, we would like to show how the prescription given in \cite{Cardona:2016bpi} applies straightforwardly to deal with any one-loop computation in theories admitting CHY description. In order to do so, we provide a proposal to build one-loop integrands from the known tree-level counterparts. Based on the fact that any integrand at tree-level can be written as products, or chain of products, in the ``distances" $z_{ab}$ between the puncture location $z_a$ and $z_b$, we build some analogous connectors-like objects on the moduli of the $n-$punctured Elliptic curves, which depends on the way how we link two points $\s_a$ and $\s_b$ on the surface of a Torus, namely by going around a $b-$cycle in one-direction, by going around a $b-$cycle in the opposite direction or by not circling any $b-$cycle at all. Doing so, we sooner realize that replacing $z_{ab}$, or even better its equivalent form $\tau_{a:b}$ in the double cover approach, by those generalized connectors have the effect of blowing-up loops in the corresponding Feynman diagrams and therefore corresponds to loop CHY-integrands.

In a sort of inverse treatment we also present a simple set of rules to build the corresponding CHY-integrad on the Torus by starting instead from a given Feynman diagram at one-loop. This rules can be thought as a one-loop generalization of the rules presentend in \cite{Baadsgaard:2015ifa}. Although the ones presented here are somehow analogous to the one-loop rules presented in \cite{Baadsgaard:2015hia}, our rules are different in nature, in particular the rules presented in this paper are fundamentally graphical.

Finally, by using the $\Lambda-$algorithm \cite{Gomez:2016bmv} we solve the given CHY-integrands at one-loop obtaining the expected results in the examples considered.

We would like to clarify that in this work we refer to integrands as the integrands in the CHY formalism and not as the integrands in the integration over the loop momenta. The later, are instead the resulting from the ``integration" of our CHY-integrands.

The remainder of this paper is organized as follows. In section \ref{Sec2} and \ref{One-Loop-section} we review the double cover formulation of CHY for the tree and one-loop level scattering amplitudes respectively. We also define connectors on $\M_{1,n}$, which are used to build  the integrands on $\M_{1,n}$. We show how to use those connectors on some examples in section \ref{EXAMPLES} and in section \ref{phi3loop} we apply it to the particular Bi-adjoint $\Phi^3$ theory, where we also provide explicit examples. The final section \ref{Discussion} is used for some discussions and possible perspectives.

\section{Tree-Level Scattering Amplitudes}\label{Sec2}

Before proceeding to the main concern of this paper, let us summarize the results of \cite{Cardona:2016bpi, Gomez:2016bmv} which will provide the background for the remaining sections. 

\subsection{Tree-Level Scattering Amplitude Prescription}\label{Tree_prescription}
The prescription for the computation of scattering amplitudes at tree-level by a double cover approach was proposed in \cite{Gomez:2016bmv}. The $n-$particle amplitude is given by the expression\footnote{Without loss of generality, we have fixed the $\{\s_{n-2},\s_{n-1},\s_n\}$ punctures and the $\{E_{n-2}^{\rm t},E_{n-1}^{\rm t},E_{n}^{\rm t}  \}$ scattering equations.}
\begin{equation}\label{treeAmplitude}
{\cal A}_n^{\rm t}(1,2,\ldots,n)=\frac{1}{2^3}\int_{\Gamma^{\rm t}} \left(\prod_{a=1}^n \frac{y_a \, d y_a}{C_a^{\rm t}}  \right)\times \left( \prod _{i=1}^{n-3} \frac{d\s_i}{E_i^{\rm t}} \right)\times \Delta^2_{\rm FP}(n-2,n-1,n)\times{\cal I}_n^{\rm t}(\s,y),
\end{equation}
where the $\Gamma^{\rm t}$ integration contour is defined by the $2n-3$ equations
\begin{equation}
C_a^{\rm t}=0,~~ a=1,\ldots, n, ~~~ E_i^{\rm t}=0, ~~ i=1,\ldots, n-3.
\end{equation}
Let us remind that, the $(2n-3)$-tupla, $(\s_1,\ldots,\s_{n-3},y_1,\ldots, y_n)$,  are the inhomogeneous coordinates  of the direct product between the moduli space of $n$-punctured Riemann spheres  (${\cal M}_{0,n}$) and the $n$-dimensional complex plane ($\mathbb{C}^n$), i.e. ${\cal M}_{0,n}\times \mathbb{C}^n$. We denote this space as $\M_{0,n}:={\cal M}_{0,n}\times \mathbb{C}^n$.

The $E_i^{\rm t}$'s correspond to the tree-level scattering equations given by\footnote{The upper index ``${\rm t}$" means tree level.} 
\begin{equation}\label{SEtree}
 E^{\rm t}_a:=\half\sum_{b=1\atop b\ne a}^n \left({y^{\rm t}_b\over y^{\rm t}_a}+1 \right) {k_a\cdot k_b \over \s_a-\s_b}=0,
\quad {\rm where } ~~ (y^{\rm t}_a)^2=\s_a-1,  
\end{equation}
and we also have used the constraints defining the double covered sphere as 
\begin{align}
C_a^{\rm t}:=(y_a^{\rm t})^2-(\s_a-1)\,.
\end{align}
The Faddeev Popov determinant, $\Delta_{\rm FP}(n-3,n-2,n)$, is defined as
\begin{equation*}
\Delta_{\rm FP}(n-2,n-1,n)=2^3\,(y^{\rm t}_{n-2}\,y^{\rm t}_{n-1}\,y^{\rm t}_n)\,
\left |
\begin{matrix}
1 & y^{\rm t}_{n-2} &(y^{\rm t}_{n-2})^2\\
1 & y^{\rm t}_{n-1} & (y^{\rm t}_{n-1})^2\\
1 & y^{\rm t}_{n} & (y^{\rm t}_{n})^2\\
\end{matrix}
\right|\, .
\end{equation*}

The ${\cal I}_n^{\rm t}(\s,y)$ integrand, which defines the theory,  is a rational function in terms of {\bf chains}.  Let us remind that we define a {\bf $k$-chain} as a  sequence of $k$-objects \cite{Cachazo:2015nwa}, in this case a $k$-chain is read as
\begin{equation}\label{chains}
\tau_{i_1:i_2} \tau_{i_2:i_3}\cdots \tau_{i_{k-1}:i_k} \tau_{i_{k}:i_1}:=(i_1: i_2:\cdots : i_k)^{\rm t},
\end{equation}
where the $\tau_{a:b}$'s are the third-kind forms given by
\begin{equation}\label{tau}
\tau_{a:b}:={1\over 2\,y^{\rm t}_a}\left({y^{\rm t}_a+y^{\rm t}_b\over\s_{ab}}\right)={1\over 2\,y^{\rm t}_a}\left({1 \over y^{\rm t}_a-y^{\rm t}_b}\right)\, ,
\end{equation}
on the support, $C_a^{\rm t}=0$.  

In this paper, we call {\bf the connectors} to the objects that shape a chain. In this case, the $\tau_{a:b}$'s third-kind forms are the connectors. 

In addition, it is useful to note that in this context the chains have a maximum length, which is the total number of particles, i.e., $n$. A $n$-chain is known as a {\bf Parker-Taylor factor}.

Finally, under the transformation, $\s_a=z_a^2+1$,  one can check that \eqref{treeAmplitude} is mapped in the original CHY prescription.

\subsection{CHY-Tree Level Graph}

Let us recall here that each ${\cal I}^{\rm t}_n(\s,y)$ integrand have  a {\bf regular graph}\footnote{A $G$ graph  is defined by the two finite sets, $V$ and $E$.  $V$ is the vertex set and $E$ is the edge set.}  (bijective map) associated, which we denoted by $G=(V_G,E_G)$ \cite{graph1,graph2,Cachazo:2015nwa}.  The vertex set of $G$ is given by the $n$-labels (punctures) 
$$
V_G=\{1,2,\ldots,n\}
$$ 
and the edge set is given by the elements 
\begin{align}
&\tau_{a:b}\,\leftrightarrow\,a ~\overline{~~~~~~~~~~~~~}~ b ~~~{\rm (Line)}  \\
& \tau^{-1}_{a:b}\,\leftrightarrow\, a\,- \, - \,  - \, -\, b ~~~     {\rm  (Anti-line)}.
 \end{align}

Since $\tau_{a:b}$ always appears into a chain,  then the graph is not a directed graph,  in the same way as in \cite{Cachazo:2015nwa}. 

\begin{figure}[h]
  \centering
  \includegraphics[width=1.2in]{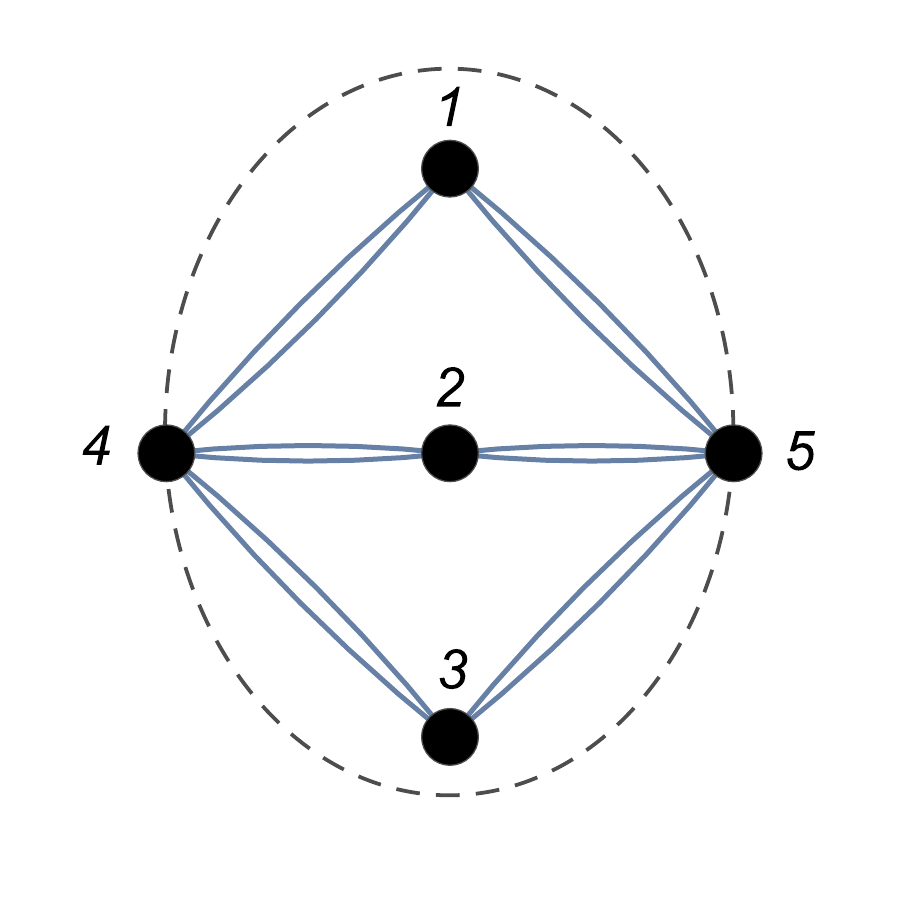}\\
  \caption{The ${\cal I}^{\rm t}_5(1,2,3,4,5)$  {\sl regular graph}.}\label{example_CHYG}
\end{figure}

For example, let us consider the integrand
\be\label{fexample}
{\cal I}^{\rm t}_5(1,2,3,4,5)=\frac{(1:5:2:4)^{\rm t} (3:4:2:5)^{\rm t}\times (1:4:2:5)^{\rm t} (3:5:2:4)^{\rm t}}{(4:5)^{\rm t}}.
\ee
This integrand is represented by the  $G$ graph in figure \ref{example_CHYG}.
\\
Note that for each vertex the number of lines minus anti-lines  must always be 4,
$$
\#\, {\rm  Lines} - \#\,{\rm Antilines}=4,
$$
this is in order to obtain $PSL(2,\mathbb{C})$ invariance.

\subsubsection{Color Code}

\begin{figure}[h]
  \centering
    \includegraphics[width=4.6in]{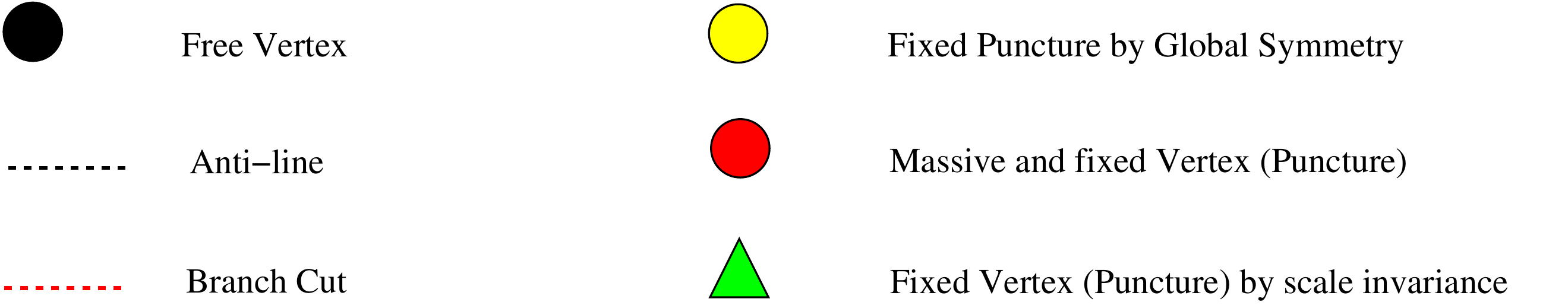}
  \caption{Color Code.}\label{color_cod}
\end{figure}

Since most of the computations are performed using the $\L$-algorithm \cite{Gomez:2016bmv}, which is a pictorial technique, we introduce the color code given in figure \ref{color_cod}, to be used quite often in the remaining of the paper.

\section{One-Loop  Scattering Amplitudes}\label{One-Loop-section}

Before proceeding to the one-loop amplitude prescription, let us remind that the whole construction is supported  on a Torus 
embedded in a $\mathbb{C}P^2$ space,  with local coordinates $(z,y)$, i.e. a Torus  described by the Elliptic curve 
\be\label{cubiccurve}
y^2= z(z-1)(z-\l)\,,
\ee
being $\l$ the complex moduli of tori. In addition, for the rest of the paper we will take the ${\rm a-cycle}$ on the upper brach cut, $y=\sqrt{z(z-1)(z-\l)}$.

\subsection{One-Loop Scattering Amplitude prescription}\label{One-Loop-prescription}

In \cite{Cardona:2016bpi} we have proposed a prescription for computing scattering amplitudes on the moduli space of $n$-punctured Elliptic curves. The prescription for the  $n-$particle amplitude at one-loop is given by the following expression,
\begin{equation}\label{GenA}
{\cal A}^1_n(1,\ldots,n) =  \int d^Dq  \int_{\Gamma^1} \frac{d\l}{\l(1-\l)}\times \left(\prod_{a=1}^{n}\, {dy_a \over C_a}\right)\times \left(\prod_{i=1}^{n-1} \, {d\s_i \over E^1_i}\right)  \left( {\Delta^2_{\rm FP}(n) \over {\cal L}\,\prod_{b=1}^n y_b} \right)H(\sigma,y),
\end{equation}
where the elements of the $2n$-tupla, $(\l,\s_1,\ldots,\s_{n-1},y_1,\ldots, y_n)$,  are the  coordinates  of the direct product between the moduli space of $n$-punctured Elliptic curves (${\cal M}_{1,n}$) and the $n$-dimensional complex plane ($\mathbb{C}^n$), i.e. ${\cal M}_{1,n}\times \mathbb{C}^n$. We denote this space as $\M_{1,n}:={\cal M}_{1,n}\times \mathbb{C}^n$.

The prescription in  \eqref{GenA}  is obtained after performing the global residue theorem, where the function
\begin{equation}\label{Ldefinition}
{\cal L}:=
\rho 
\oint_{\rm a-cycle} \left[q^\mu + \frac{1}{2}\sum_{a=1}^n \frac{k_a^\mu}{z-\s_a}(y_a+y)\right]^2{dz \over y}, ~~ {\rm with}~~ \frac{1}{\rho}:=\int_{\rm a-cycle} \frac{dz}{y},
\end{equation}
becomes to be part of the integrand, so it does not define  a integration cycle anymore\footnote{As it  was shown in \cite{Cardona:2016bpi}, the ${\cal L}$ function  becomes to be the square loop momentum  after integration over the moduli $\lambda$.}. The integration contour, $\Gamma^1$,  is defined by the $2n$ equations\footnote{Let us remember that the integration around the poles, $\l=\{0,1,\infty\}$, is the same. Therefore, it is enough just to integrate around $\l=0$. } \cite{Cardona:2016bpi}
\begin{equation}\label{cont}
\l=0,~~    C_a=0, ~a=1,\ldots,n,~~  E_i^1=0,~ i=1,\ldots, n-1,
\end{equation}
where the $C_a$'s  are the constraints on the punctures over the Elliptic curve,
\begin{equation}\label{Ccurves1}
C_a=y_a^2-\s_a(\s_a-1)(\s_a-\l),
\end{equation}
and the $E_i^1$'s are the {\bf Elliptic scattering equations} defined as
\begin{equation}\label{esa}
 E^1_a:={q\cdot k_a \over y_a}+ \half\sum_{b=1\atop b\ne a}^n \left({y_b\over y_a}+1 \right) {k_a\cdot k_b \over \s_a-\s_b}=0,
\qquad a\in \{1,2,...,n \},
\end{equation}
which are the  genus, $g=1$,  generalization of the tree level scattering equations \eqref{SEtree}.
The Faddeev Popov determinant, $\Delta_{\rm FP}^2(n)=(y_n)^2$, is given by fixing\footnote{$\s_n$ is a constant  such that $\s_n\neq \{0,1,\infty\}$. Note that $\{0,1,\infty\}$ are the branch points.} the $\s_n$ puncture and the $E_n^1$ scattering equation.

The $d^Dq$ measure  is the integration over the freedom to add a global holomorphic form, where $q^\mu$ is related, by a shift, with the physical loop-momentum. 

Finally, the $H(\s,y)$ function defines the theory at one-loop that one would be considering and it is the main topic of our discussion in this paper. 

Let us then move now to the simplest possible $H(\s,y)$ at one-loop, namely $H(\s,y)={\it constant}$.\footnote{In this paper we are going to consider only functions $H(\s,y)$ analytic in $y$ variable.}

\subsection{The {\rm n-gon} and its CHY-Graph}\label{CHYGNGON}

In the particular case when $H(\s,y)=1$,  the integral in \eqref{GenA}  becomes 
\begin{equation}\label{Angon}
{\rm A}^{\rm n-gon}_n(1,\ldots,n) =  \int d^Dq  \int_{\Gamma^1} \frac{d\l}{\l(1-\l)}\times \left(\prod_{a=1}^{n}\, {dy_a \over C_a}\right)\times \left(\prod_{i=1}^{n-1} \, {d\s_i \over E^1_i}\right)  \left( {\Delta^2_{\rm FP}(n) \over {\cal L}\,\prod_{b=1}^n y_b} \right).
\end{equation}
It was shown in \cite{Cardona:2016bpi} that this integral, in fact, corresponds to the ${\rm n-gon}$. 

Performing the integration  over the $\l$ variable, i.e $\l=0$,  the integral in \eqref{Angon}  can be written as a tree level amplitude in the double cover prescription  \cite{Gomez:2016bmv,Cardona:2016bpi}, such as in \eqref{treeAmplitude}
\begin{equation}
{\rm A}^{\rm n-gon}_n(1,\ldots,n) =  \int {d^D\ell  \over \ell^2}\,\, {\cal I}_{\rm n-gon}^{\rm t}(1,\ldots,n|-\ell,\ell),
\end{equation}
where the loop momentum $\ell^\mu$ is defined as a shift of $q^\mu$,
\begin{equation}\label{loopM}
\ell^\mu :=(-{\rm I})
\left(q^\mu-\frac{1}{2}\sum_{b=1}^n y_b^{\rm T}\, k^\mu_b\right),\qquad {\rm with}~~ {\rm I}:=\sqrt{-1},
\end{equation}
the $1/{\cal L}$ becomes
\begin{equation}\label{Lmomentum}
{\cal L}\Big|_{\l=0}=\rho \oint _{|z|=\epsilon}\left[q^\mu+\half \sum_{a=1}^n\frac{k_a^\mu}{z-\s_a}(y_a+y)
\right]^2_{\l=0}\frac{dz}{y}=-\ell ^2,
\end{equation}
and the ${\cal I}_n^{\rm t}(1,\ldots,n|-\ell,\ell)$ integrand is read as
\begin{align}\label{HyperSmatrixForwardtau}
&{\cal I}_{\rm n-gon}^{\rm t}(1,\ldots,n|-\ell,\ell) \\
&=\int_{\Gamma^t} \left( \prod_{a=1}^n
{ y^{\rm t}_a\,d y^{\rm t}_a \over C^{\rm t}_a}\right)
\left(
\prod_{i=1}^{n-1}\,{d\s_i \over E^{ \rm t}_i}\right)\Delta^2_{\rm FP}(n,n+1,n+2)\times
 {\prod_{a=1}^n(a:n+1)^{\rm t}(a:n+2)^{\rm t} \over [(n+1:n+2)^{\rm t}]^{(n-2)}}\nonumber,
\end{align}
with $(\s_{n+1},y^{\rm t}_{n+1}):=(\s_{\ell},y^{\rm t}_{\ell})=(0,{\rm I})$, $(\s_{n+2},y^{\rm t}_{n+2}):=(\s_{-\ell},y^{\rm t}_{-\ell})=(0,-{\rm I})$, $k_{n+1}^\mu:=\ell^\mu$ and $k_{n+2}^\mu:=-\ell^\mu$. 

\begin{figure}[h]
  \centering
  \includegraphics[width=1.5in]{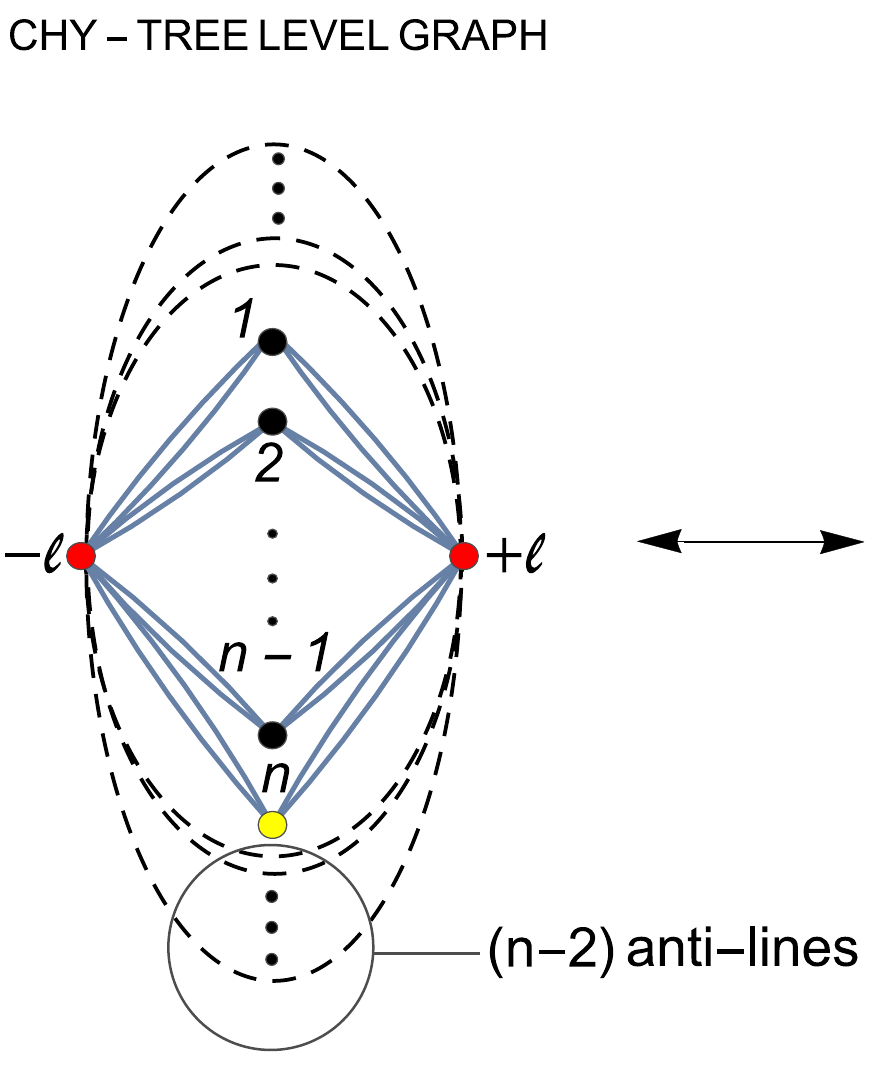}
    \includegraphics[width=1.9in]{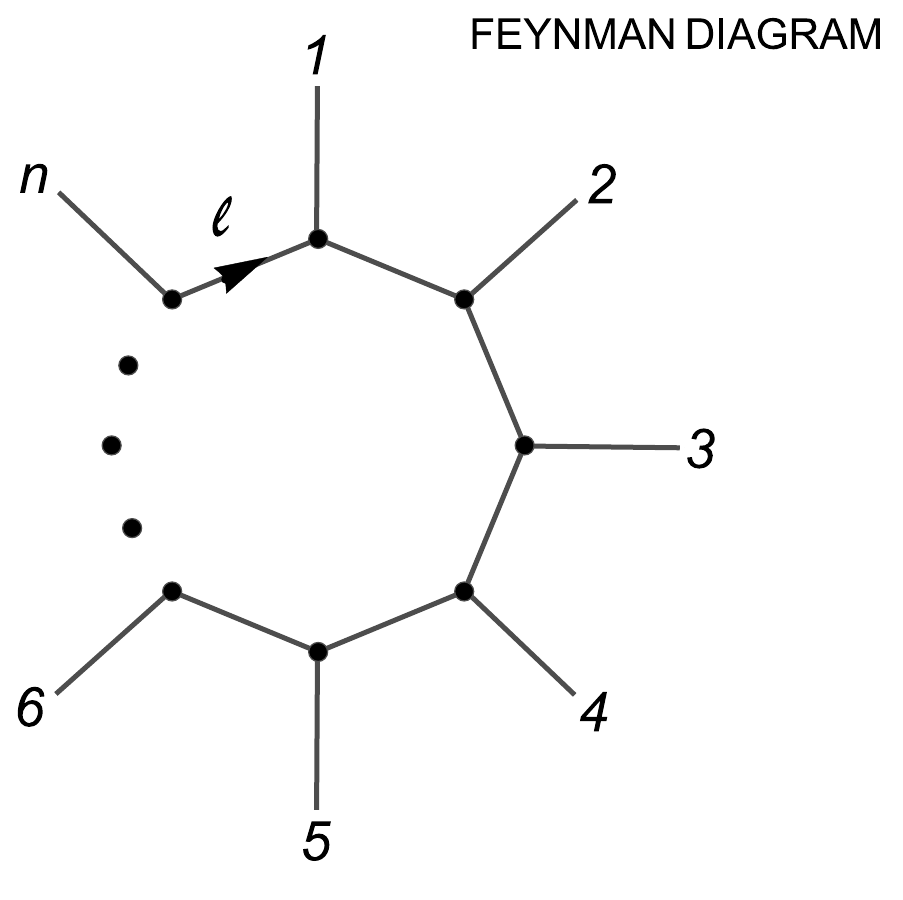}\\
  \caption{{\sl n-gon} representation. (1) CHY-graph on a sphere (up to $\ell^2$ overall factor). (2) Feynman diagram.}\label{CHY-ngon}
\end{figure}

In figure \ref{CHY-ngon}, we have drawn the CHY-graph on a sphere (tree-level) that represents the ${\cal I}_{\rm n-gon}^{\rm t}(1,\ldots,n|\ell,-\ell)$ integrand, where the yellow vertex denotes the puncture fixed by the gauge symmetry on a Torus and the red vertices denotes fixed punctures such that $\ell^2\neq 0$ (for details on the color code see figure \ref{color_cod}). A natural question arises, {\sl  What is a n-gon CHY-graph on a Torus?}   

We will give an answer to this question in next section.

\subsubsection{CHY-Graph on a Torus}\label{CHY-graph-ngon}

Before giving a general graph interpretation on a Torus, let us consider again the expression in \eqref{Angon}, which can be rewritten as
\begin{equation}\label{Angon2}
{\rm A}^{\rm n-gon}_n=  \int d^Dq \int_{\Gamma^{1}} \frac{d\l}{\l(1-\l)}\times \left(\prod_{a=1}^{n}\, {y_a\,dy_a \over C_a}\right)\times \left(\prod_{i=1}^{n-1} \, {d\s_i \over E^1_i}\right)  \Delta^2_{\rm FP}(n) \times \left(\frac{1}{{\cal L}\,\prod_{b=1}^n y^2_b}\right).
\end{equation}
Comparing \eqref{Angon2} with the tree level double cover prescription in \eqref{treeAmplitude}, one can read the last term as an integrand for the n-gon, i.e.
\begin{equation}
{\cal I}_{\rm n-gon}^1(1,\ldots,n)=\frac{1}{{\cal L}}\left(\frac{1}{ y_1\,y_2\,\cdots \,y_n}\right)\times \left(\frac{1}{ y_1\,y_2\,\cdots \,y_n}\right).
\end{equation} 
The $1/{\cal L}$ factor, which comes from one the scattering equations,  is just interpreted as the propagator, $1/\ell^2$, by \eqref{Lmomentum}.

Now the question is, how to interpret the $\prod_{b=1}^n y^{-2}_b$ factor? 

Let us remember that the CHY-graph on the sphere in figure \ref{CHY-ngon} was obtained performing the integral around $\l=0$ in \eqref{Angon}, i.e.  pinching the  a-cycle. After this procedure two new off-shell punctures arises on different sheets, which are conected by anti-lines. 

\begin{figure}[h]
  \centering
    \includegraphics[width=2.2in]{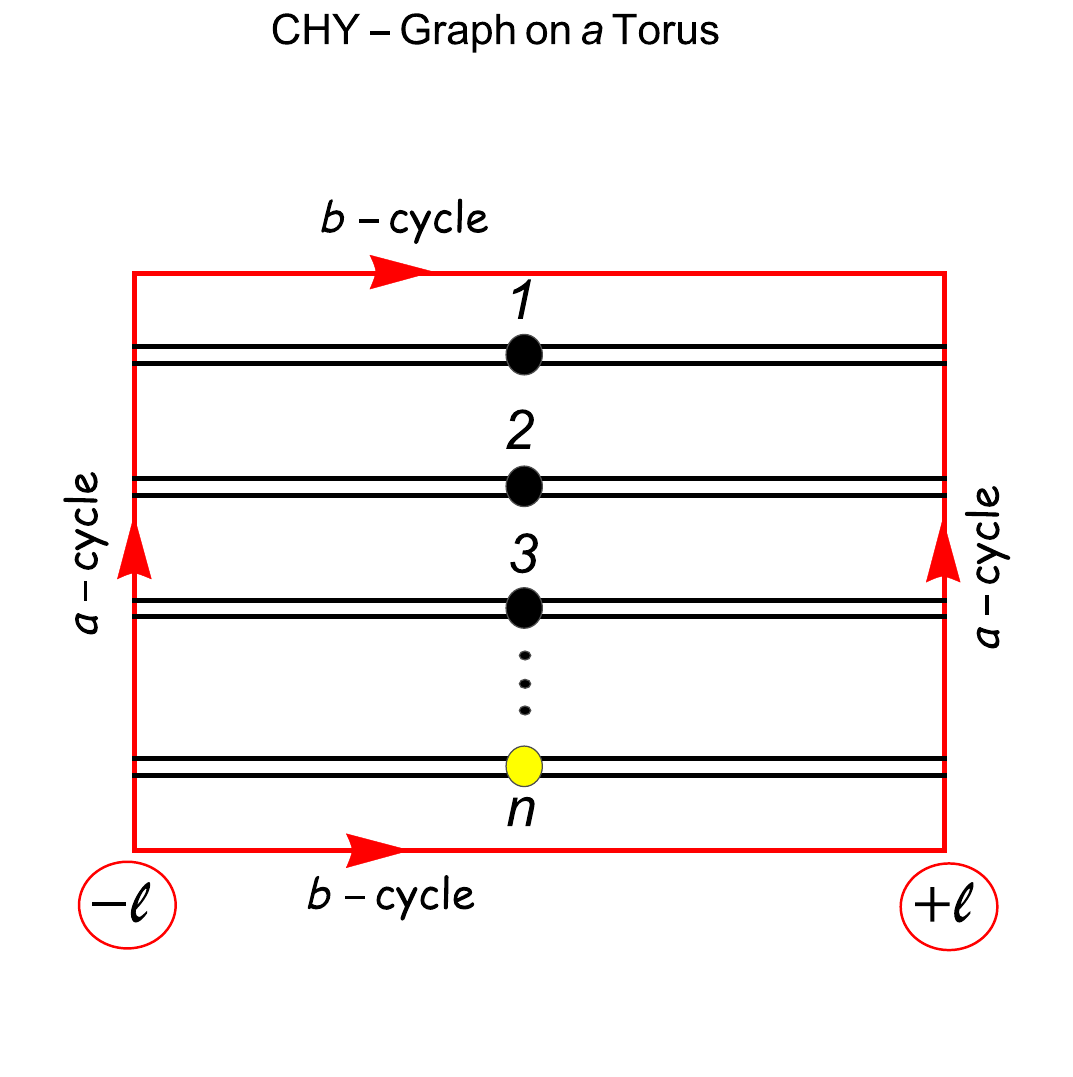}
       \includegraphics[width=1.9in]{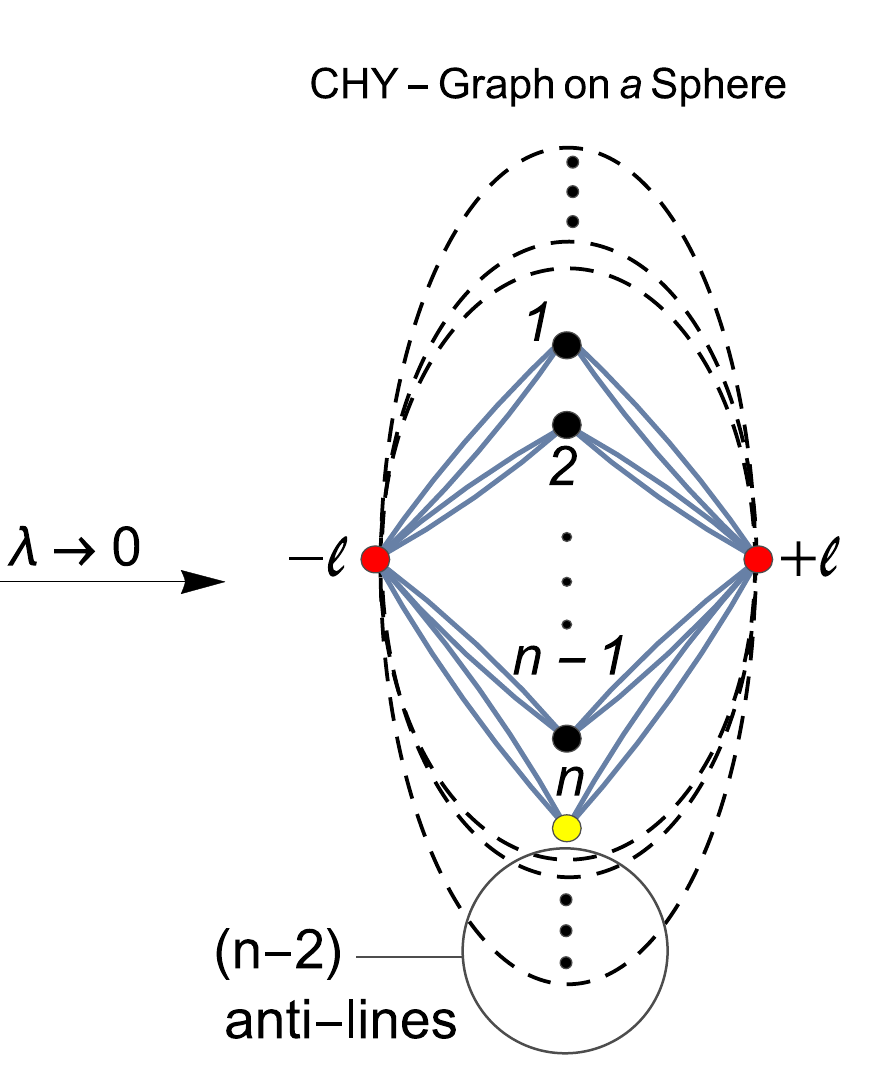} \\
  \caption{CHY Torus representation for the n-gon. After pinching the {\rm a-cycle} one obtains the CHY-tree level graph. One can think that the anti-lines among $\s_\ell$ and $\s_{-\ell}$ arose in order to obtain $PSL(2,\mathbb{C})$ invariance on the CHY-tree level graph.}\label{CHY-1loop-ngon}
\end{figure}

The whole process is shown graphically in figure \ref{CHY-1loop-ngon}, where the red rectangle represents a Torus. At this moment, we are able  to give the following interpretation to the $1/y_a$ factor
\begin{equation}\label{Haa}
\boxed{
\begin{matrix}
H_{a:a} :=\frac{1}{y_a}\\
\text{\small Loop around b-cycle connecting the $\s_a$ puncture. }
\end{matrix}
}
\end{equation}

Clearly, this kind of mathematical object does not have an analog at tree level or on $\M_{0,n}$.  The reason is because on the sphere there is not a non-trivial homological cycle. Hence, unlike to the tree level scattering amplitudes, where all possibles  CHY-graphs are regular graphs,  here on a Torus, the self-connections (loops) are allowed.   

Let us remind what happens with $H_{a:a}$ when $\l=0$,
\begin{equation}\label{HL0}
H_{a:a}\Big|_{\l=0}=\frac{1}{\s_a\, y_a^{\rm t}},
\end{equation}  
on the support of the Elliptic curve, $C_a\Big|_{\l=0}=y_a^2-\s_a^2(\s_a-1)=\s_a^2C_a^{\rm t}=0$. As it was shown in \cite{Cardona:2016bpi}, the expression in \eqref{HL0} can be written as
\begin{equation}\label{Htau}
H_{a:a}\Big|_{\l=0}=(2^2\, {\rm I})\tau_{a:n+1}\tau_{n+2:a} = -(2^2\, {\rm I})\tau_{a:n+2}\tau_{n+1:a},
\end{equation}
where $(\s_{n+1},y^{\rm t}_{n+1}):=(\s_{\ell},y^{\rm t}_{\ell})=(0,{\rm I})$, $(\s_{n+2},y^{\rm t}_{n+2}):=(\s_{-\ell},y^{\rm t}_{-\ell})=(0,-{\rm I})$. The $2^2$  extra factor in \eqref{Htau} arises from the connector that links the fixed off-shell punctures,
\begin{equation}
\tau_{n+1:n+2}=\tau_{n+2:n+1}=\frac{1}{2\,y^{\rm t}_{n+1}}\left( \frac{1}{y^{\rm t}_{n+1}-y^{\rm t}_{n+2}}\right)=-\frac{1}{2^2}.
\end{equation}
Therefore, \eqref{HL0} can be read as  
\begin{align}\label{Htautau}
H_{a:a}\Big|_{\l=0}&=(-{\rm I})\frac{\tau_{a:n+1}\tau_{n+2:a}}{\tau_{n+2,n+1}} = ({\rm I})\frac{\tau_{a:n+2}\tau_{n+1:a}}{\tau_{n+1,n+2}}\\
&=(-{\rm I})\frac{(a:n+1:n+2)^{\rm t}}{(n+1:n+2)^{\rm t}}=({\rm I})\frac{(a:n+2:n+1)^{\rm t}}{(n+1:n+2)^{\rm t}}.
\end{align}

\subsection{Generalizing Integrands}\label{General_Integrand}

\begin{figure}[h]
  \centering
    \includegraphics[width=2in]{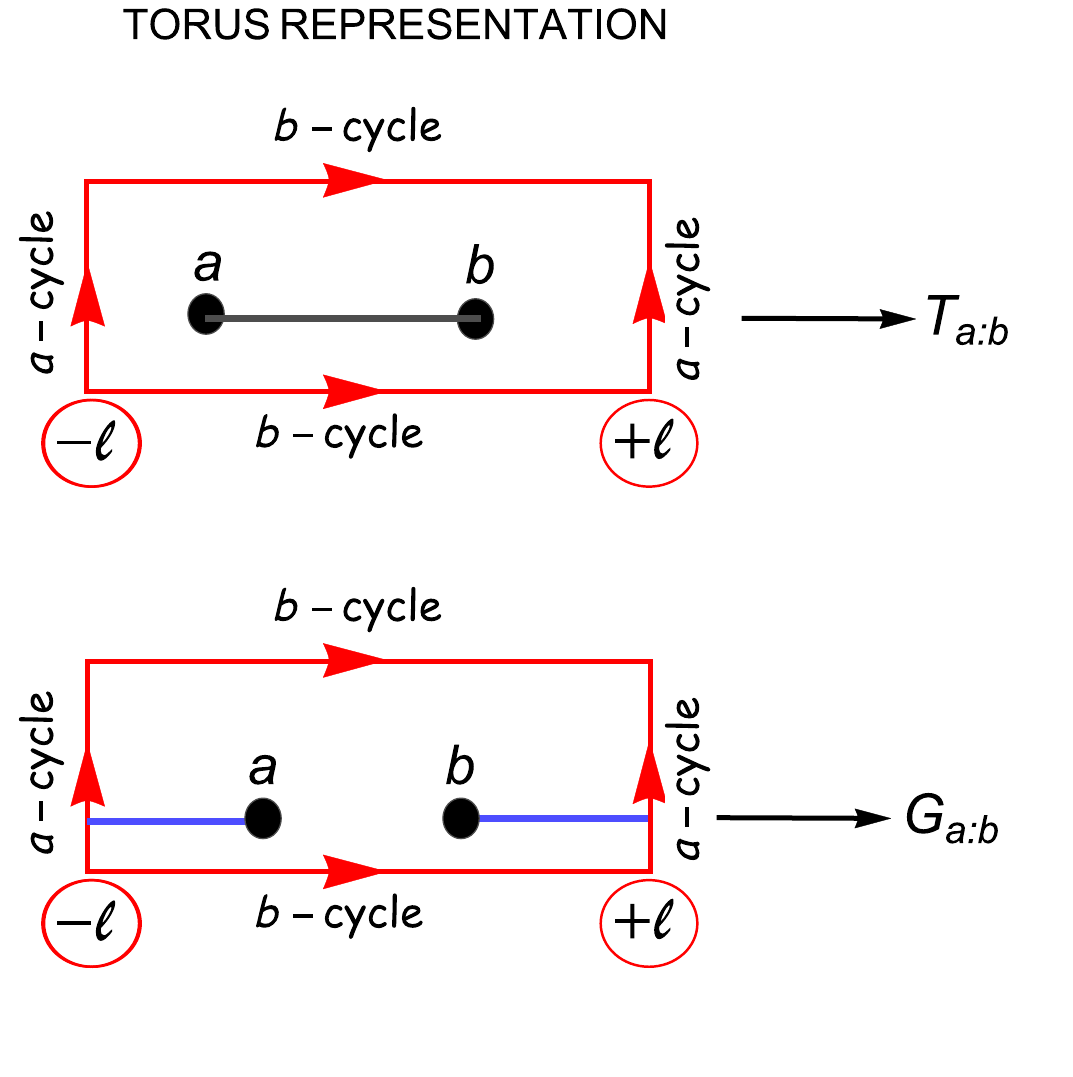}
  \caption{Two different options to connect $\s_a$ with $\s_b$ on a Torus.}\label{TandG}
\end{figure}

In order to construct any integrands on $\M_{1,n}$ (one-loop), we would like to generalize the idea presented in section \ref{CHY-graph-ngon}, where we learned  how to connect a puncture with itself through the form $H_{a:a}$ (connector).  Now, we must build a form connecting two punctures at different locations on a Torus, namely $\s_a$ with $\s_b$.  Nevertheless, at this point we have two options, which are given in figure \ref{TandG}.

The first option, which we have called $T_{a:b}$, is the simplest one. This object has the particularity that the {\sl line} connecting $\s_a$ with $\s_b$ does not go around the b-cycle, similar to tree level, i.e. $\tau_{a:b}$. So, in order to construct the $T_{a:b}$ form we impose the following condition 
\begin{equation}\label{Tcondition}
T_{a:b}\Big|_{\l=0}=\tau_{a:b},
\end{equation}
on the support of $C_a=C_b=0$. Given this constraint we propose the form
\begin{align}\label{Tab}
\boxed{
\begin{matrix}
T_{a:b} :=\frac{1}{2\,y_a}\left( \frac{y_a+y_b}{\s_{ab}}+\frac{y_b}{\s_b}  \right)\\
\text{\small Connecting $\s_a$ with $\s_b$ without encircle the b-cycle.}
\end{matrix}
}
\end{align}
It is straightforward to check that $T_{a:b}$ satisfies the condition in \eqref{Tcondition} on the support of $C_a=C_b=0$. 

\begin{figure}[h]
  \centering
    \includegraphics[width=2in]{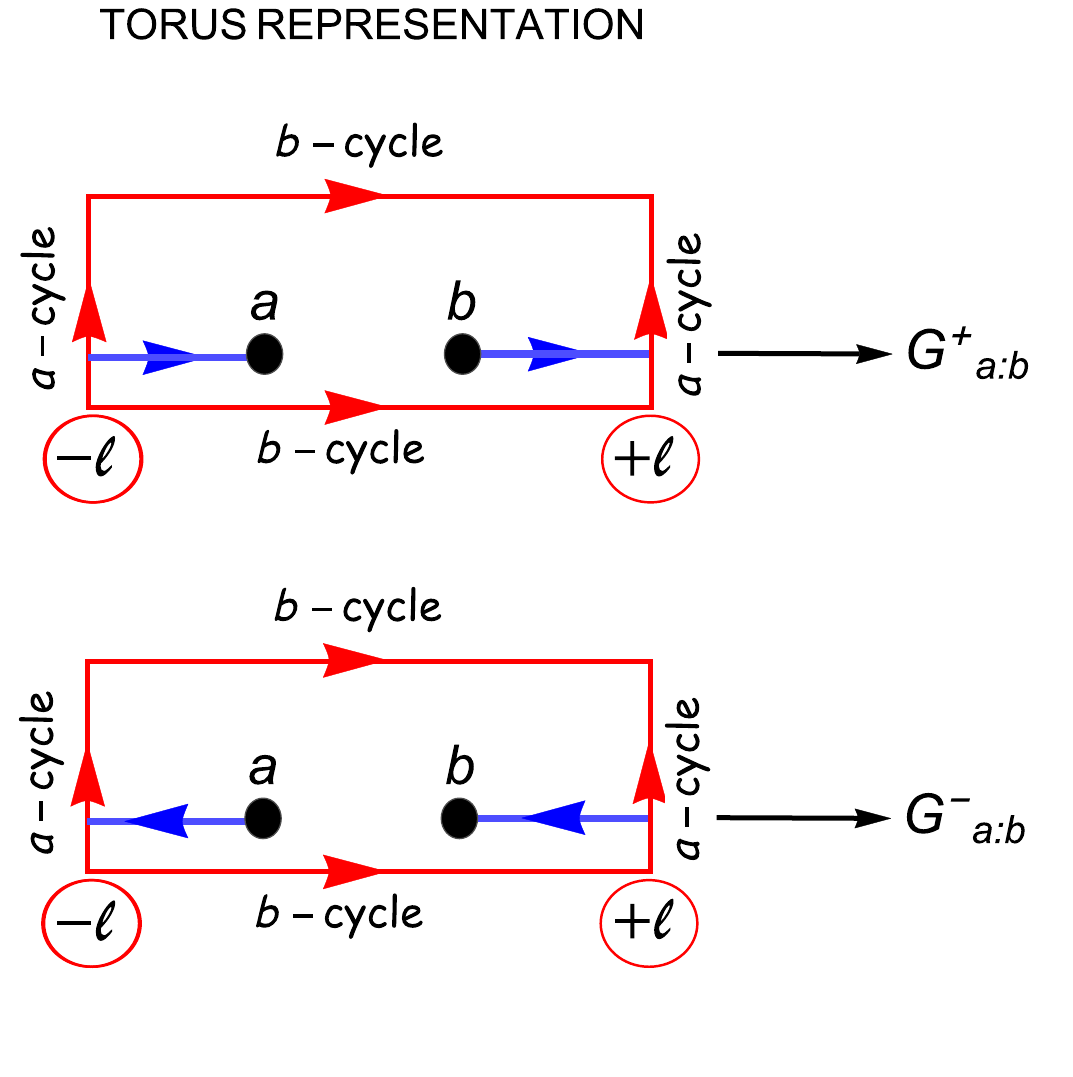}
  \caption{Two different ways to encircle the b-cycle. $G^+_{a:b}$ is on the right hand and $G^-_{a:b}$ is on the left hand.}\label{G+andG-}
\end{figure}

The next step is to construct the $G_{a:b}$ form. Note that there are two possibilities to assemble $G_{a:b}$, or in other words, there are two possible directions to encircle the  b-cycle, which are given in figure \ref{G+andG-}. 
So as to build $G_{a:b}^+$, we impose the following two constraints on the support of $C_a=C_b=0$, 
\begin{align}\label{ConstraintG+}
& G^+_{a:b}\Big|_{\l=0}=(-{\rm I})\frac{\tau_{a:n+1}\, \tau_{n+2:b}}{\tau_{n+2:n+1}}=(-{\rm I})\frac{\tau_{a:n+1}\, \tau_{n+1:n+2}\,   \tau_{n+2:b}}{(n+1:n+2)^{\rm t}},\\
& G^+_{a:a}\Big|_{\l=0} =  H_{a:a}\Big|_{\l=0}\nonumber,
\end{align}
where\footnote{Let us remind that $\tau_{n+1:n+2}=\tau_{n+2:n+1}$.} $(\s_{n+1},y^{\rm t}_{n+1}):=(\s_{\ell},y^{\rm t}_{\ell})=(0,{\rm I})$, $(\s_{n+2},y^{\rm t}_{n+2}):=(\s_{-\ell},y^{\rm t}_{-\ell})=(0,-{\rm I})$. These two conditions are natural from the figure \ref{G+andG-} and the limit in \eqref{Htautau}.  Our proposal for $G^+_{a:b}$ is
\begin{align}\label{G+ab}
\boxed{
\begin{matrix}
G^+_{a:b} :=\frac{1}{\,y_a}\left[ \frac{(y_a+{\rm I}\,\s_a)\,\,(y_b - {\rm I}\,\s_b)}{\s_a\,\s^2_{b}}  \right]\\
\text{\small Connecting $\s_a$ with $\s_b$ and encircling the b-cycle from the right.}
\end{matrix}
}
\end{align}
It is simple to check that $G^+_{a:b}$ in \eqref{G+ab} satisfies \eqref{ConstraintG+}. In a similar way, the conditions for $G^-_{a:b}$ are
\begin{align}\label{ConstraintG-}
& G^-_{a:b}\Big|_{\l=0}=({\rm I})\frac{\tau_{a:n+2}\, \tau_{n+1:b}}{\tau_{n+1:n+2}}=({\rm I})\frac{\tau_{a:n+2}\, \tau_{n+2:n+1}\,  \tau_{n+1:b}}{(n+1:n+2)^{\rm t}},\\
& G^-_{a:a}\Big|_{\l=0} =  H_{a:a}\Big|_{\l=0}\nonumber.
\end{align}
So, our proposal for $G^-_{a:b}$ is
\begin{align}\label{G-ab}
\boxed{
\begin{matrix}
G^-_{a:b} :=\frac{1}{\,y_a}\left[ \frac{(y_a- {\rm I}\,\s_a)\,\,(y_b+ {\rm I}\,\s_b)}{\s_a\,\s^2_{b}}  \right]\\
\text{\small Connecting $\s_a$ with $\s_b$ and encircling the b-cycle from the left.}
\end{matrix}
}
\end{align}
It is straightforward to see that in fact $G^-_{a:b}$ satisfies the constraints  in \eqref{ConstraintG-}.

Although we have been able to build the forms $T_{a:b}$, $G^+_{a:b}$ and $G^-_{a:b}$, which are the {\bf connectors}, we do not know yet whether they are unique or not. 

With this basic ingredients we are now ready to build integrands on $\M_{1,n}$.

\subsubsection{Integrands}\label{Integrands}

The connectors just obtained, $\{ H_{a:a}, T_{a:b}, G^{\pm}_{a:b} \}$, become the building blocks of the integrands on $\M_{1,n}$. Before to proceed on some examples let us do an important remark on them.

For the sake of clarity, let us recall the {\rm n-gon} amplitude,
\begin{equation}\label{Angon3}
{\rm A}^{\rm n-gon}_n=  \int d^Dq \int_{\Gamma^{1}}\frac{d\l}{\l(1-\l)}\times d\mu\,\Delta^2_{\rm FP}(n) \times {\cal I}_{\rm n-gon}^1(1,\ldots,n),
\end{equation}
where $d\mu$ is the analogous measure to the one on $\M_{0,n}$  (tree-level),
\begin{equation}
d\mu:= \left(\prod_{a=1}^{n}\, {y_a\,dy_a \over C_a}\right)\times \left(\prod_{i=1}^{n-1} \, {d\s_i \over E^1_i}\right),  
\end{equation}
and the integrand is given by
\begin{equation}
{\cal I}_{\rm n-gon}^1(1,\ldots,n)=\frac{ \left( H_{1:1} H_{2:2} \cdots H_{n:n} \right)^2 }{{\cal L}}.
\end{equation} 
It is very interesting to note that the {\rm n-gon} can also be written as the following  product of two $n$- chains (Parker-Taylor $\times$ Parker-Taylor)
\begin{equation}\label{PTPTngon}
{\cal I}_{\rm n-gon}^1(1,\ldots,n)=\frac{1 }{{\cal L}}\left( G^+_{1:2} G^-_{2:3}G^+_{3:4} \cdots G^{\pm}_{n-1:n} G^{\mp}_{n:1} \right)\times  \left( G^-_{1:2} G^+_{2:3}G^-_{3:4} \cdots G^{\mp}_{n-1:n} G^{\pm}_{n:1} \right).
\end{equation} 
For example, the box and pentagon are written as
\begin{align}
{\cal I}_{\rm 4-gon}^1(1,2,3,4)&=\frac{1 }{{\cal L}}\left( G^+_{1:2} G^-_{2:3}G^+_{3:4}  G^{-}_{4:1} \right)\times  \left( G^-_{1:2} G^+_{2:3}G^-_{3:4} G^{+}_{4:1} \right),\\
{\cal I}_{\rm 5-gon}^1(1,2,3,4,5)&=\frac{1 }{{\cal L}}\left( G^+_{1:2} G^-_{2:3}G^+_{3:4}  G^{-}_{4:5}G^{+}_{5:1} \right)\times  \left( G^-_{1:2} G^+_{2:3}G^-_{3:4} G^{+}_{4:5} G^{-}_{5:1} \right).\nonumber
\end{align} 
Using the identities given in \eqref{Htautau}, \eqref{ConstraintG+} and \eqref{ConstraintG-}, it is trivial to check
\begin{align}
\left( G^+_{1:2} G^-_{2:3}G^+_{3:4}  G^{-}_{4:1} \right)\times  \left( G^-_{1:2} G^+_{2:3}G^-_{3:4} G^{+}_{4:1} \right)\Big|_{\l=0}&=\left( H_{1:1} H_{2:2} H_{3:3}H_{4:4}\right)^2 \Big|_{\l=0}\\
\left( G^+_{1:2} G^-_{2:3}G^+_{3:4}  G^{-}_{4:5} G^{+}_{5:1} \right)\times  \left( G^-_{1:2} G^+_{2:3}G^-_{3:4} G^{+}_{4:5} G^{-}_{5:1}\right)\Big|_{\l=0}&=\left( H_{1:1} H_{2:2} H_{3:3}H_{4:4}H_{5:5} \right)^2 \Big|_{\l=0}\nonumber.
\end{align}

In \eqref{PTPTngon}  we have written the {\rm n-gon} integrand as a product of two Parke-Taylor, $({\rm Parke}$-${\rm Taylor})^2$, which is a well defined integrand on $\M_{1,n}$. So, from this example,  we can say that a well defined integrand on $\M_{1,n}$ must satisfies the same condition as at tree level, namely,
\\

{\bf Proposition 1}\\
{\it
A well defined integrand on $\M_{1,n}$  is given by product of chains such that the difference among lines and anti-lines on each vertex is always 4,}
\begin{equation}
\#{\rm Lines} - \#{\rm Antilines}=4.
\end{equation} 

Note that $H_{a:a}$ is a self-chain, in fact, it is simple to see 
\begin{equation}
G^+_{a:b}G^-_{b:a}\Big|_{\l=0}=H_{a:a}H_{b:b} \Big|_{\l=0}.
\end{equation}

Now, from the connectors set, $\{ H_{a:a}, T_{a:b}, G^{\pm}_{a:b} \}$, and following the proposition \ref{pro1},  we can build integrands on $\M_{1,n}$. In the next sections we will give some particular examples.

\section{Simple Examples and The $\L$-Algorithm}\label{EXAMPLES}

So as to clarify the ideas presented previously, in this section we shall give some simple examples and perform some explicit computations.

\subsection{A very simple example}\label{simple_example}

\begin{figure}[h]
  \centering
    \includegraphics[width=1.7in]{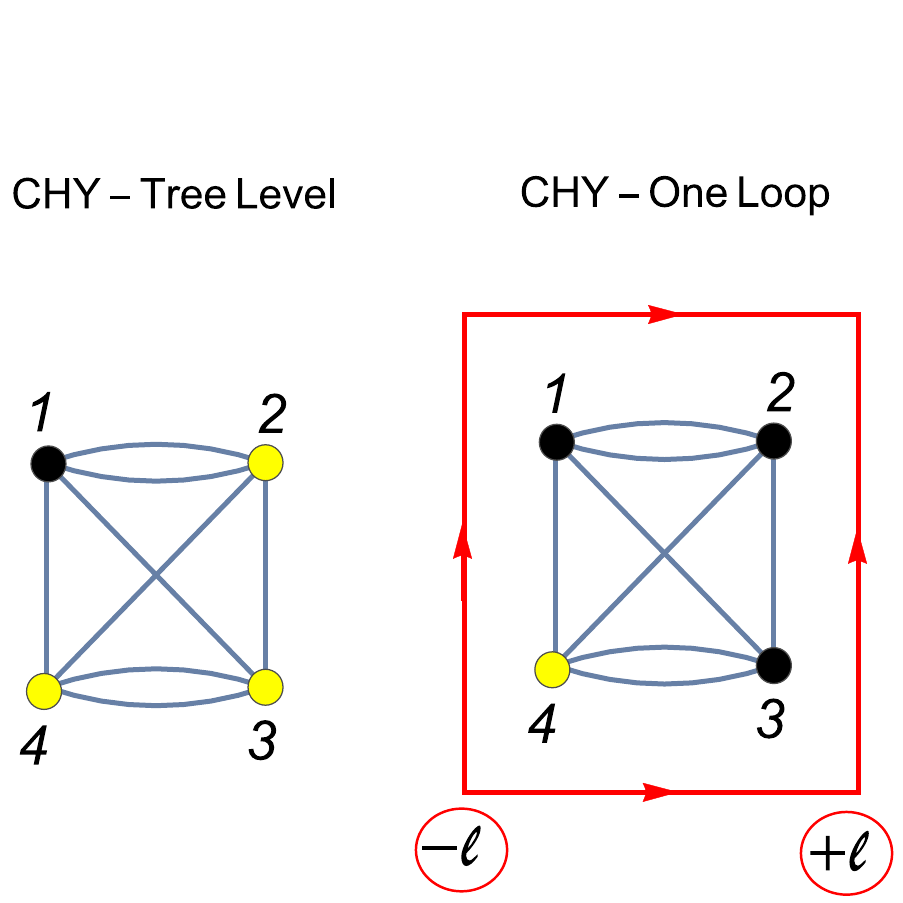}  
  \caption{CHY-graph on a sphere. CHY-graph on a Torus.}\label{CHY_tree_loop_ex1}
\end{figure}

The first simple example that we wish to consider is the following one
\begin{equation}\label{CHY_tree_loop_INT}
\begin{matrix}
{\rm Tree-Level} ~ (\M_{0,n}) & {\rm One-Loop} ~ (\M_{1,n})\\
{\cal I}_4^{\rm t}=(\tau_{1:2}\tau_{2:3}\tau_{3:4}\tau_{4:1})\times (\tau_{1:2}\tau_{2:4}\tau_{4:3}\tau_{3:1}) ~& ~{\cal I}_4^{1}=(T_{1:2}T_{2:3}T_{3:4}T_{4:1})\times (T_{1:2}T_{2:4}T_{4:3}T_{3:1})\,.
\end{matrix}
\end{equation}
Where we have trade the $\tau_{a:b}$'s at tree-level only by $T_{a:b}$'s at loop-level, i.e, none $G^{\pm}$ has been used.
\\
From now on, we will omit the global factor , $1/{\cal L}$, in the one-loop integrands, please bear it in mind. 
The CHY-graphs on a sphere (tree-level) and a Torus (one-loop) corresponding to the integrands in \eqref{CHY_tree_loop_INT}, are given by the left and right drawings at figure \ref{CHY_tree_loop_ex1}, respectively, where the yellow vertices denote fixed punctures.
As it is very well known from \cite{Cachazo:2013iea},  the tree-level Feynman diagram, corresponding to the CHY  integrand  on the left side of \eqref{CHY_tree_loop_INT},  is just $1/s_{12}$. Now, we would like to know what is the result for the CHY integrand on $\M_{1,n}$, which was just obtained replacing the $\tau$'s for $T$ 's, as it is shown in \eqref{CHY_tree_loop_INT} and in figure  \ref{CHY_tree_loop_ex1}. 

\begin{figure}[h]
  \centering
    \includegraphics[width=1.7in]{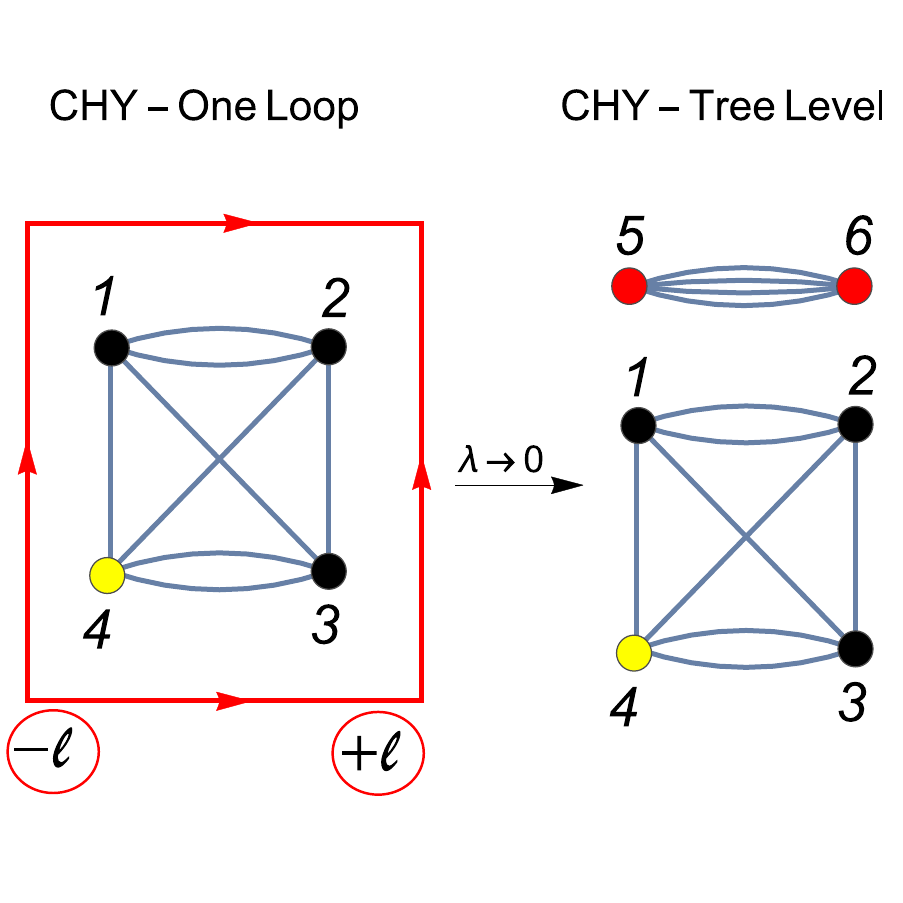}
  \caption{From CHY-graph on a Torus to CHY-graph on a sphere. The yellow vertices mean fixed punctures and the red vertices mean fixed off-shell punctures.}\label{CHY_loop_to_tree}
\end{figure}

In order to solve 
 it, the first step to follow is to integrate the $\l$ variable, i.e  $\l=0$. After integrating  $\l$, it is straightforward to see that the one-loop amplitude prescription becomes
\begin{align}
d^Dq\,d\mu\Big|_{\l=0}&= d^D\ell\,\left(\prod_{a=1}^{4}\, {y^{\rm t}_a\,dy^{\rm t}_a \over C^{\rm t}_a}\right)\times \left(\prod_{i=1}^{3} \, {d\s_i \over E^{\rm t}_i}\right)\, ,\label{Measuretree}  \\
\Delta^2_{\rm FP}(4)\Big|_{\l=0}& =-\left[(5:6)^{\rm t}\right]^2\Delta^2_{\rm FP}(4,5,6)\, ,\label{FPtree}\\
{\cal I}_4^{1}\Big|_{\l=0}&=(1:2:3:4)^{\rm t} \times (1:2:4:3)^{\rm t}={\cal I}_4^{\rm t}\, ,\label{Intree}
\end{align}
where $\ell^\mu$ is as in \eqref{loopM}, $(\s_{5},y^{\rm t}_{5}):=(\s_{\ell},y^{\rm t}_{\ell})=(0,{\rm I})$, $(\s_{6},y^{\rm t}_{6}):=(\s_{-\ell},y^{\rm t}_{-\ell})=(0,-{\rm I})$, $k_{5}^\mu:=\ell^\mu$, $k_{6}^\mu:=-\ell^\mu$ and the scattering equations are reduced to the tree-level ones in the following way
\begin{equation}\label{SEreduction} 
 E^{\rm t}_i=
 \half\sum_{j=1\atop j\ne i}^{4}  {k_i\cdot k_j \over \s_{ij}} \left({y^{\rm t}_j\over y^{\rm t}_i}+1 \right)  + \half   {k_i\cdot k_5 \over \s_{i5}} \left({y^{\rm t}_5\over y^{\rm t}_i}+1 \right)  + \half  {k_i\cdot k_6 \over \s_{i6}}  \left({y^{\rm t}_6\over y^{\rm t}_i}+1 \right),\,\, i=1,2,3.
\end{equation}
From \eqref{Measuretree}, \eqref{FPtree}, \eqref{Intree} and \eqref{SEreduction}, it is clear that after performing the integral over $\l$, we obtain a 6-point tree-level amplitude (on $\M_{0,n}$), with integrand given by
\begin{equation}
{\cal I}^{\rm t}_6(1,2,3,4|5,6)=(1:2:3:4)^{\rm t}(5:6)^{\rm t} \times (1:2:4:3)^{\rm t}(5:6)^{\rm t},
\end{equation}
such as it have been drawn in figure \ref{CHY_loop_to_tree}.

\begin{figure}[h]
  \centering
    \includegraphics[width=1.7in]{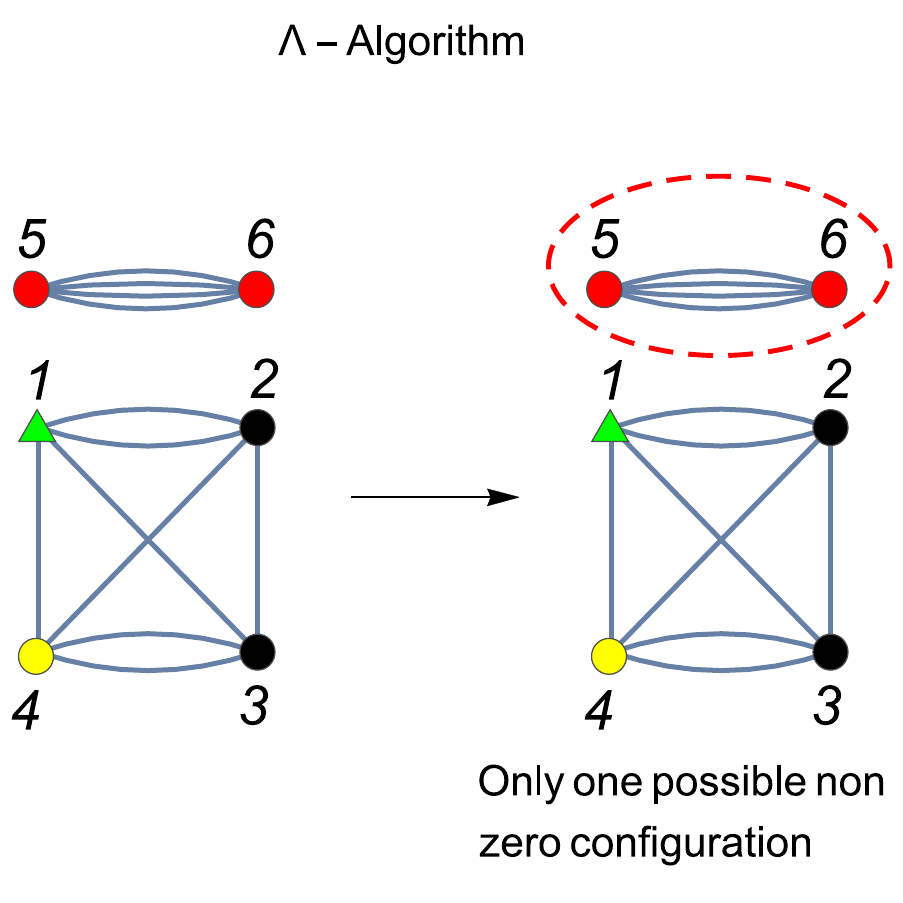}
  \caption{Applying the $\L$-algorithm. There is only one possible non zero configuration.}\label{Lalgorithm_six1}
\end{figure}

Naively, the graph in figure \ref{CHY_loop_to_tree} looks simple to solve,
nevertheless one should be very careful. For example, using the $\L$-algorithm over this graph \cite{Gomez:2016bmv}, where we have fixed the $\s_1$ puncture by the scale invariance (green vertex in figure  \ref{Lalgorithm_six1}), there is only one possible non-zero configuration, as it is shown in figure \ref{Lalgorithm_six1}.

However, from the expressions obtained in \eqref{Measuretree}, \eqref{FPtree}, \eqref{Intree} and \eqref{SEreduction}, the off-shell punctures $(\s_{5},y^{\rm t}_{5}):=(\s_{\ell},y^{\rm t}_{\ell})=(0,{\rm I})$ and  $(\s_{6},y^{\rm t}_{6}):=(\s_{-\ell},y^{\rm t}_{-\ell})=(0,-{\rm I})$, are fixed on different branch cuts and so, the following proposition is immediate
\\

{\bf Proposition 2. The $\L$-Algorithm at One-Loop}\\
{\it
Let $G$ be a CHY-graph on a sphere,  which is coming from a CHY-graph on a Torus, then
the $\L$-algorithm over $G$ has one restriction}
\begin{itemize}
\item  {\it All configurations, where the punctures $\s_\ell$ and $\s_{-\ell}$ are located {\bf alone} on the branch cut, are forbidden.}
\end{itemize}

\begin{figure}[h]
  \centering
    \includegraphics[width=1.7in]{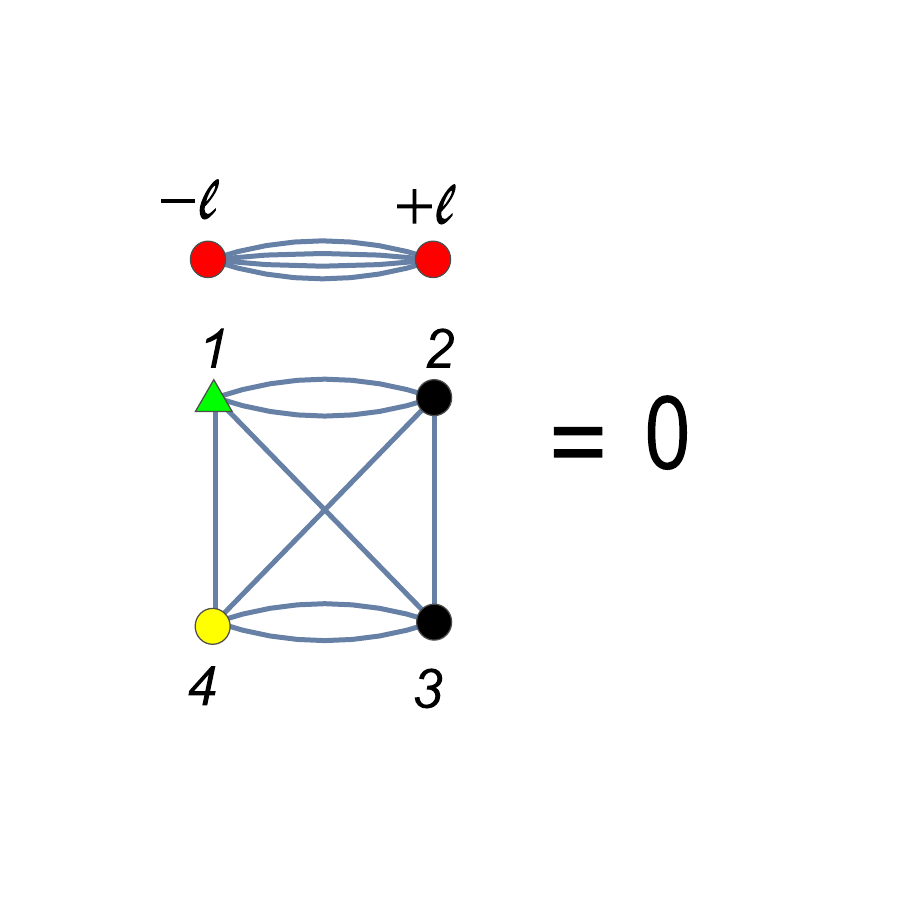}
  \caption{Final result for the integral of $(T_{1:2}T_{2:3}T_{3:4}T_{4:1})\times (T_{1:2}T_{2:4}T_{4:3}T_{3:1})$ .}\label{Sol_ex1}
\end{figure}

Therefore, from the figure \ref{Lalgorithm_six1} and following the {\bf Proposition 2}, one has the integral for the one-loop expression in \eqref{CHY_tree_loop_INT} vanishes (see  figure \ref{Sol_ex1}), i.e.
\begin{equation}
{\cal A}^{1}_4=  \int d^Dq \int_{\Gamma^{1}}\frac{d\l}{\l(1-\l)}\times d\mu\,\,\Delta^2_{\rm FP}(n)\, \frac{(T_{1:2}T_{2:3}T_{3:4}T_{4:1})\times (T_{1:2}T_{2:4}T_{4:3}T_{3:1})}{\cal L}=0.
\end{equation}

Finally, although we solved the CHY-graph on a Torus given in figure \ref{CHY_tree_loop_ex1},  we do not know what is its physical meaning (Feynman diagram).  Nevertheless,  since that the loop momentum, $\ell^\mu$, is not connect to any vertex, figure \ref{CHY_loop_to_tree}, we believe that this amplitude is just a tadpole.

\subsection{A more Interesting example}\label{Example2}

In this section we  compute another simple yet non-trivial example.

\begin{figure}[h]
  \centering
    \includegraphics[width=2in]{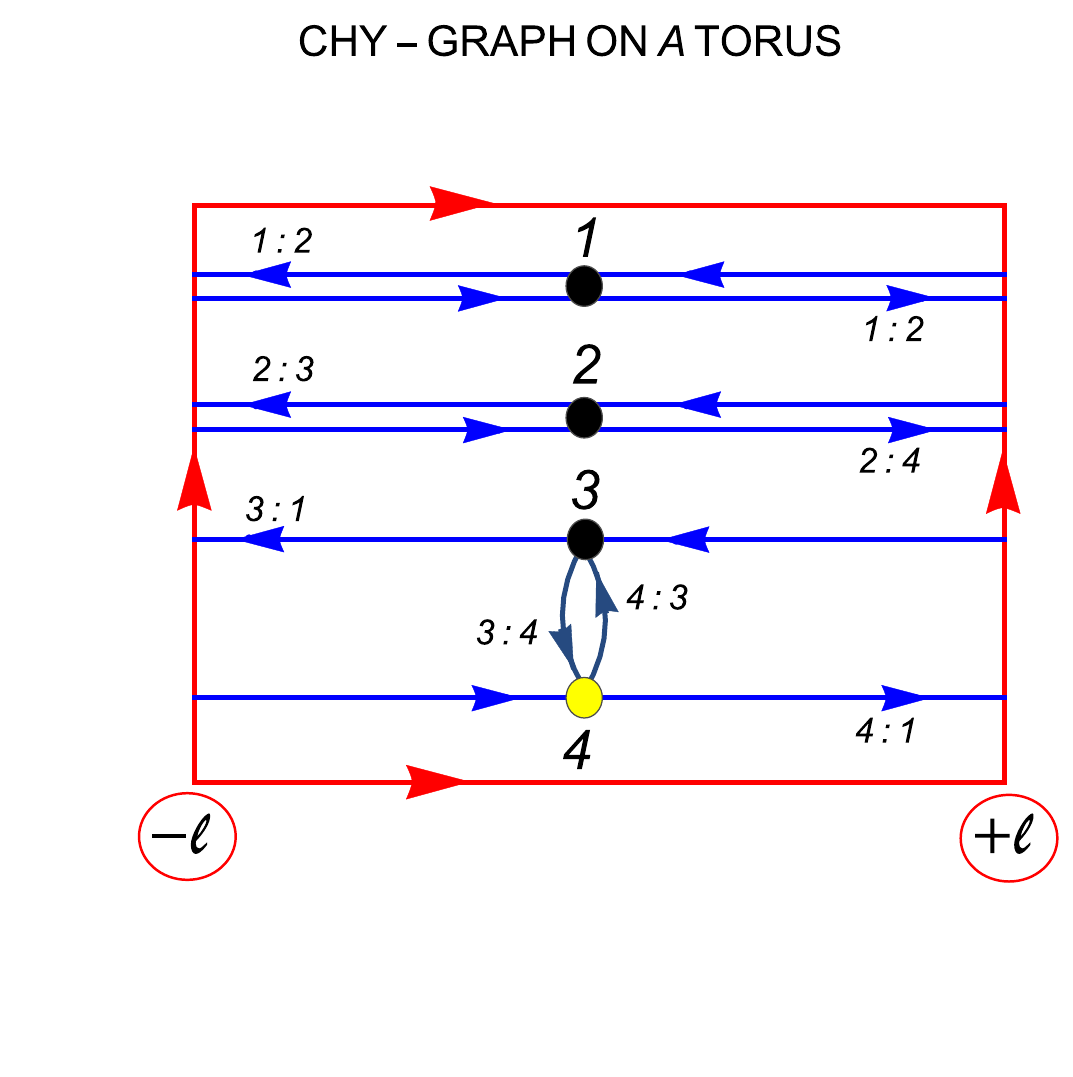}
  \caption{CHY-Graph on a Torus.  The blue lines and labels, $a:b$, mean $G^{\pm}_{a:b}$. $G^+_{a:b}$ enters in the right hand and $G^-_{a:b}$ enters in the left hand.}\label{CHY_loop_ex2}
\end{figure}

Let us consider the same tree level integrand as in section \ref{simple_example}, i.e. ${\cal I}_4^{\rm t}=(1:2:3:4)^{\rm t}\times (1:2:4:3)^{\rm t}$. Similarly as in the previous section,  from this $\ph^3$ tree level integrand we construct an integrand on $\M_{1,n}$, but now, using the $G^+_{a:b}$ and $G_{a:b}^-$ connectors.  Let ${\cal I}_{4}^1(1,2,3,4)$ be the integrand on $\M_{1,n}$ given by 
\begin{equation}\label{example2}
{\cal I}_{4}^1(1,2,3,4)=(G_{1:2}^+G_{2:3}^-T_{3:4}G_{4:1}^+) \times (G_{1:2}^-G_{2:4}^+T_{4:3}G_{3:1}^-).
\end{equation}
Its CHY-graph is represented in figure \ref{CHY_loop_ex2}. 
\\
Integrating the $\l$ variable,  the one loop integrand in \eqref{example2} becomes
\begin{align}
{\cal I}_{4}^1(1,2,3,4)\Big|_{\l=0}&=(G_{1:2}^+G_{2:3}^-T_{3:4}G_{4:1}^+) \times (G_{1:2}^-G_{2:4}^+T_{4:3}G_{3:1}^-)\Big|_{\l=0}\nonumber\\
&=\frac{(1:5:6)^{\rm t} (2:6)^{\rm t} (3:4:5)^{\rm t} \times (1:6:5)^{\rm t} (2:5)^{\rm t} (3:6:4)^{\rm t}  }{[(5:6)^{\rm t}]^4},
\end{align}
with  $(\s_{5},y^{\rm t}_{5}):=(\s_{\ell},y^{\rm t}_{\ell})=(0,{\rm I})$, $(\s_{6},y^{\rm t}_{6}):=(\s_{-\ell},y^{\rm t}_{-\ell})=(0,-{\rm I})$, $k_{5}^\mu:=\ell^\mu$, $k_{6}^\mu:=-\ell^\mu$. This integrand together with the  Faddeev-Popov determinant, 
$$
\Delta^2_{\rm FP}(4)\Big|_{\l=0}=-\left[(5:6)^{\rm t}\right]^2\Delta^2_{\rm FP}(4,5,6),
$$
results in a new six-point tree level integrand,
\begin{align}\label{integrand_tree_2}
{\cal I}^{\rm 3}_6(1,2|3,4|5,6)=\frac{(1:5:6)^{\rm t} (2:6)^{\rm t} (3:4:5)^{\rm t} \times (1:6:5)^{\rm t} (2:5)^{\rm t} (3:6:4)^{\rm t}  }{[(5:6)^{\rm t}]^2},
\end{align}
where the upper index ``3" is due to the similarity with the ${\rm 3-gon}$ graph. It is not a coincidence and in the next section we will explain it. We must now compute a six-point tree level amplitude with integrand given by \eqref{integrand_tree_2}.

\begin{figure}[h]
  \centering
    \includegraphics[width=2in]{CHY_loop_ex2-eps-converted-to.pdf}
      \includegraphics[width=1.7in]{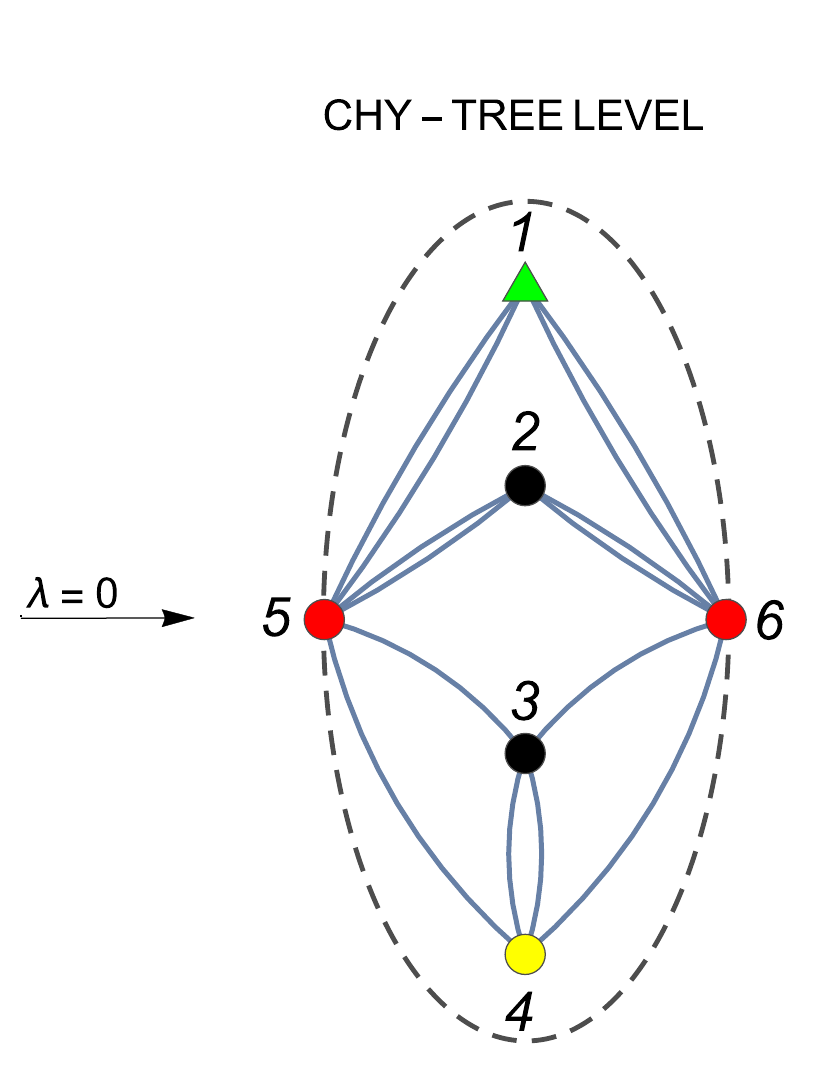}
  \caption{From CHY-Graph on a Torus to CHY-Graph on a sphere. Here $(\s_{5},y^{\rm t}_{5}):=(\s_{\ell},y^{\rm t}_{\ell})=(0,{\rm I})$, $(\s_{6},y^{\rm t}_{6}):=(\s_{-\ell},y^{\rm t}_{-\ell})=(0,-{\rm I})$, $k_{5}^\mu:=\ell^\mu$ and $k_{6}^\mu:=-\ell^\mu$.}\label{CHY_loop_tree_ex2}
\end{figure}

In figure \ref{CHY_loop_tree_ex2}, we represent  the CHY-tree level graph of this six-point integrand, where we have fixed the $\s_1$ puncture by scale symmetry in order to use the $\L$-algorithm \cite{Gomez:2016bmv}.

So as to find the answer of this six point CHY-graph,  the next step is to apply the $\L$-algorithm over it.

\subsubsection{The $\L$-Algorithm }\label{Lalg}


Before applying the $\L$-algorithm over the six-point graph in figure \ref{CHY_loop_tree_ex2}, it is useful to introduce the following  notation
\begin{align}
&k_{a_1\ldots a_m}:=\sum_{a_i<a_j}^m k_{a_i}\cdot k_{a_j},\\
& [a_1,a_2,\ldots, a_m]=k_{a_1}+k_{a_2}+\cdots + k_{a_m}.
\end{align}

\begin{figure}[h]
  \centering
    \includegraphics[width=1.6in]{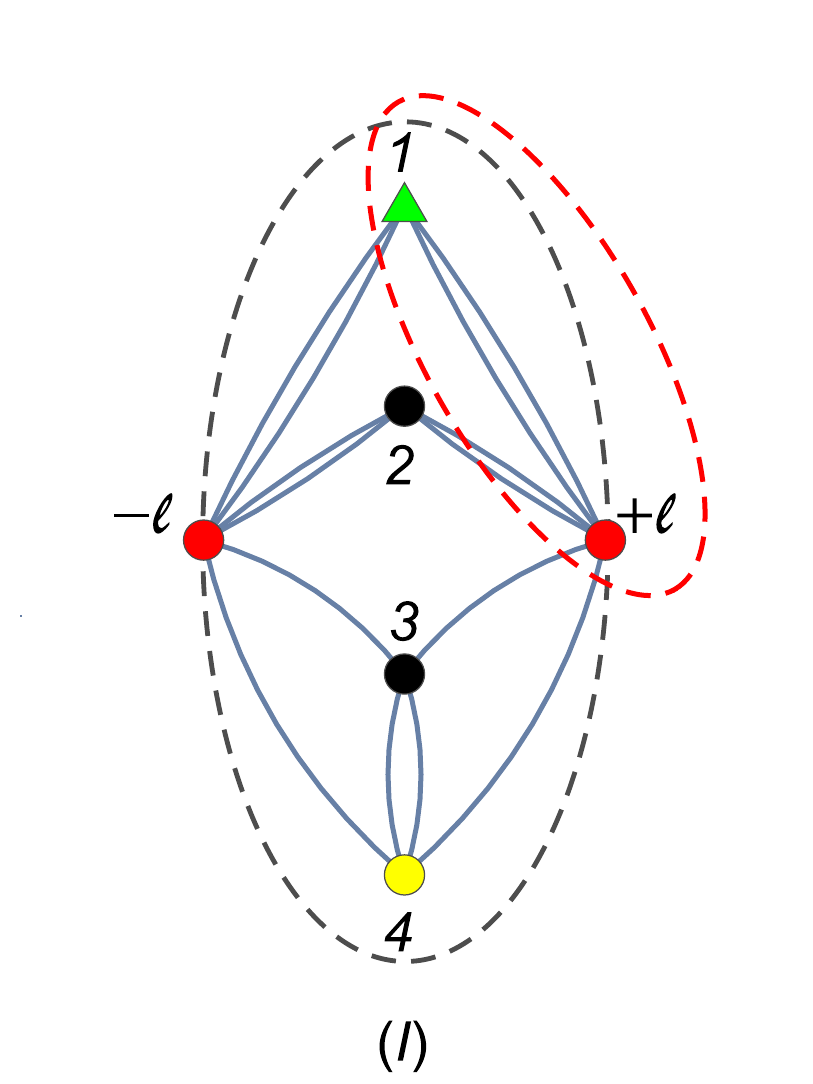}\qquad\qquad\qquad
      \includegraphics[width=1.6in]{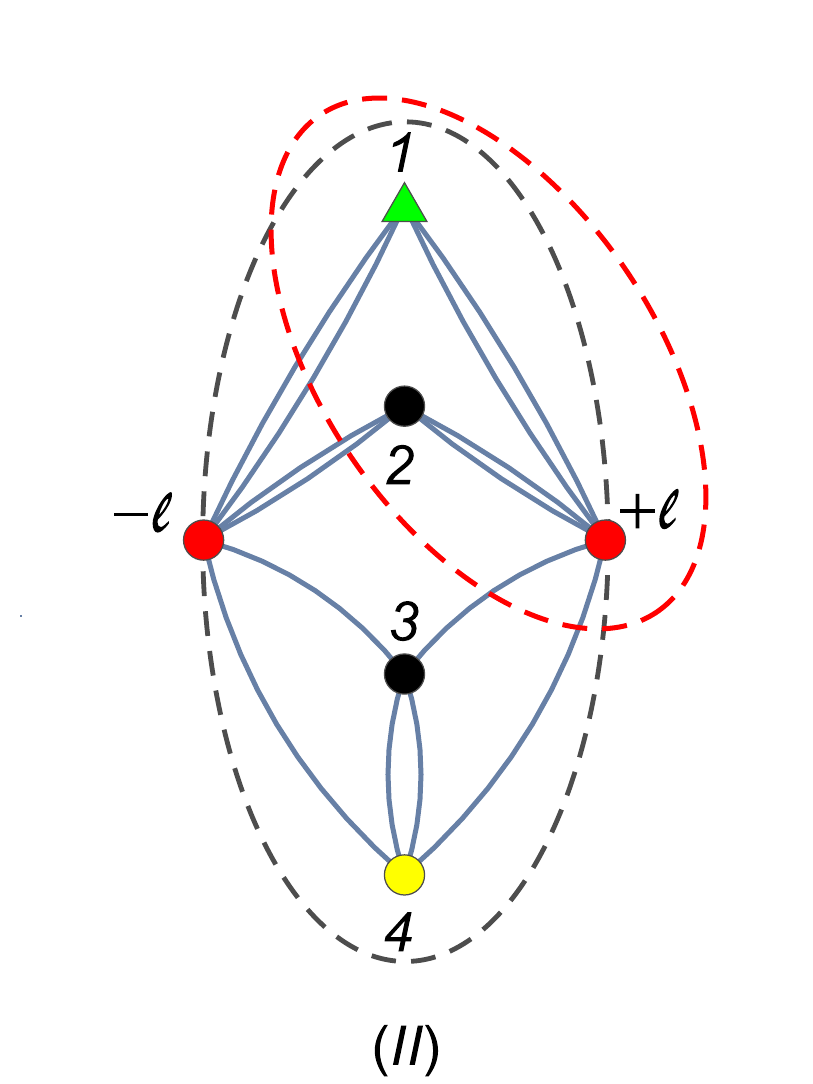}
  \caption{All non-zero allowable configurations, up to $\ell\,  \leftrightarrow \, -\ell$ symmetry.}\label{ex2_configurations}
\end{figure}

Now we are ready to use the $\L$-algorithm (for more details, please see \cite{Gomez:2016bmv}).
The six-point graph in figure \ref{CHY_loop_tree_ex2} has only two non-zero allowable configurations, up to $\ell \,  \leftrightarrow \, -\ell$, which  are drawn in figure \ref{ex2_configurations}.

These two configurations are straightforward to carry out, and their results are given by the expressions 
\begin{align}
(I)=\frac{{\cal I}_5^2(2|3,4|[1,\ell],-\ell)}{k_{\ell 1}},\qquad
(II)=\frac{{\cal I}^2_4(2,1|\ell,[3,4,-\ell])}{k_{\ell 12}}\times {\cal I}_{4}^{\rm t}(3,4,[1,2,\ell],-\ell),
\end{align}
where ${\cal I}^2_5(a|b,c|i,j)$, ${\cal I}^2_4(a,b|i,j)$ and ${\cal I}_4^{\rm t}(a,b,c,d)$ are read in figure \ref{I_5(a,b,c,i,j)} (the upper index ``2" means ${\rm 2-gon}$, \cite{Cardona:2016bpi})
\begin{figure}[h]
  \centering
    \includegraphics[width=1.9in]{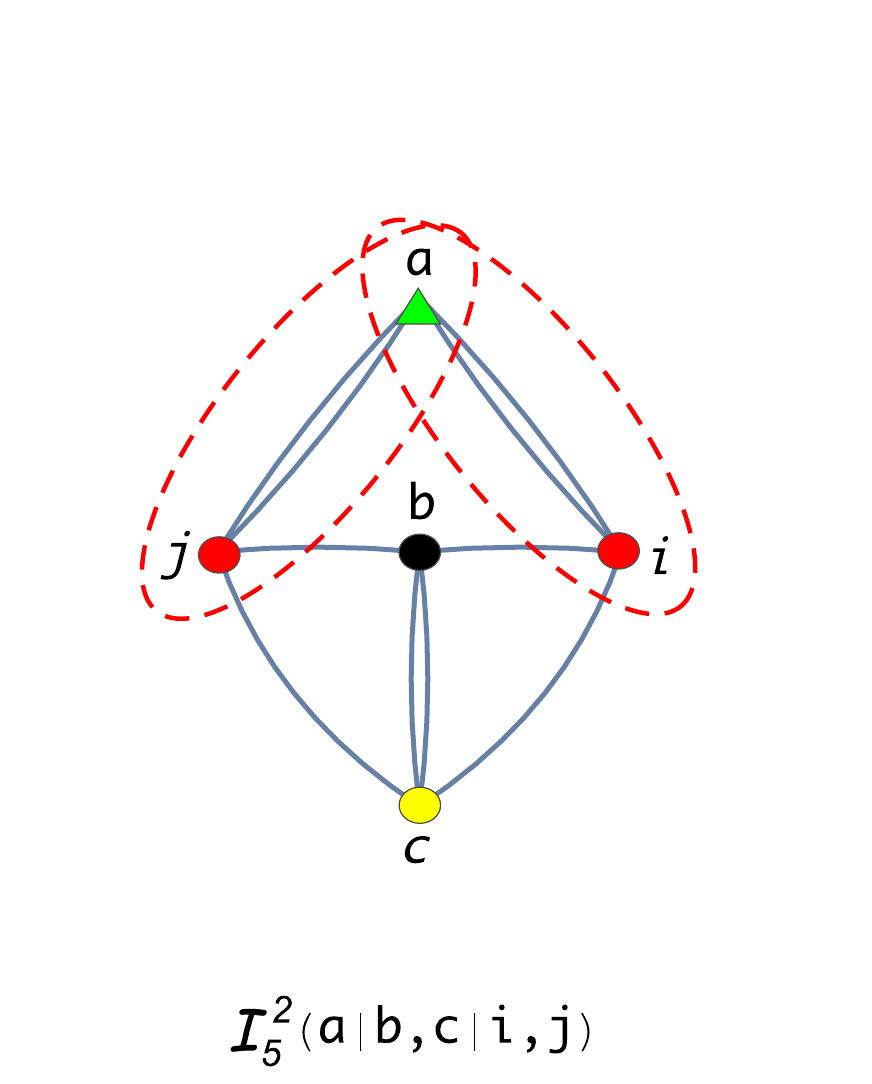}
    \includegraphics[width=1.9in]{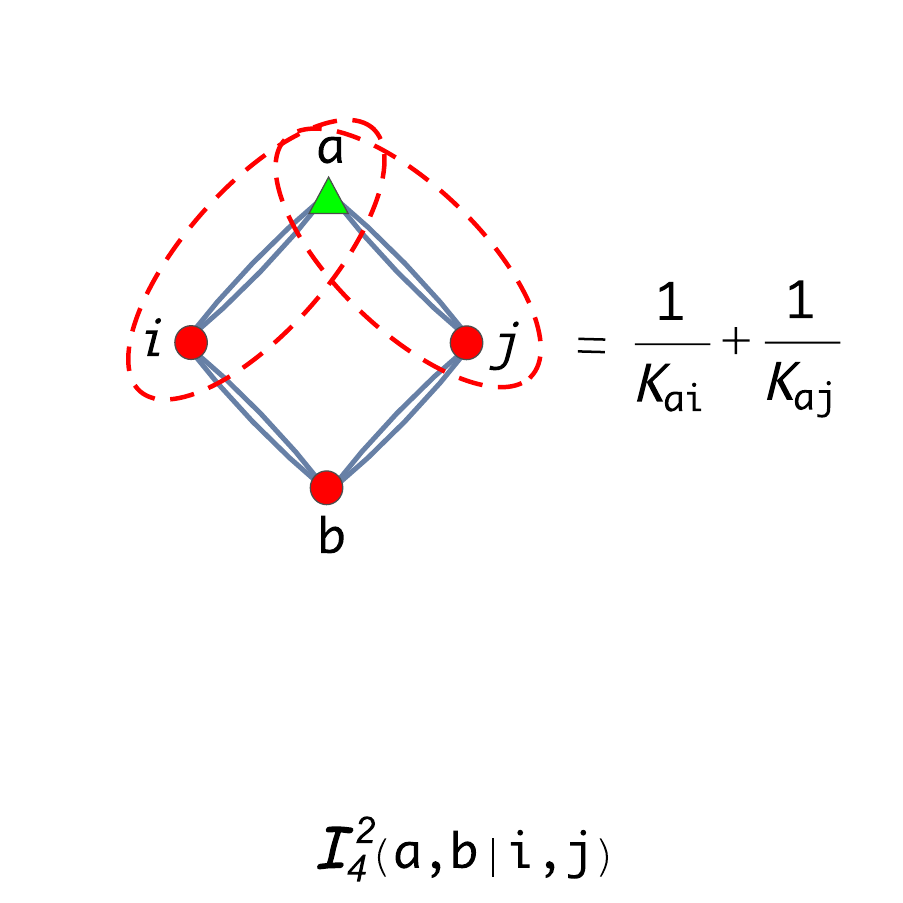}
    \includegraphics[width=1.9in]{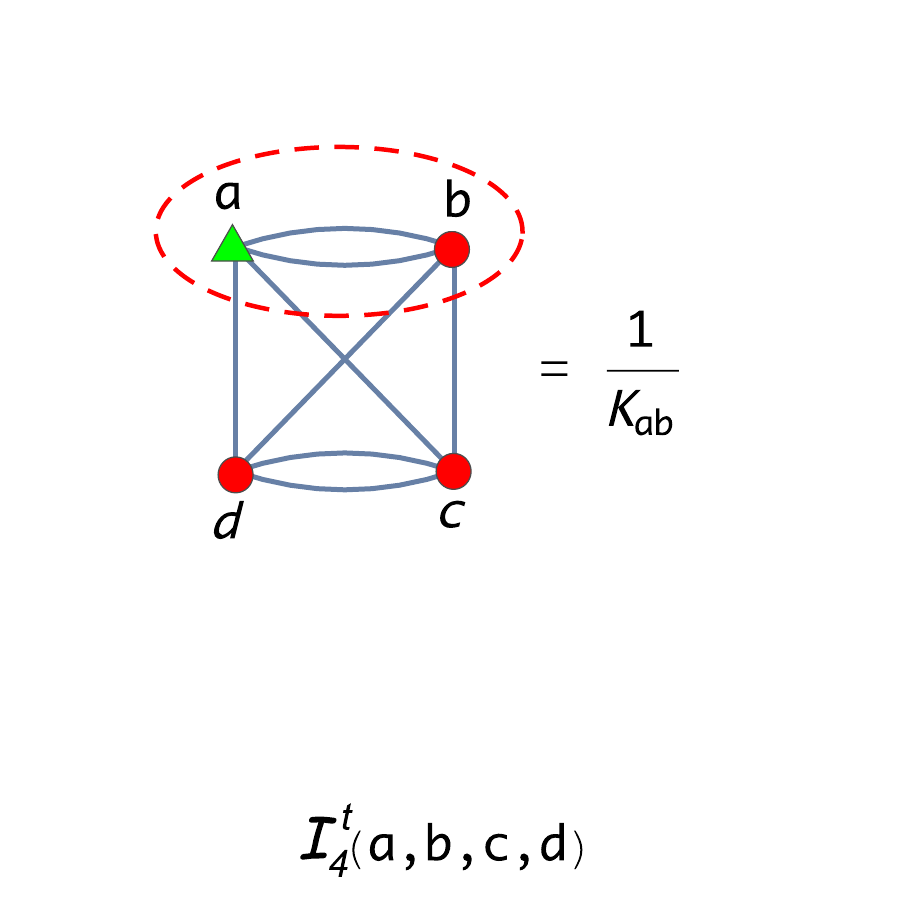}
  \caption{CHY-graphs of ${\cal I}^2_5(a|b,c|i,j)$, ${\cal I}^2_4(a,b|i,j)$, ${\cal I}_4^{\rm t}(a,b,c,d)$ and their allowable configurations.}\label{I_5(a,b,c,i,j)}
\end{figure}

\begin{figure}[h]
  \centering
    \includegraphics[width=1.8in]{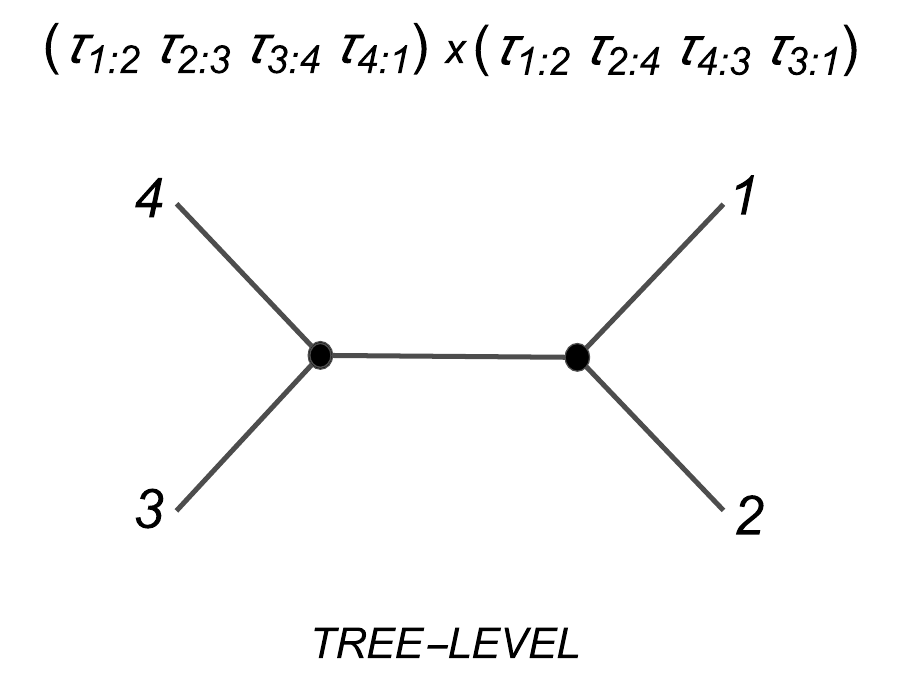}~~~~~~~~\qquad\quad
        \includegraphics[width=1.7in]{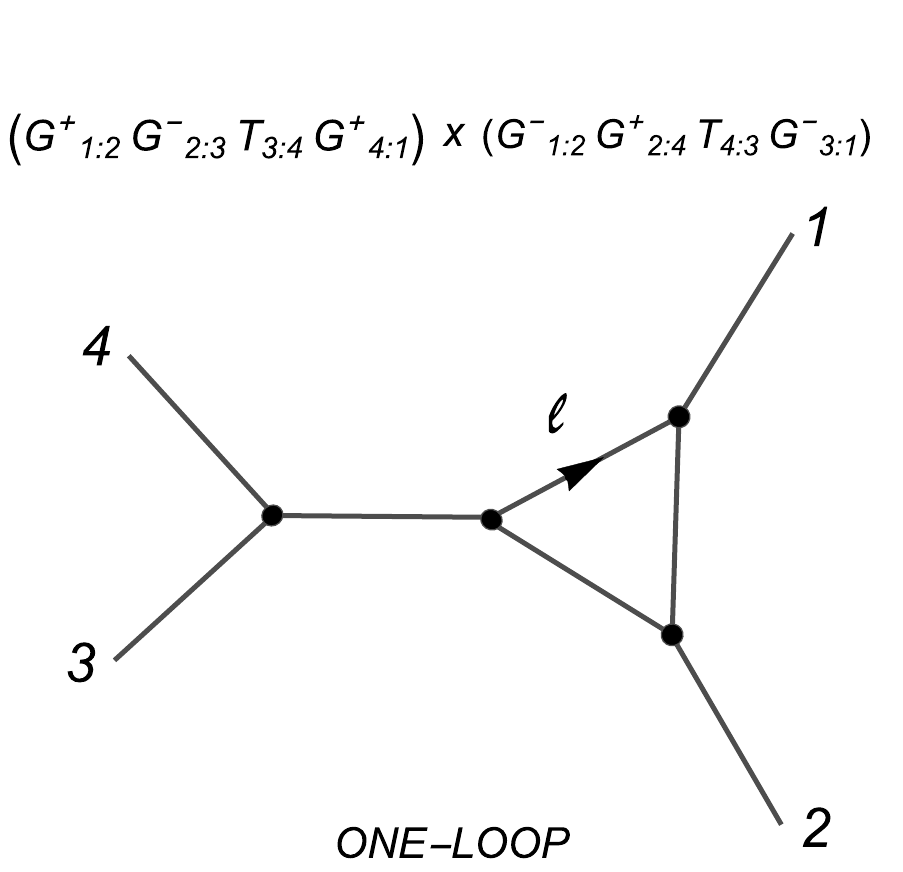}
  \caption{From tree-level to one -loop integrand.}\label{FEYNMAN_ex2}
\end{figure}

For the CHY-graph, ${\cal I}^2_5(a|b,c|i,j)$ we get rapidly,
\begin{equation}
{\cal I}^2_5(a|b,c|i,j)=\frac{{\cal I}_{4}^{\rm t}(b,c,[a,i],j)}{k_{ai}}+\frac{{\cal I}_{4}^{\rm t}(b,c,[a,j],i)}{k_{aj}}=\frac{1}{k_{bc}}\left(\frac{1}{k_{ai}}+\frac{1}{k_{aj}}\right).
\end{equation} 
Therefore, using the above result along with the expressions given in figure \ref{I_5(a,b,c,i,j)} the final result for ${\cal I}^3_6(1,2|3,4|5,6)$ is given by
\begin{align}\label{RI52}
{\cal I}^3_6(1,2|3,4|\ell,-\ell) & =(I)+(II)+ ( \ell \,   \leftrightarrow       \, -\ell)\\
&  =\frac{1}{\ell^2}\left\{
\frac{{\cal I}_5^2(2|3,4|[1,\ell],-\ell)}{k_{\ell 1}}  + \frac{{\cal I}^2_4(2,1|\ell,[3,4,-\ell])   \times {\cal I}_{4}^{\rm t}(3,4,[1,2,\ell],-\ell)  }{ k_{\ell 12} } \right. \nonumber\\
&~~~~~\quad+\left.
\frac{{\cal I}_5^2(2|3,4|[1,-\ell],\ell)}{k_{-\ell 1}}  + \frac{{\cal I}^2_4(2,1|-\ell,[3,4,\ell])   \times {\cal I}_{4}^{\rm t}(3,4,[1,2,-\ell],\ell)  }{k_{-\ell12}}  
\right\} 
\nonumber \\
&=\frac{1}{\ell^2\,k_{34}} 
\left\{ 
\frac{1}{k_{\ell 1}} \left(  
\frac{1}{k_{-\ell,2}} + \frac{1}{k_{\ell,2}+k_{12}}
 \right) 
+ 
 \frac{1}{k_{\ell 12}} \left(  
\frac{1}{k_{\ell,2}} + \frac{1}{k_{-\ell,2}+k_{23}+k_{24}}
 \right) 
 \right. \nonumber\\
&~~~~~\quad+
\left. 
\frac{1}{k_{-\ell 1}} \left(  
\frac{1}{k_{\ell,2}} + \frac{1}{k_{-\ell,2}+k_{12}}
 \right) 
+ 
 \frac{1}{k_{-\ell 12}} \left(  
\frac{1}{k_{-\ell,2}} + \frac{1}{k_{\ell,2}+k_{23}+k_{24}}
 \right) 
 \right\} \nonumber  ,
\end{align}
where we have introduced the overall factor, $1/\ell^2$, which is coming from the term, $1/{\cal L}$. 
In order to interpret \eqref{RI52}, let us consider the one-loop Feynman diagram in the right hand side of figure \ref{FEYNMAN_ex2}.  
After using the partial fractions identity \cite{Feynman:1963},
\begin{equation}\label{partialfrac}
{1\over \prod_{i=1}^n D_i}=\sum_{i=1}^{n}{1\over D_i\prod_{j\neq i}(D_j-Di)}\, ,
\end{equation}
the one-loop Feynman diagram in figure \ref{FEYNMAN_ex2} becomes
\begin{equation}\label{Ftriangle}
2^2\,{\cal I}_{\rm Feynman}=\frac{1}{\ell^2\,\,k_{34}}\sum_{\s \in P_3} \frac{1}{\,\, k_{\ell\s_1}\,\, k_{\ell\s_1\s_2}},
\end{equation}
where $P_3$ is the permutation group defined as 
\begin{align}
P_3:&=\{ \{a_1,a_2,a_3\} , \{a_2,a_3,a_1\}, \{a_3,a_1,a_2\}, \{a_2,a_1,a_3\}, \{a_1,a_3,a_2\}, \{a_3,a_2,a_1\}\},\\
& ~~{\rm with}~~a_1=1,a_2=2,a_3=34\nonumber ,
\end{align}
for example, $k_{\ell a_2 a_3}=k_{\ell234}$.
 
The expression found in \eqref{RI52} is exactly the same as that given in \eqref{Ftriangle}. So, one can say in fact that the integral on $\M_{1,n}$ with integrand 
\begin{equation}\label{integrandTORUS}
{\cal I}_{4}^1(1,2,3,4)=\frac{1}{\cal L}\left\{(G_{1:2}^+G_{2:3}^-T_{3:4}G_{4:1}^+) \times (G_{1:2}^-G_{2:4}^+T_{4:3}G_{3:1}^-)\right\},
\end{equation}
(see figures \ref{CHY_loop_ex2} and \ref{CHY_loop_tree_ex2}), it is just the one-loop $\ph^3$ amplitude given by the Feynman diagram at the right side of figure \ref{FEYNMAN_ex2}.

This is an encouraging result and in the next sections we will elaborate more on it.

\section{$\ph^3$ Theory at One-Loop}\label{phi3loop}\label{PHI3theory}

The connectors set, $\{H_{a:a}, T_{a:b}, G^{\pm}_{a:b}   \}$, are our main ingredients in order to construct integrands on $\M_{1,n}$. With them, we will be able to build a wide range of integrands. 

In this section we give a systematic way to obtain the $\ph^3$ one-loop Feynman diagrams from integrals on $\M_{1,n}$, and conversely we also provide a set rules to obtain the integral on $\M_{1,n}$ corresponding to a given Feynman diagram.

\subsection{From CHY-Integrands to  One-Loop $\ph^3$-Feynman Diagrams.}

Previously, in section \ref{Example2}, we have given a simple but illustrative example, where we have begun with a $\ph^3$ integrand on $\M_{0,n}$ and, replacing the $\tau_{a:b}$'s connectors on a sphere by connectors $T_{a:b}$'s and $G^{\pm}_{a:b}$'s, we have obtained a  $\ph^3$ integrand on $\M_{1,n}$, it is explicitly shown in figure \ref{FEYNMAN_ex2}. In this subsection we would like to consider the general construction.

As it is very well known from \cite{Cachazo:2013iea}, that the integral of a product of two Parker-Taylor ($\rm PT$) factors over $\M_{0,n}$, is just the sum over  $\ph^3$ (Bi-adjoint) tree-level Feynman diagrams. Conversely, for any $\ph^3$ (Bi-adjoint) tree-level Feynman diagram, there is at least a product of two Parker-Taylor factors, such that its integral over $\M_{0,n}$ is exactly that $\ph^3$ amplitude \cite{Baadsgaard:2015ifa}.

Following this idea, let us consider integrals made from products of two Parker-Taylor factors, but now on $\M_{1,n}$. 

\begin{figure}[h]
  \centering
    \includegraphics[width=1.3in]{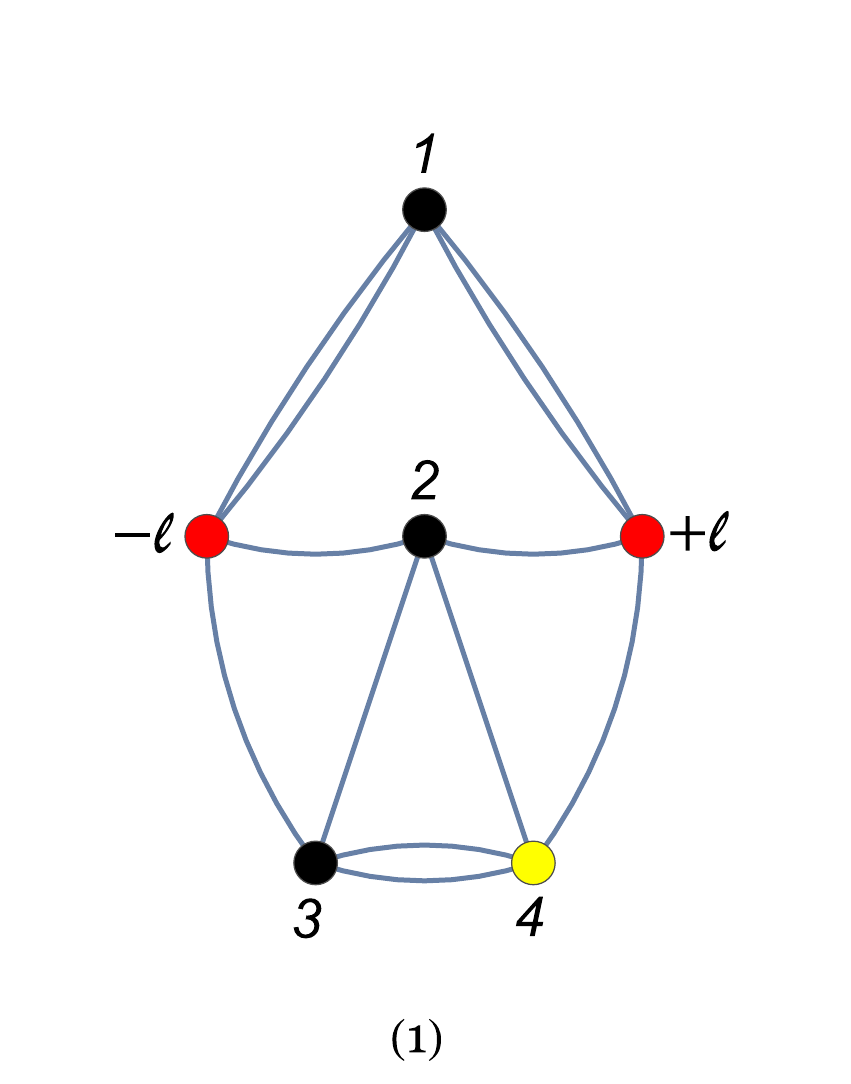}
      \includegraphics[width=1.3in]{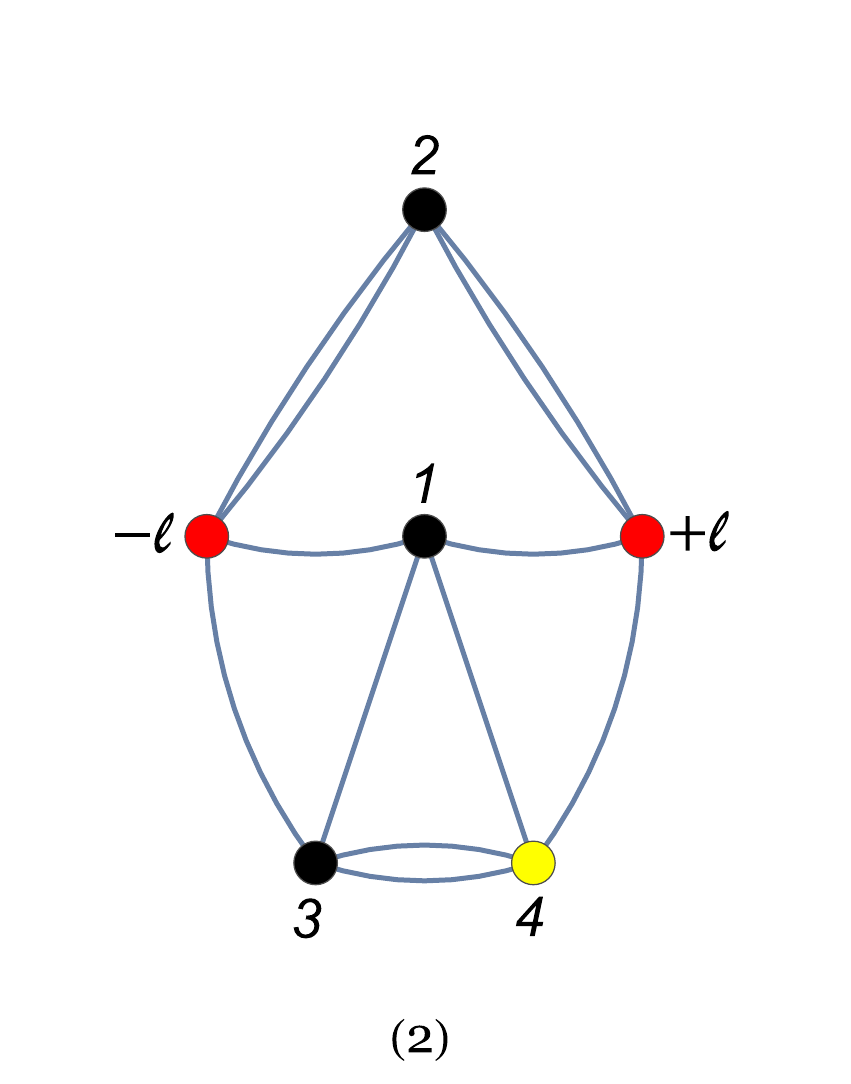}
        \includegraphics[width=1.3in]{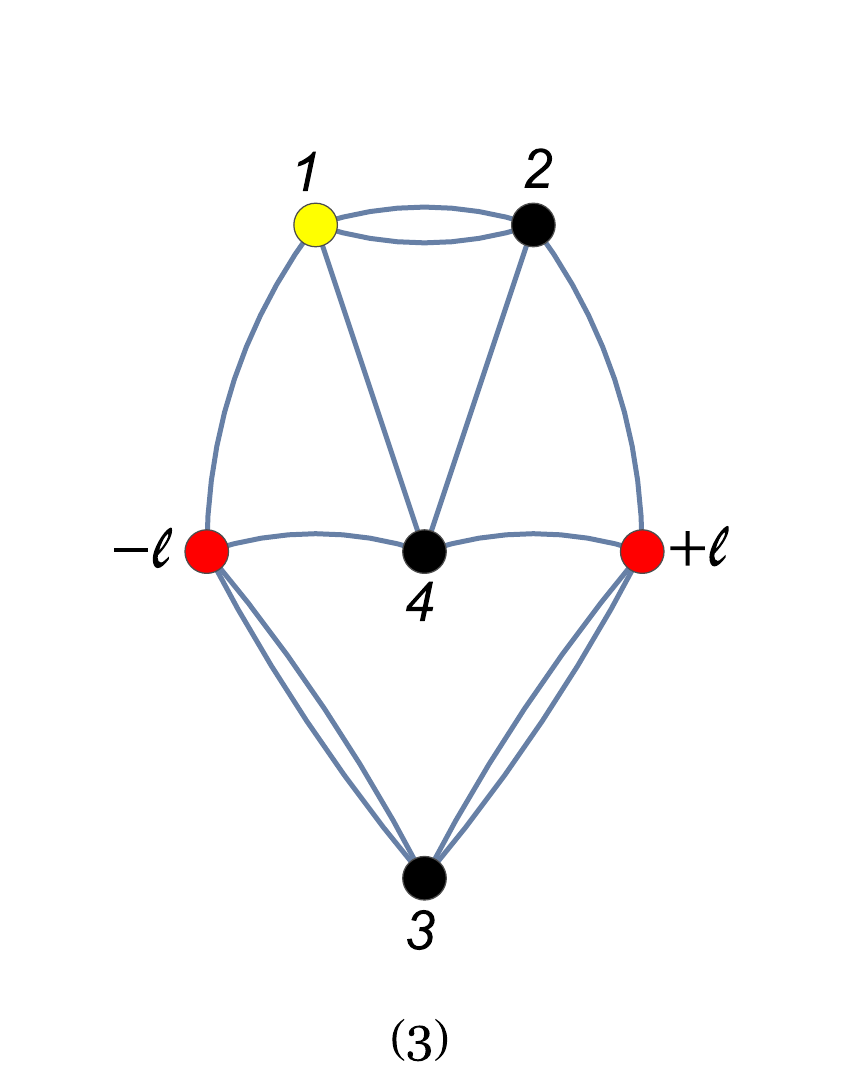}
          \includegraphics[width=1.3in]{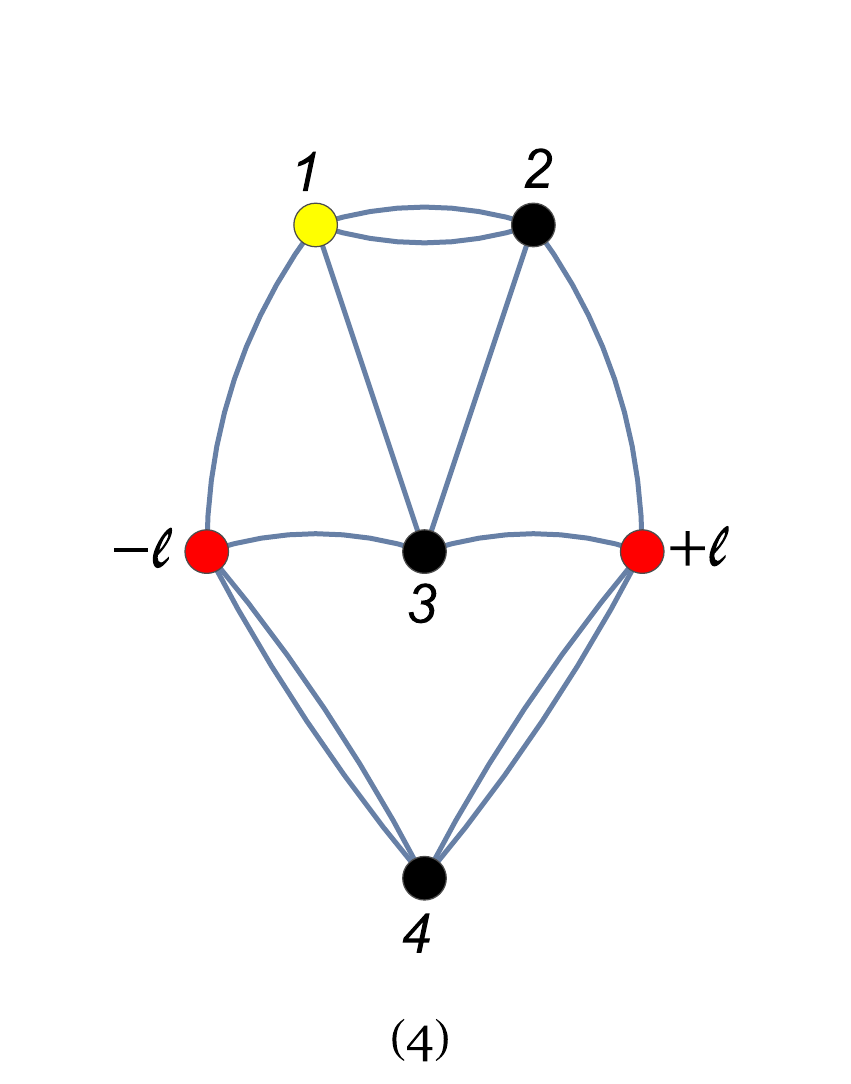}
                \includegraphics[width=1.3in]{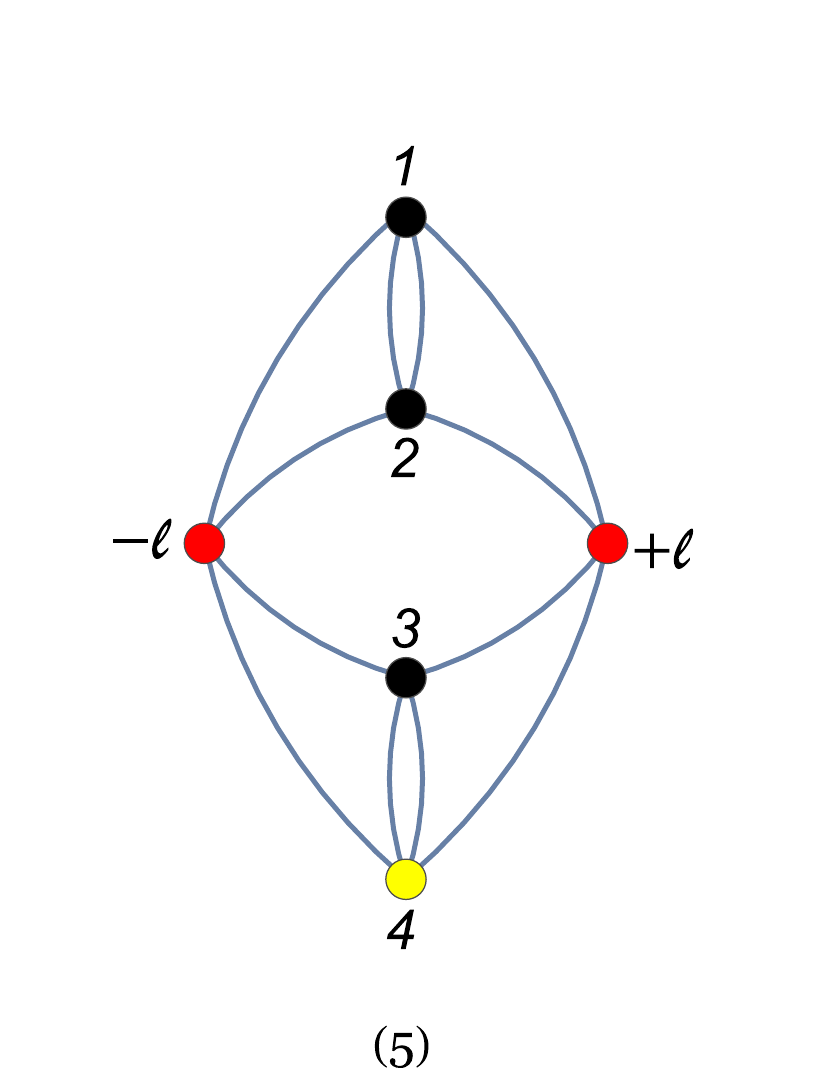}
                        \includegraphics[width=1.3in]{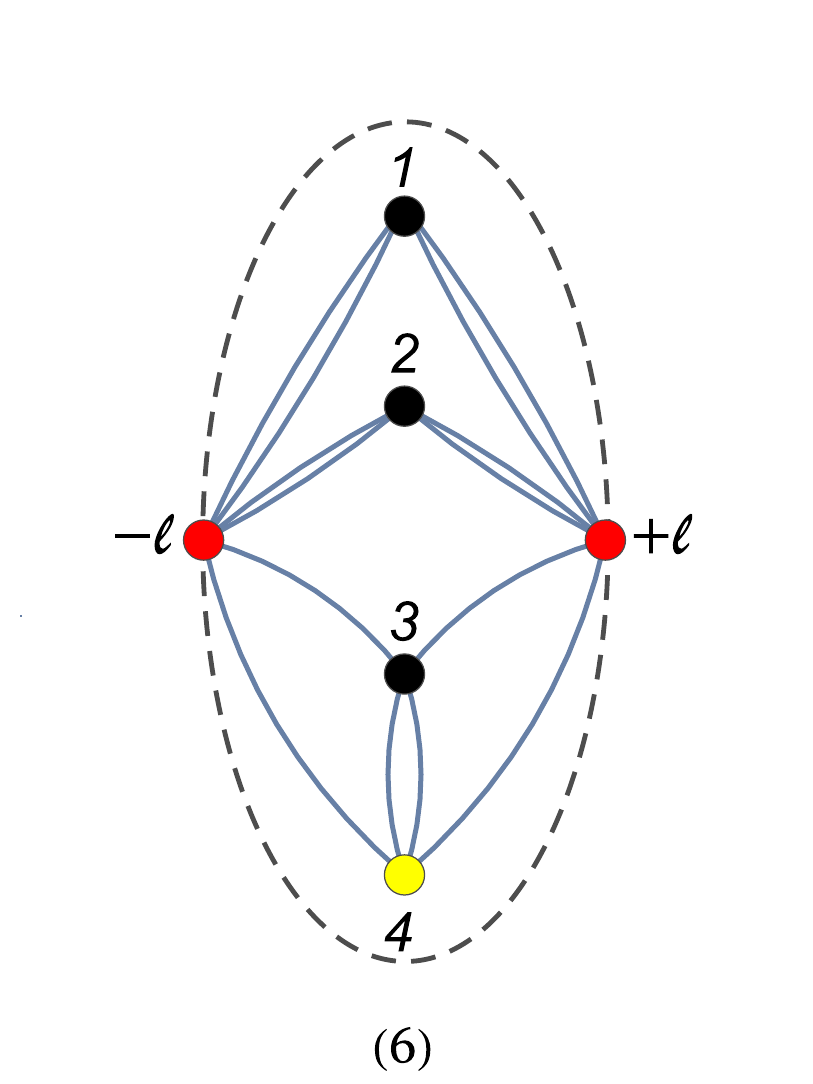}
                          \includegraphics[width=1.3in]{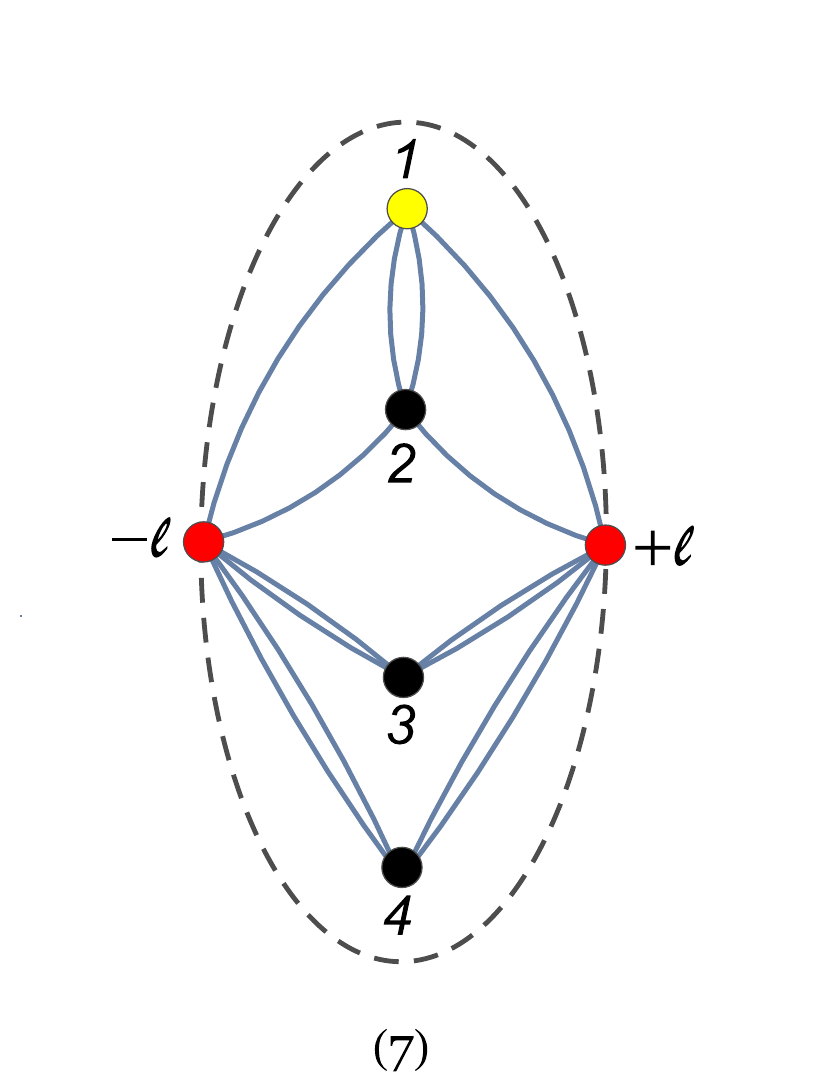}
                                \includegraphics[width=1.3in]{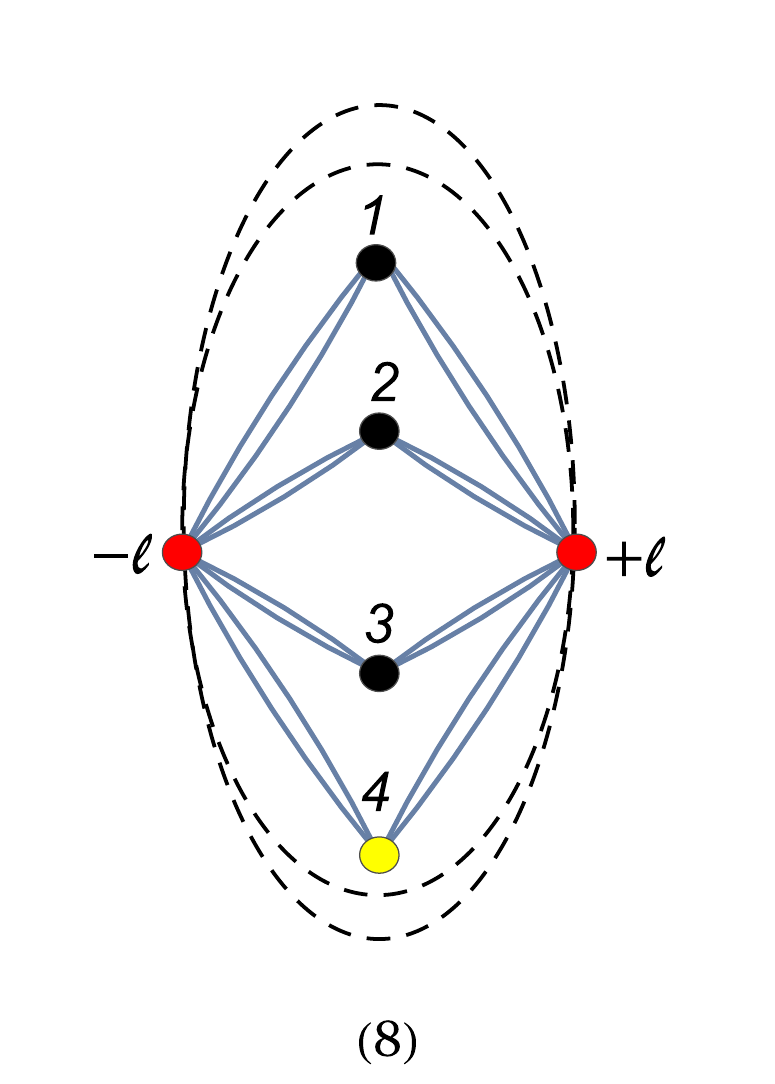}
  \caption{CHY-graphs on a Sphere for the 8 CHY  integrands given in {\bf Table (I)}.}\label{CHY_8}
\end{figure}

For example, let us come back to the $\ph^3$  integrand on $\M_{0,n}$ in section \ref{EXAMPLES}, i.e. ${\cal I}^{\rm t}_4(1,2,3,4)=(1:2:3:4)^{\rm t}\times (1:2:4:3)^{\rm t}$. Over  this integrand, we perform the following eight replacements in order to obtain a well defined expression on $\M_{1,n}$,
\begin{equation}
\begin{matrix}
{\rm Tree-Level~~(\M_{0,n})}  \quad &\quad {\rm One-Loop~~(\M_{1,n})}\\
&\\
(\tau_{1:2}\, \tau_{2:3}\, \tau_{3:4}\, \tau_{4:1})^{\rm t}\times (\tau_{1:2}\, \tau\, \tau_{4:3}\, \tau_{3:1})^{\rm t}  \quad  &\quad  (G^{+}_{1:2}\,T_{2:3}\,T_{3:4}\,G^-_{4:1})\times (G^{+}_{1:2}\,T_{2:4}\,T_{4:3}\,G^-_{3:1}),~~(1)\\
\quad  &\quad  (G^{+}_{1:2}\,G^-_{2:3}\,T_{3:4}\,T_{4:1})\times (G^{+}_{1:2}\,G^-_{2:4}\,T_{4:3}\,T_{3:1}),~~(2)\\
\quad  &\quad  (T_{1:2}\,G^+_{2:3}\,G^-_{3:4}\,T_{4:1})\times (T_{1:2}\,T_{2:4}\,G^+_{4:3}\,G^-_{3:1}),~~(3)\\
\quad  &\quad  (T_{1:2}\,T_{2:3}\,G^+_{3:4}\,G^-_{4:1})\times (T_{1:2}\,G^+_{2:4}\,G^-_{4:3}\,T_{3:1}),~~(4)\\
&\\
\quad  &\quad  (T_{1:2}\,G^+_{2:3}\,T_{3:4}\,G^-_{4:1})\times (T_{1:2}\,G^+_{2:4}\,T_{4:3}\,G^-_{3:1}),~~(5)\\
&\\
\quad  &\quad  (G^+_{1:2}\,G^-_{2:3}\,T_{3:4}\,G^+_{4:1})\times (G^-_{1:2}\,G^+_{2:4}\,T_{4:3}\,G^-_{3:1}),~~(6)\\
\quad  &\quad  (T_{1:2}\,G^+_{2:3}\,G^-_{3:4}\,G^+_{4:1})\times (T_{1:2}\,G^-_{2:4}\,G^+_{4:3}\,G^-_{3:1}),~~(7)\\
&\\
\quad  &\quad  (G^-_{1:2}\,G^+_{2:3}\,G^-_{3:4}\,G^+_{4:1})\times (G^+_{1:2}\,G^-_{2:4}\,G^+_{4:3}\,G^-_{3:1}),~~(8)
\end{matrix}\nonumber
\end{equation}
\begin{center}
{\small {\bf Table (I).}}
\end{center}
Using a similar procedure to the one presented in section  \ref{Example2}, we  integrate the $\l$ variable over the eight integrals given in {\bf Table (I)}. It is simple to check that after integrating $\l$, the eight  CHY-graphs obtained are respectively given by the ones in the figure \ref{CHY_8}. 

Note that the CHY-graphs, $\{(1),(2),(3),(4),(5) \}$ in figure \ref{CHY_8}, look  pretty similar to the $\rm {2-gon}$  in figure \ref{CHY-ngon}, while $\{ (6),(7) \} $ look like a  ${\rm 3-gon}$. The last graph in figure \ref{CHY_8} is the actually a ${\rm 4-gon}$.  This is an important observation and it will be discussed later.

\begin{figure}[h]
  \centering
    \includegraphics[width=1.45in]{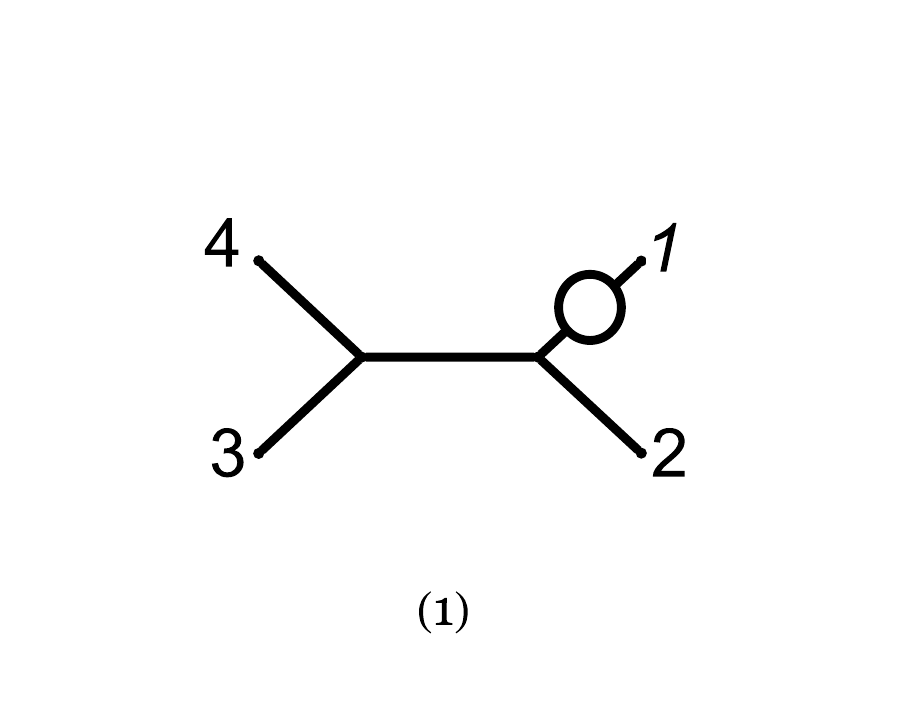}
      \includegraphics[width=1.45in]{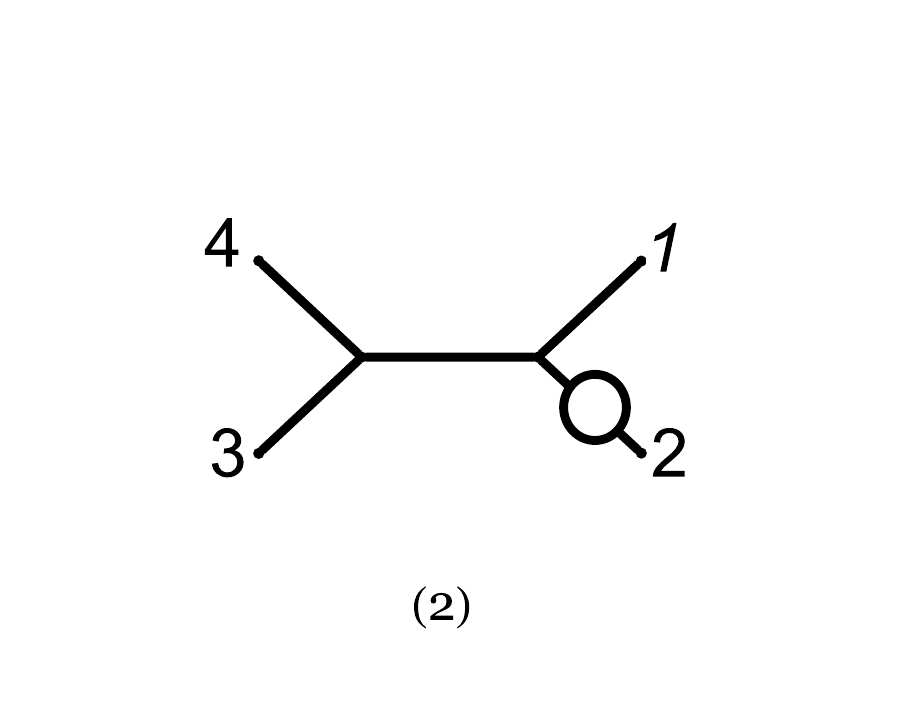}
        \includegraphics[width=1.45in]{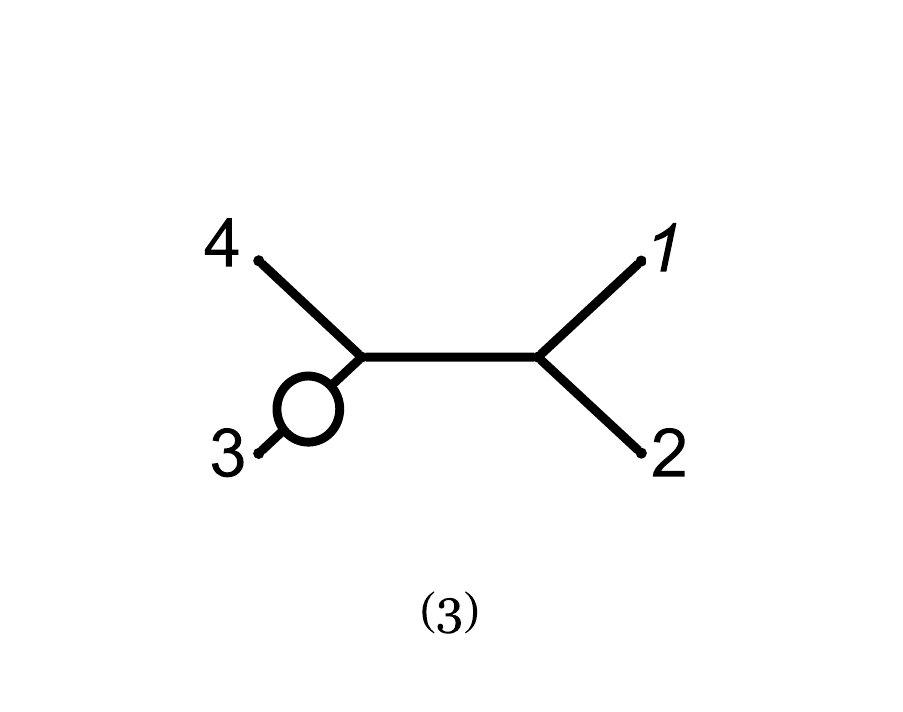}
          \includegraphics[width=1.45in]{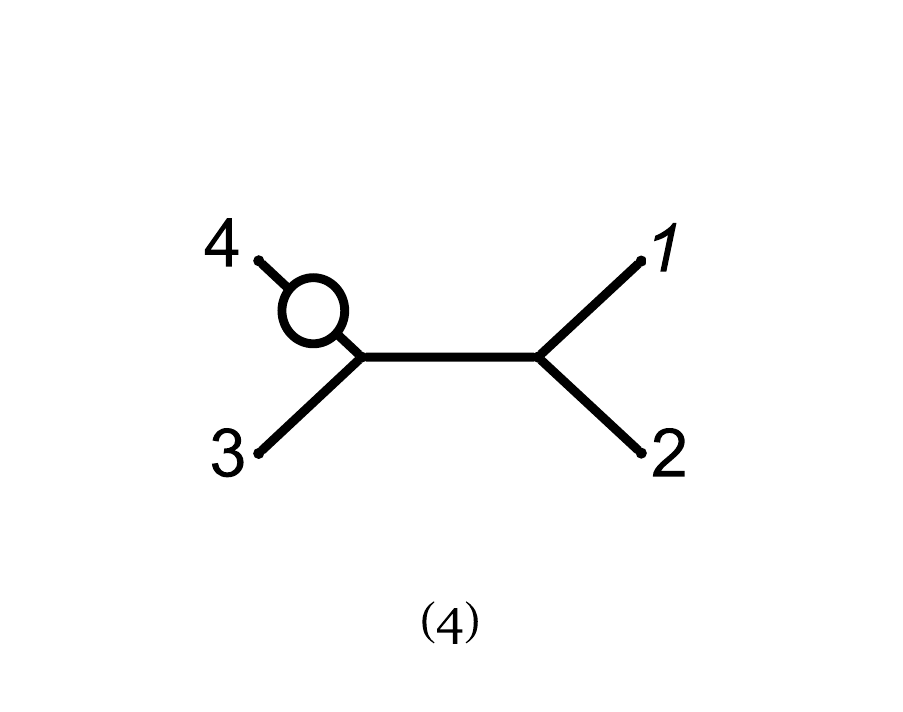}
                \includegraphics[width=1.45in]{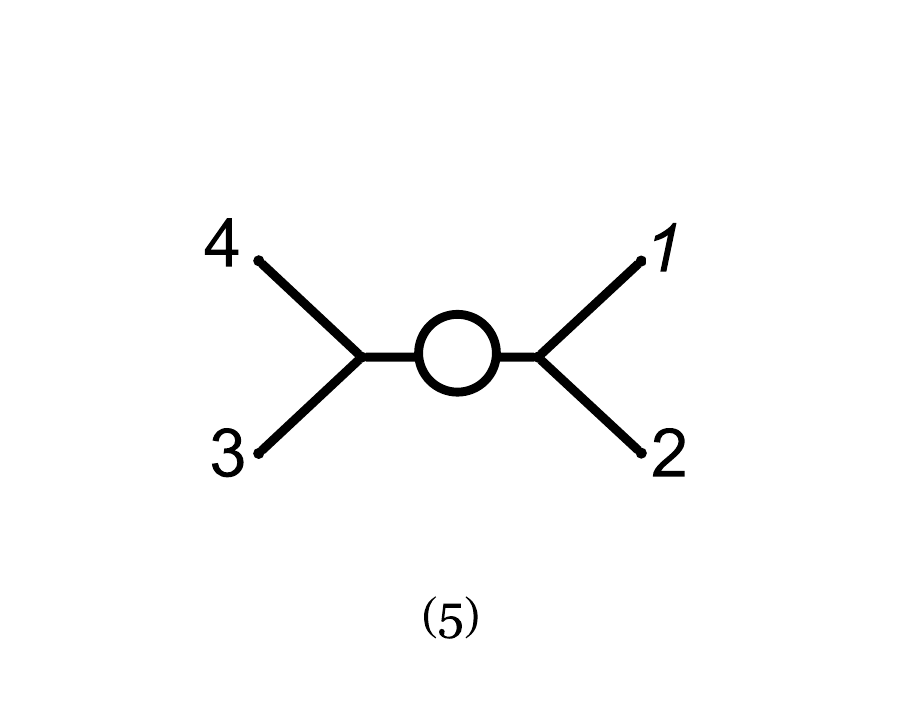}
                        \includegraphics[width=1.45in]{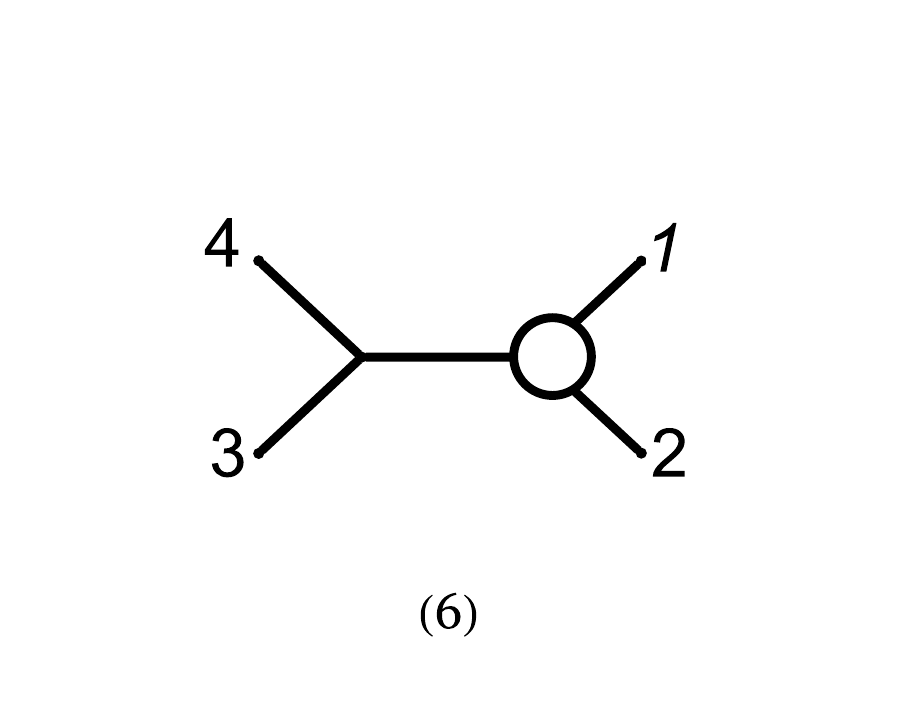}
                          \includegraphics[width=1.45in]{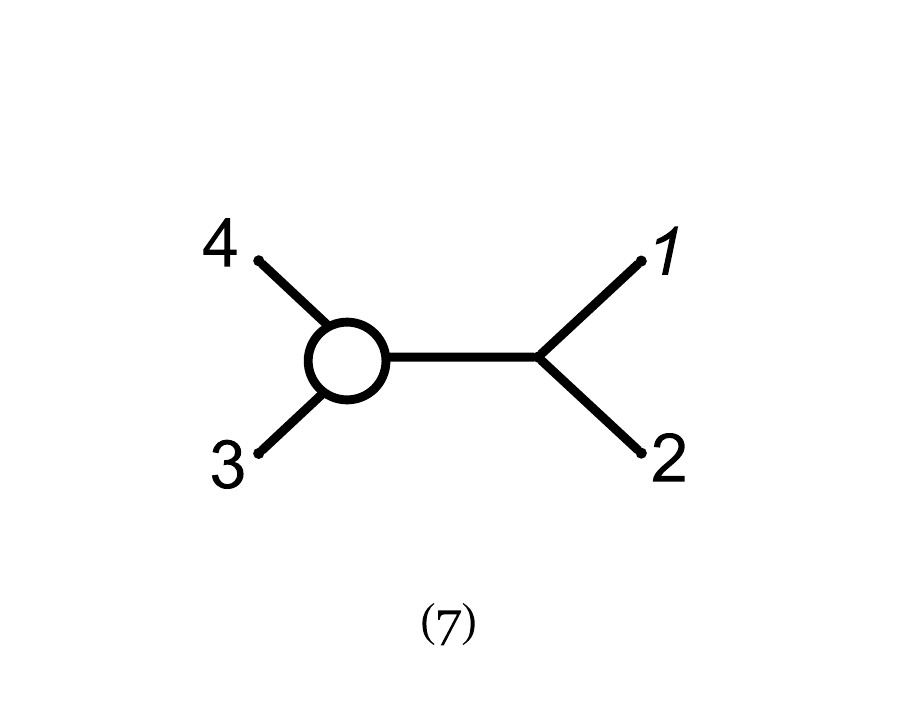}
                                \includegraphics[width=1.45in]{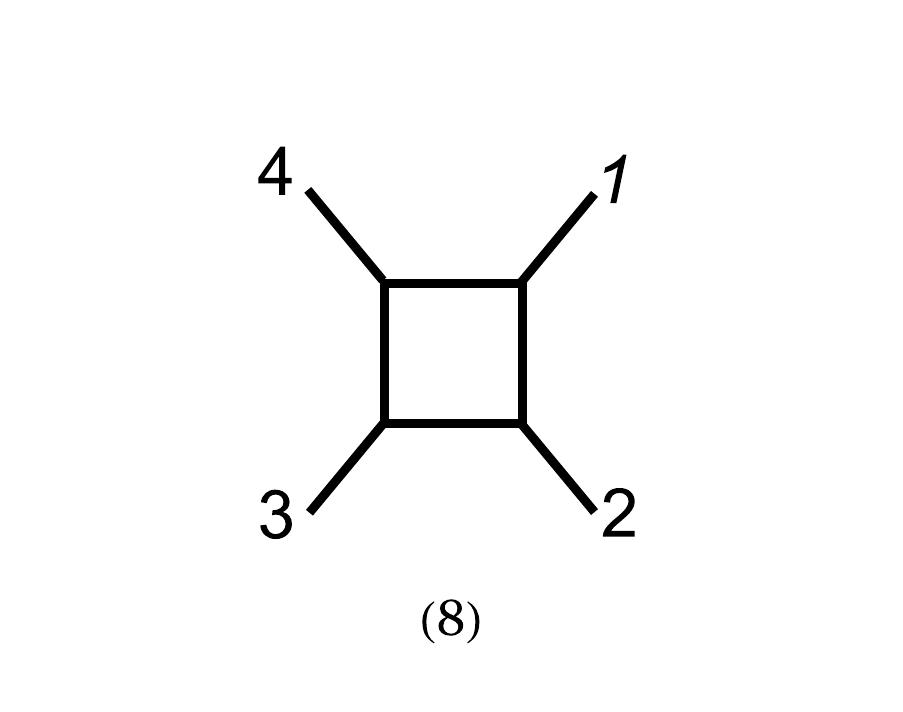}
  \caption{Feynman diagrams for the 8 CHY integrands in  {\bf Table (I)}.}\label{Feynman_8}
\end{figure}

As in section  \ref{Example2}, the CHY-graphs  in figure \ref{CHY_8} are straightforwardly computed by the $\L$-algorithm. The results from the given evaluations are the $\ph^3$ one-loop Feynman diagrams displayed in figure \ref{Feynman_8}.
From this result, we are able to see that the similarity between the CHY-graphs in figure \ref{CHY_8} and the ${\rm n-gon}$ is not a mere coincidence. 

We would like to use this example as a tool to identify a pattern that allow us to construct $\ph^3$ CHY-integrands  on $\M_{1,n}$  from integrands on  $\M_{0,n}$.

\subsection{Construction Rule}\label{ruleone}

In this section, we formulate a simple rule to build $\ph^3$ integrands on $\M_{1,n}$.  This rule is not a necessary condition but it is sufficient.

Let ${\cal I}^{\rm t}_n$ be a $\ph^3$ integrand on $\M_{0,n}$, i.e.
\begin{equation}
{\cal I}^{\rm t}_n:=(\a(1):\a(2):\cdots : \a(n))\times (\b(1):\b(2):\cdots : \b(n)).
\end{equation}
where $\a$ and $\b$ are a particular ordering. The integral of   ${\cal I}^{\rm t}_n$  is just the sum over all $\ph^3$ Feynman diagrams compatible with the $\a$ and $\b$ ordering. Now, the rule is the following
\begin{itemize}
\item {\bf Rule (I)}\\
From a $\ph^3$ integrand on $\M_{0,n}$, ${\cal I}^{\rm t}_n$, we obtain a  $\ph^3$ integrand on $\M_{1,n}$,  with a\footnote{In this context, ``Loop" is referred to the $\ph^3$ one-loop Feynman diagram obtained after performing the integral over  $\M_{1,n}$. Therefore, the sets, $A_1,\ldots,A_p$, are just the external trees of the loop.} ``Loop'' connecting the sets,  $A_1=\{a_1,a_2,\ldots, a_k\}, A_2=\{b_1,b_2,\ldots, b_m\}$, $\ldots ,A_p=\{p_1,p_2,\ldots, p_l\}$, where $A_1\cup A_2\cup\cdots \cup A_p=\{1,2,\ldots,n\}$, if after doing the following replacements
\begin{align}\label{replacement}
\tau_{a:b}~~\rightarrow~~
\left\{
\begin{matrix}
&G^{\pm}_{a:b},\quad {\rm If}~~ \{a,b\}\nsubseteq A_1,~ \{a,b\}\nsubseteq A_2,~\ldots,~ \{a,b\}\nsubseteq A_p \\
&T_{a:b},\quad {\rm Otherwise,}
\end{matrix}
\right.
\end{align}
such that $G^+_{a:b}$ and $G^-_{a:b}$ are put in an alternating way, the condition
\begin{equation}\label{Gcondition}
\#G^+_{a:b} - \#G^-_{c:d}=0,\qquad \text{({\bf Condition (I)})},
\end{equation}
is satisfied.
\end{itemize}

Note that the integrands in {\bf Table (I)} satisfy the {\bf Rule (I)} and {\bf Condition (I)}, therefore they are $\ph^3$ integrands as it is confirmed by the resulting Feynman diagrams in figure \ref{Feynman_8}.

We give now an example where after applying the replacements in \eqref{replacement}, the {\bf Condition (I)} is not satisfied. 
Let us consider again the same tree-level integrand as in {\bf Table (I)}
\begin{equation}\label{rule1tree}
{\cal I}^{\rm t}_4(1,2,3,4)=
(\tau_{1:2}\, \tau_{2:3}\, \tau_{3:4}\, \tau_{4:1})^{\rm t}\times (\tau_{1:2}\, \tau_{2:4}\, \tau_{4:3}\, \tau_{3:1})^{\rm t}. 
\end{equation}
We would like to obtain a loop connecting the particle sets, $A_1=\{2\}, A_2= \{3\}$ and  $A_3= \{1,4\}$.
Applying the  replacements in \eqref{replacement}, the new integrand is read as
\begin{equation}\label{Exrule1}
(\tau_{1:2}\, \tau_{2:3}\, \tau_{3:4}\, \tau_{4:1})^{\rm t}\times (\tau_{1:2}\, \tau_{2:4}\, \tau_{4:3}\, \tau_{3:1})^{\rm t}~\rightarrow~
(G^+_{1:2}\, G^-_{2:3}\, G^+_{3:4}\, T_{4:1})\times (G^-_{1:2}\, G^+_{2:4}\, G^-_{4:3}\, G^+_{3:1}).
\end{equation}
Obviously,  the expression in \eqref{Exrule1} does not satisfy the {\bf Condition (I)},  hence this  is not a $\ph^3$ integrand.  Physically is simple to understand why: the tree-level integrand in \eqref{rule1tree} is just the Feynman diagram given in the left hand side of figure \ref{FEYNMAN_ex2}, from which is not possible to blow a loop connecting the sets (trees), $A_1=\{2\}$, $A_2=\{3\}$ and $A_3=\{1, 4\}$.

Clearly, the sets,  $A_1,\, A_2,\ldots,\, A_p$, are just the external trees of the $\ph^3$ one-loop Feynman diagrams.

In addition, the {\bf Rule (I)} satifies a selection rule, which we explain in the next section.

\subsection{Selection Rule}

In order to formulate the selection rule, it is useful to give the following definition
\begin{itemize}
\item {\bf Compatibility (From a Tree to a Loop)}\\
We say that a Tree-level Feynman diagram is {\bf compatible} with the loop  connected to the trees  $A_1, A_2, \ldots ,A_p$, if there is a common vertex linking these sets.
\end{itemize}

\begin{figure}[h]
  \centering
   \includegraphics[width=1.45in]{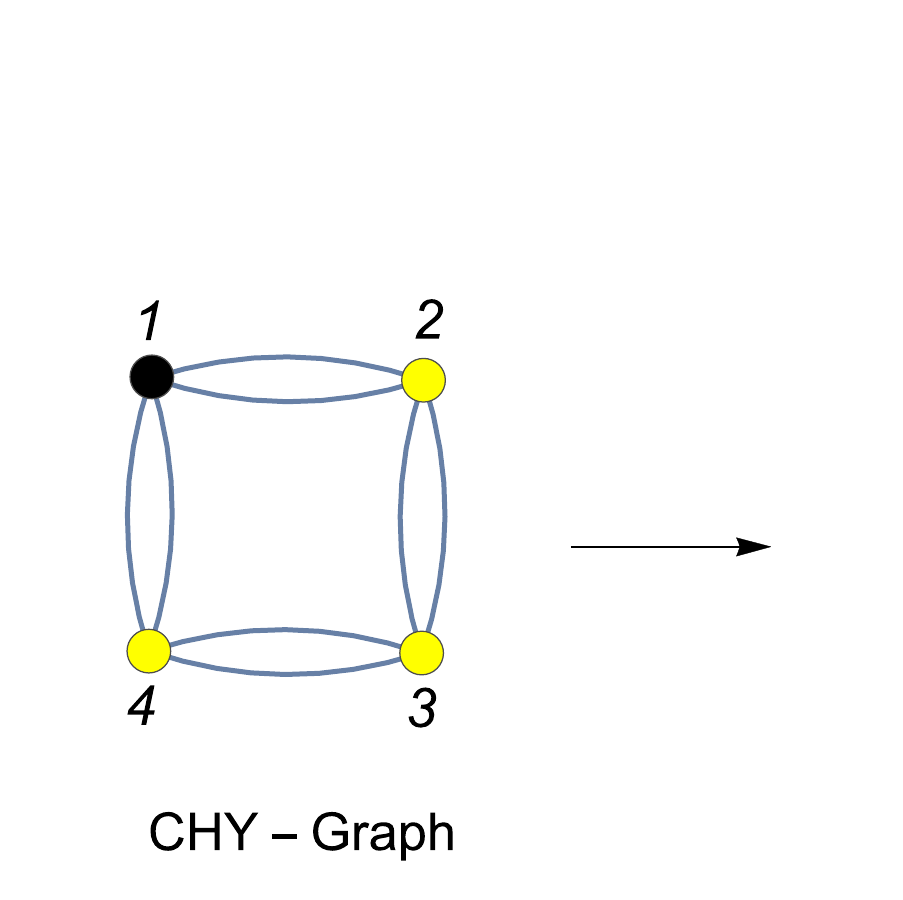}
      \includegraphics[width=1.5in]{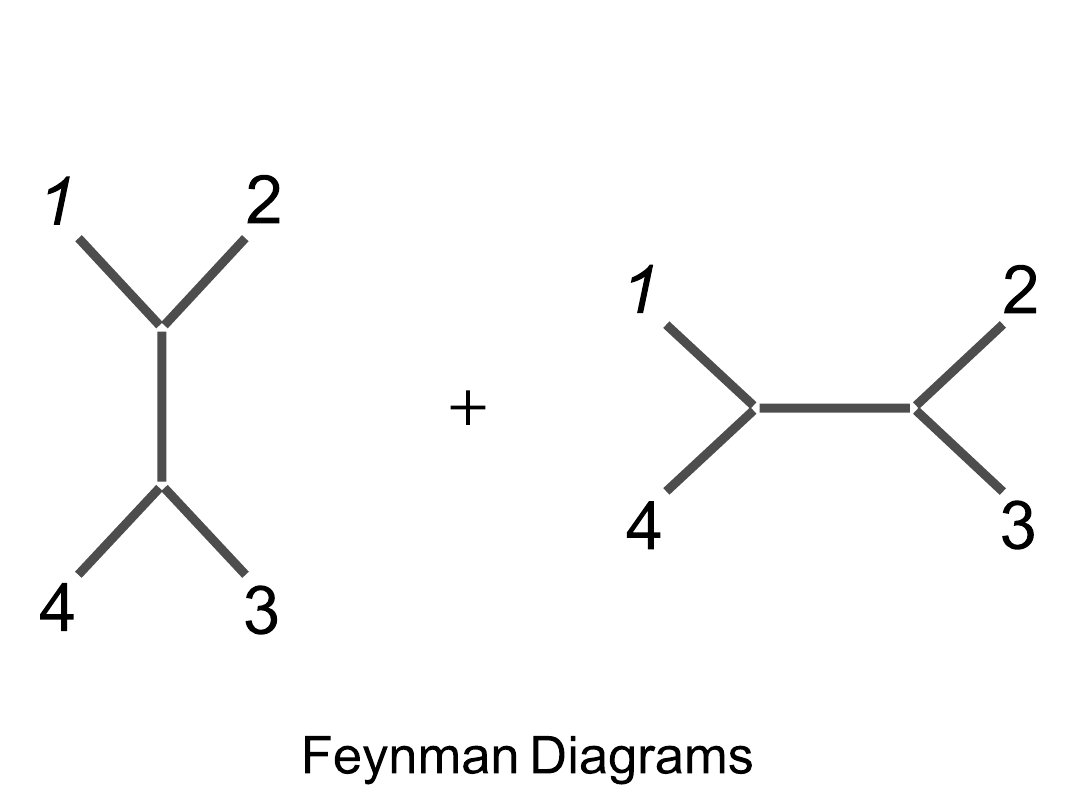}
  \caption{CHY-graph and its Feynman diagram representation.}\label{FEY_selection}
\end{figure}

So as to illustrate  the selection rule we give a simple example. 
Let us consider the CHY-tree level integrand
\begin{equation}
{\cal I}^{\rm t}_4=(1:2:3:4)^{\rm t}\times (1:2:3 : 4)^{\rm t}.
\end{equation}
It is well known that, the integral of ${\cal I}^{\rm t}_4$ over $\M_{0,n}$  is just the sum over two tree-level Feynman diagrams given in figure \ref{FEY_selection}.

Now, we would like to obtain a loop among the particle sets, $A_1=\{1\}, A_2= \{2\}$ and  $A_3= \{3,4\}$. Applying the {\bf Rule (I)}, the new integrand is read as
\begin{equation}
(\tau_{1:2}\, \tau_{2:3}\, \tau_{3:4}\, \tau_{4:1})^{\rm t}\times (\tau_{1:2}\, \tau_{2:3}\, \tau_{3:4}\, \tau_{4:1})^{\rm t}~\rightarrow~
(G^+_{1:2}\, G^-_{2:3}\, T_{3:4}\, G^+_{4:1})\times (G^-_{1:2}\, G^+_{2:3}\, T_{3:4}\, G^-_{4:1}).
\end{equation}

\begin{figure}[h]
  \centering
  \qquad \qquad  ~~ \includegraphics[width=1.2in]{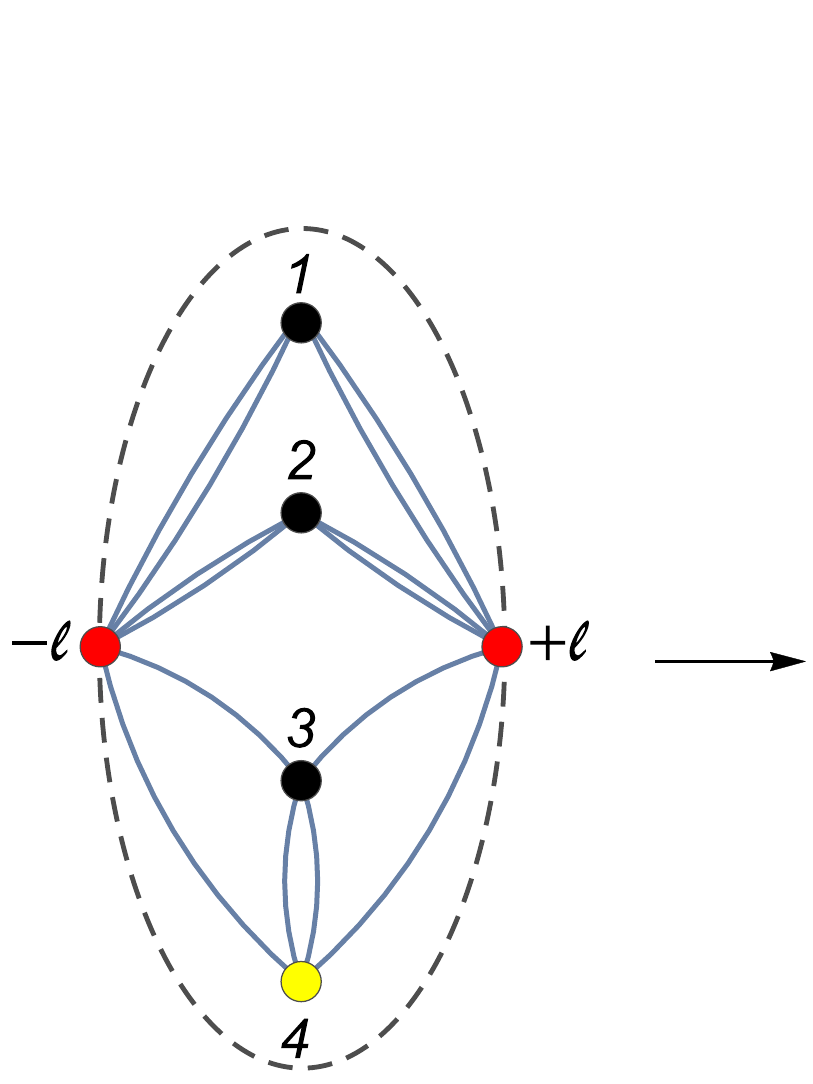}
      \includegraphics[width=1.5in]{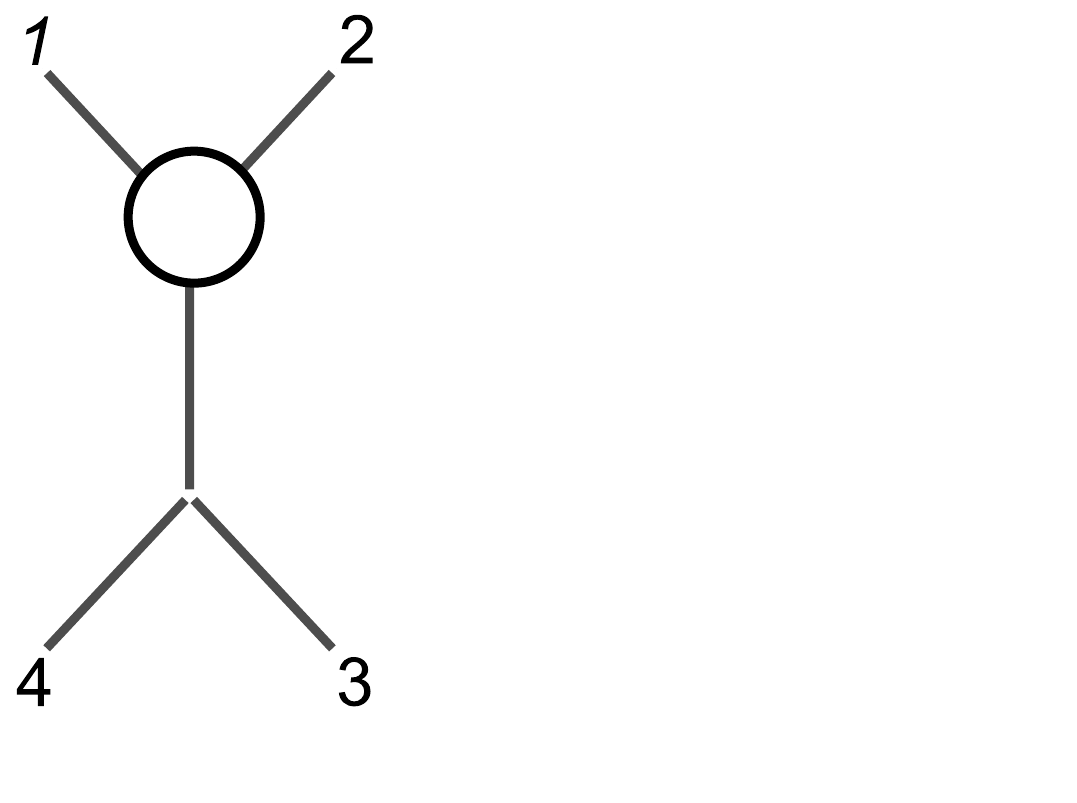}
  \caption{CHY-graph and its Feynman diagram representation for the $(G^+_{1:2}\, G^-_{2:3}\, T_{3:4}\, G^+_{4:1})\times (G^-_{1:2}\, G^+_{2:3}\, T_{3:4}\, G^-_{4:1})$ integrand.}\label{FEY_loop_s}
\end{figure}

Performing the integral over $\l$, it is not hard to check that one obtains the CHY-graph in figure \ref{FEY_loop_s} (left), which, as it has been already shown, is just the $\ph^3$ one-loop Feynman diagram given in figure \ref{FEY_loop_s} (right).

This simple example shows clearly what is happening. After applying the {\bf Rule (I)},  the second Feynman diagram in figure \ref{FEY_selection} is discarded. In other words, the {\bf Rule (I)} selected the tree-level Feynman diagrams compatibles with the loop connected to the trees, $A_1=\{1\}, A_2= \{2\}$ and  $A_3= \{3,4\}$.

Now, we are ready to formulate the selection rule

\begin{itemize}
\item {\bf Selection Rule}\\
To apply the {\bf Rule (I)} to the sets, $A_1, A_2, \ldots ,A_p$,  it picks up  only the Tree-level Feynman diagrams that are compatibles with the loop connected to the trees, $A_1, A_2, \ldots ,A_p$.
\end{itemize}

In the next section we present an inverse method, i.e. given a 1-loop Feynman diagram, we construct its corresponding $\ph^3$ CHY integrand on $\M_{1,n}$.

\subsection{From One-Loop $\ph^3$-Feynman Diagrams to Integrands Over a Torus.}\label{from_FEY_CHY}

\begin{figure}[h]
  \centering
    \includegraphics[width=2.3in]{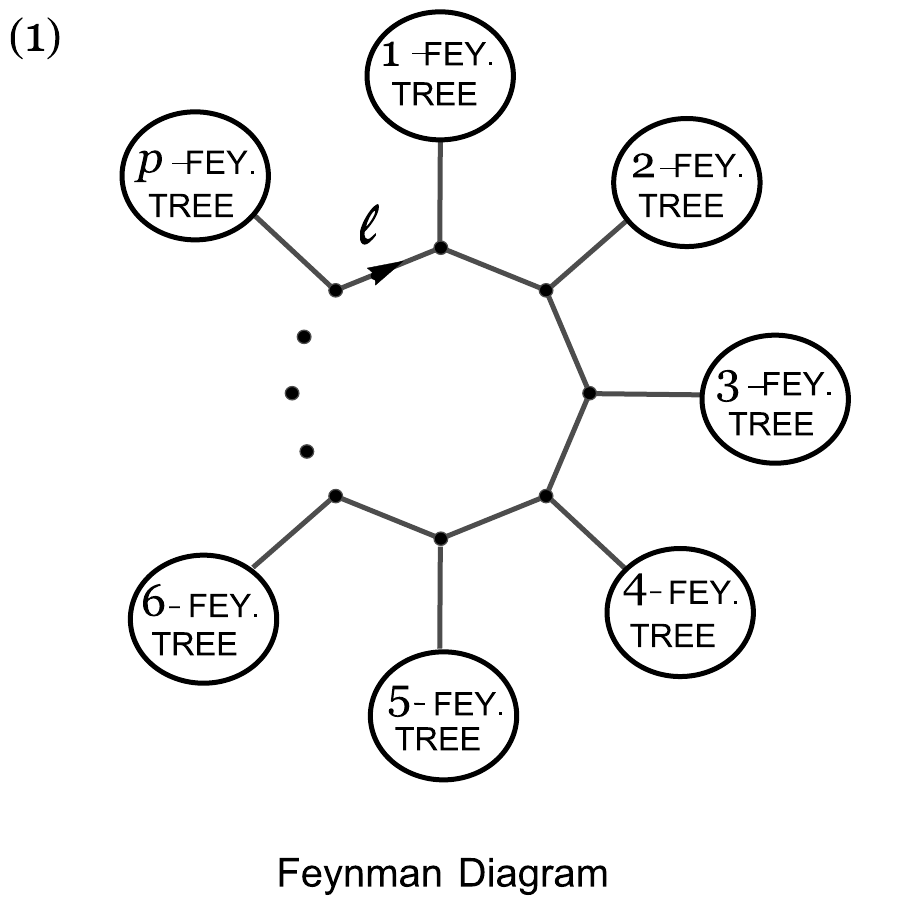}
      \includegraphics[width=1.4in]{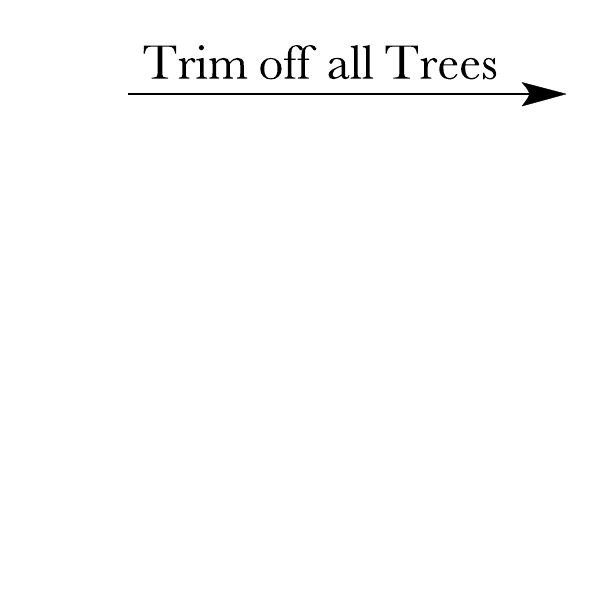}
        \includegraphics[width=2.3in]{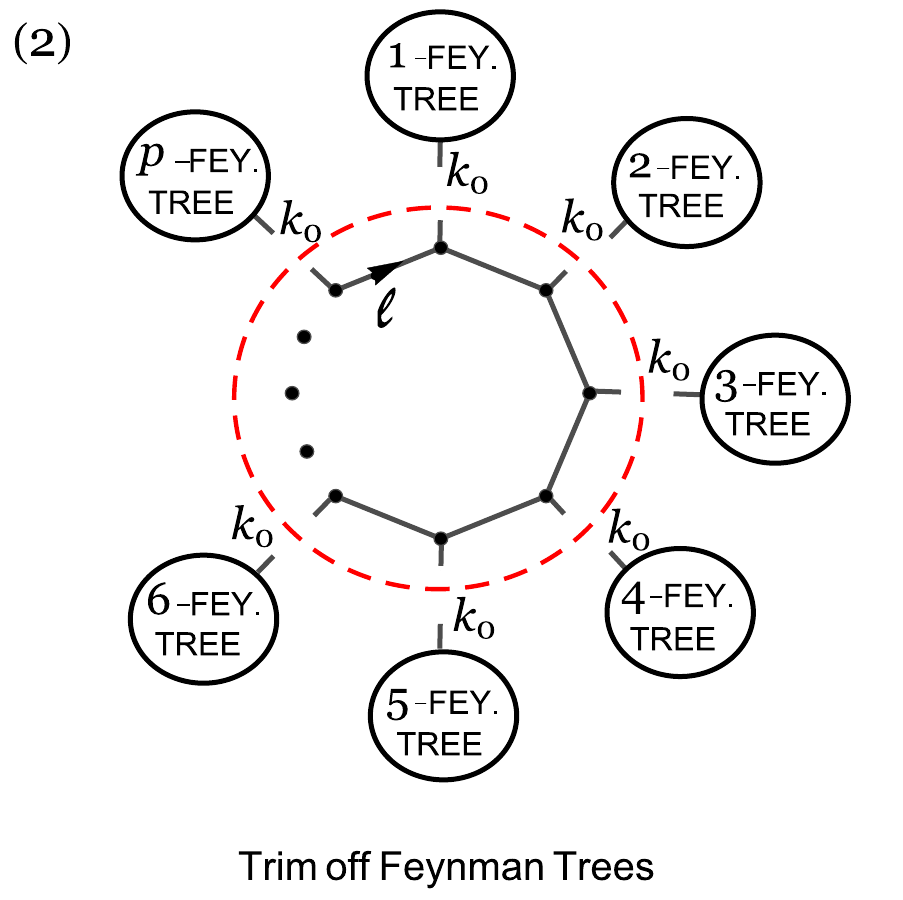}
  \caption{Generic 1-loop Feynman diagram and triming off of its trees.}\label{FEY_pgon_tree}
\end{figure}

In what follows, by mixing the rules presented in \cite{Baadsgaard:2015ifa} with the graphical form of the CHY $n-$gon, we intent to present a graphical prescription to build integrands on $\M_{1,n}$ from a given Feynman diagram. In some sense we shall build the inverse operation we already defined in the previous sections. A one-loop Feynman diagram with $n$ external particles is built from a set of disjoint tree diagrams attached to a $p-$gon, schematically represented in the left hand side of Figure \ref{FEY_pgon_tree} for the $\ph^3$ interaction.

Starting with a given Feynman diagram at one-loop, the recipe consist in the following steps,
\begin{itemize}
\item 1) Trim off all the trees attached to the loop as shown in right hand side of Figure \ref{FEY_pgon_tree}.
\item 2) To every tip arising from a cut assign a momentum equal to $k_0$.\footnote{The whole construction is independent of the momentum $k_0$ as long as it is on-shell, but for definiteness we can think of it as equal to zero.}
\item 3) To each tree sub-diagram  in the right hand side of figure \ref{FEY_pgon_tree}, draw the corresponding CHY-graph by following the Baadsgaard, Bohr, Bourjaily and Damgaard  rules (B.B.B.D) in \cite{Baadsgaard:2015ifa}, as schematically shown in Figure \ref{Gon}. 
\begin{figure}[h]
  \centering
 \includegraphics[width=2.5in]{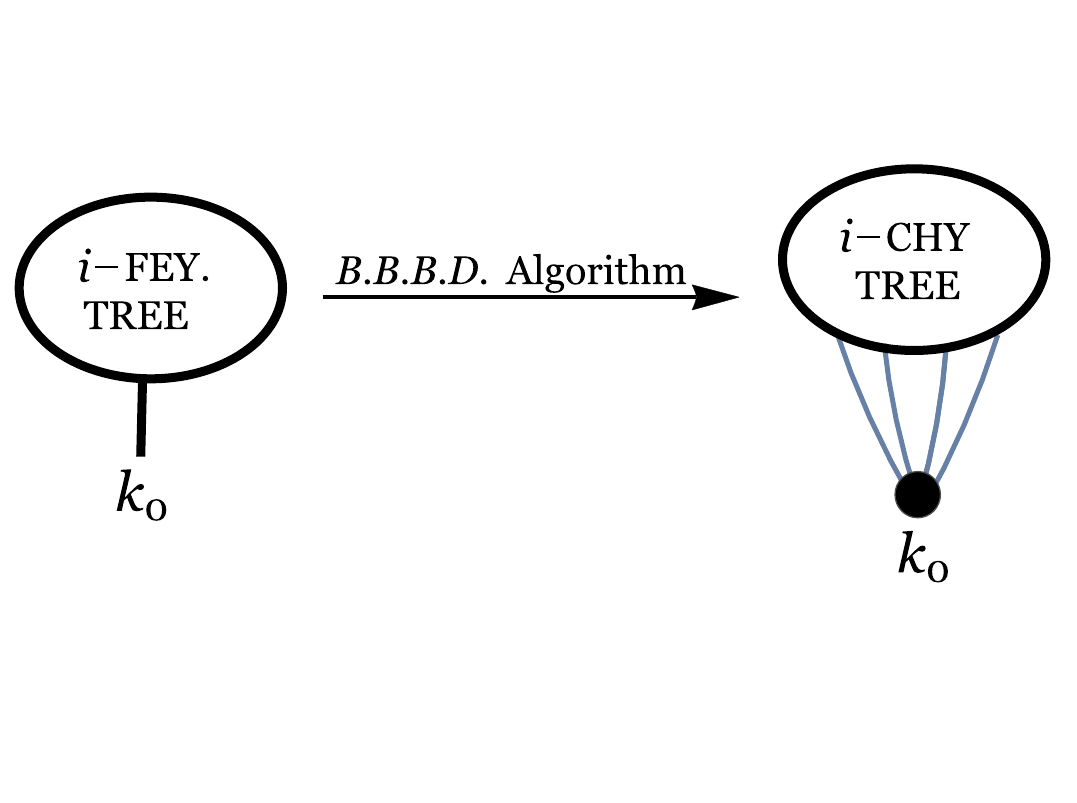}
  \caption{B.B.B.D. construction of CHY-graph  from a tree-level Feynman diagram.}\label{Gon}
\end{figure}

In the particular case when only one leg is trimmed, as in the traditional n-gon given in figure \ref{CHY-ngon},  the Feynman diagram is just a  propagator and we represented it for the CHY-graph in figure \ref{propagator-CHY}. 
\begin{figure}[h]
  \centering
 \includegraphics[width=1.6in]{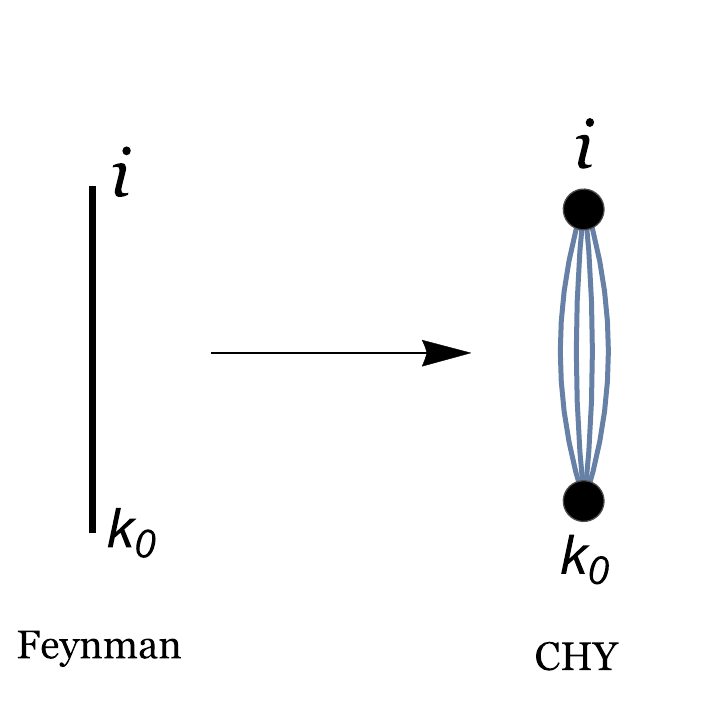}
  \caption{CHY representation for a single propagator.}\label{Gon}\label{propagator-CHY}
\end{figure}
\end{itemize}
Although it would not have any physical meaning,  we use this correspondence as a tool in order to obtain a general algorithm.

\begin{itemize}
\item 4) For each tree-level CHY-graph, split the puncture associated to the particle of momentum $k_0$ as two punctures with momentum $\ell$ and $-\ell$ respectively.  For each vertex previously connected to the puncture of momentum $k_0$, it must now be connected to the puncture $\ell$ through a single edge as well as to $-\ell$, as in the tree graphs in figure \ref{TreeOpen} and \ref{FEY_ex2_loop_cut_CHY_1}. If there is only one edge connecting $k_0$ with a vertex, then this edge can go to $\ell$ or $-\ell$, it does not matter.

For the particular case given in figure \ref{propagator-CHY}, two edges out of four should go to $\ell$ and the other two go to  $-\ell$, see figure \ref{FEY_ex2_loop_cut_CHY_1}. 
\begin{figure}[h]
  \centering
    \includegraphics[scale=0.55]{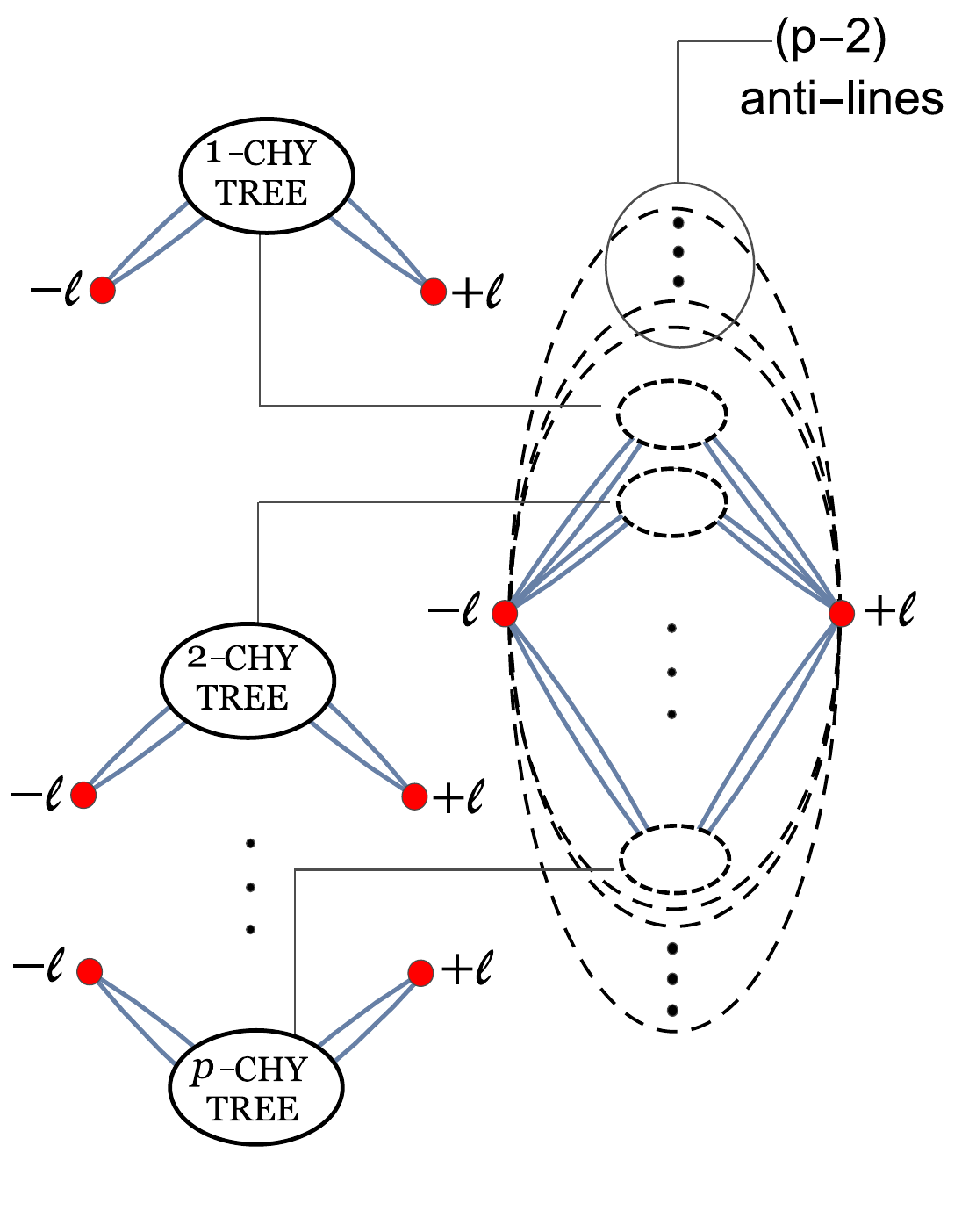}
  \caption{Schematic representation of rule 4.}\label{TreeOpen}
\end{figure}

\item 5) Glue all vertex with momentum $\ell$ together as well as  all vertex with momentum -$\ell$, as is shown in the right drawing at figure \ref{CHY_pgon_tree}. 
\begin{figure}[h]
  \centering
      \includegraphics[width=2.3in]{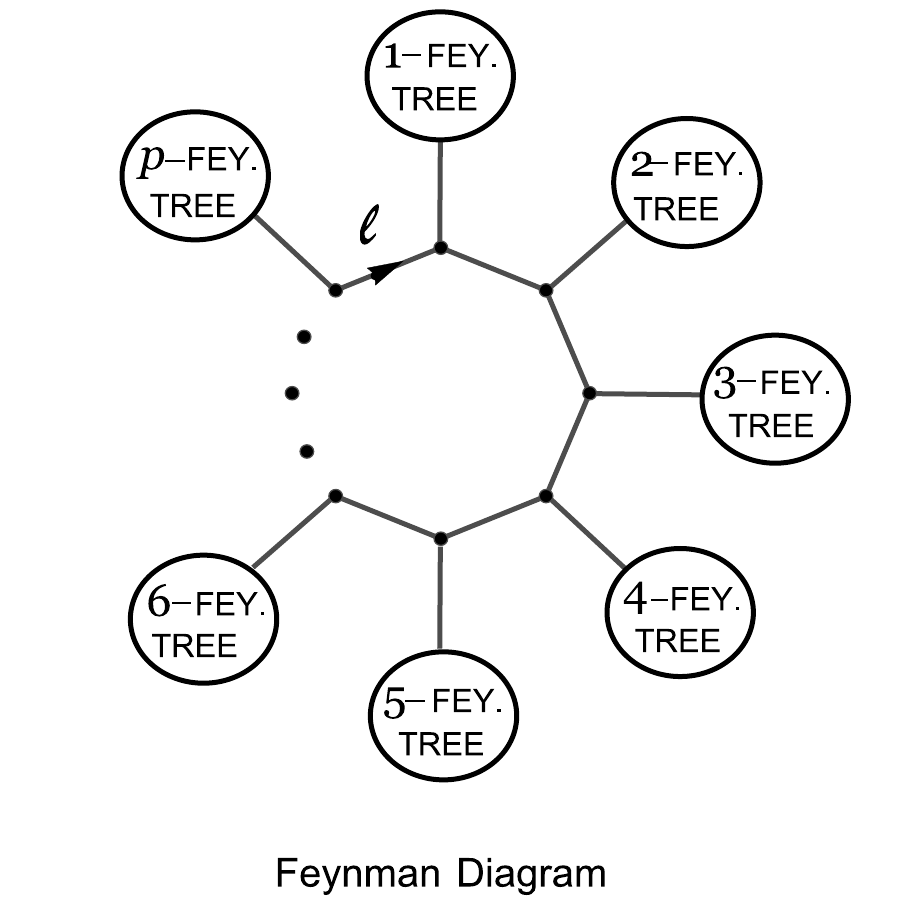}
    \includegraphics[width=2.1in]{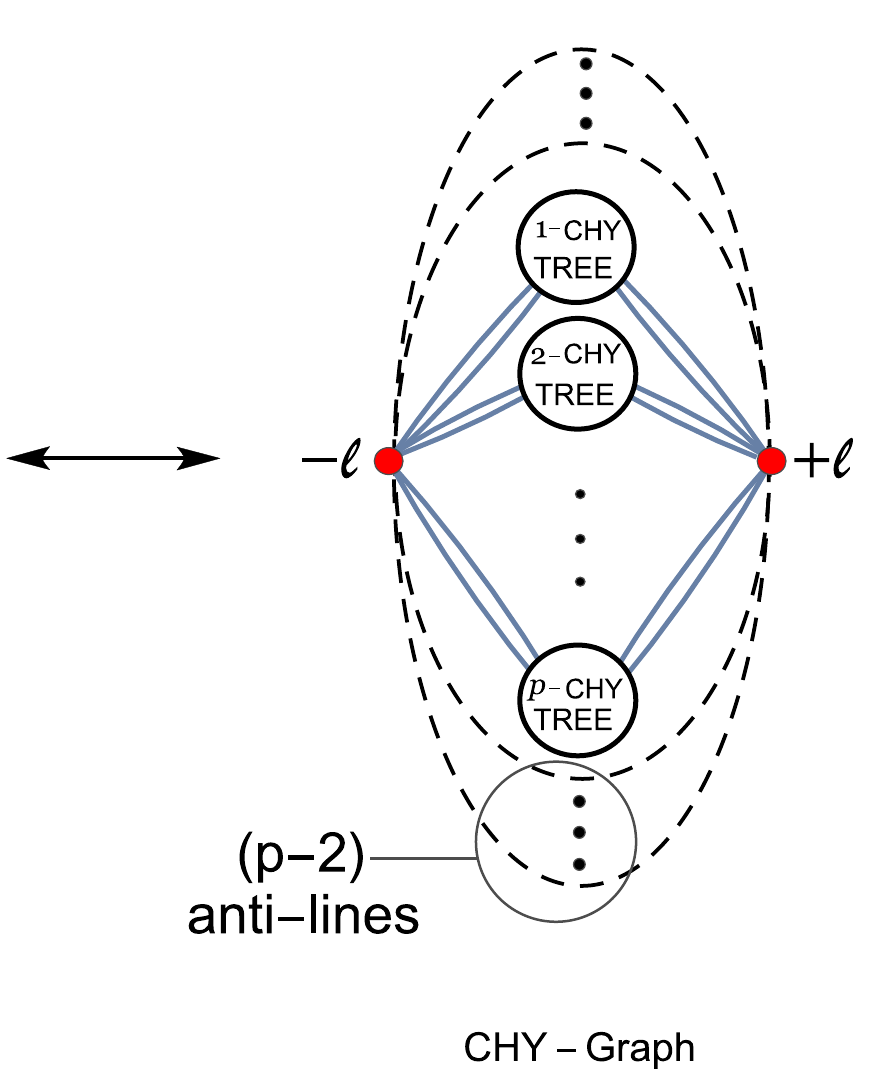}
  \caption{Final construction of the CHY-graph corresponding to a 1-loop Feynman diagram.}\label{CHY_pgon_tree}
\end{figure}
The anti-lines connecting  $\ell$  and $-\ell$ must be introduced in order to have $PSL(2,\mathbb{C})$ invariance.  In addition, one of the $n$ vertices is colored with Yellow, so as to fix the $PSL(2,\mathbb{C})$ symmetry.

\end{itemize}

Up to this point, the resulting CHY-graph corresponds to an integrand on  $\M_{0,n}$. Now, we are able to   build the graph on a Torus and so, to find an integrand on $\M_{1,n}$  by using the connectors $\{T_{a:b}, G^{\pm}_{a:b}\}$.

In order to do so, we add the following three simple rules:
\begin{itemize}
\item 6) Stretch out the points $\ell$ and $-\ell$ forming a line (${\rm a-cycle}$).

\item 7) Assign directions to the edges in such a way that the {\bf Rule (I)} in section \ref{ruleone} is satisfied.

\item 8) For every edge connecting a point $a$ to a point $b$ and wrapping the b-cycle from left to right use the chain element $G^+_{a:b}$ \eqref{G+ab}. For every line connecting a point $a$ to a point $b$ and wrapping the b-cycle from right to left use the chain element $G^-_{a:b}$ \eqref{G-ab}.Finally,  for every line connecting a point $a$ to a point $b$ no wrapping the b-cycle use the chain $T_{a:b}$ \eqref{Tab}.

The main idea is to construct a product of two Parker-Taylor, which satisfies the {\bf Rule (I)} in section \ref{ruleone}.

\end{itemize}

\subsection{Lower point examples}\label{Sec8}

In this subsection we would like to apply the rules above to a couple of simple non-trivial examples in order to clarify the procedure.

\subsubsection{Four-Point and Triangle-Loop}

\begin{figure}[h]
  \centering
 \includegraphics[width=1.4in]{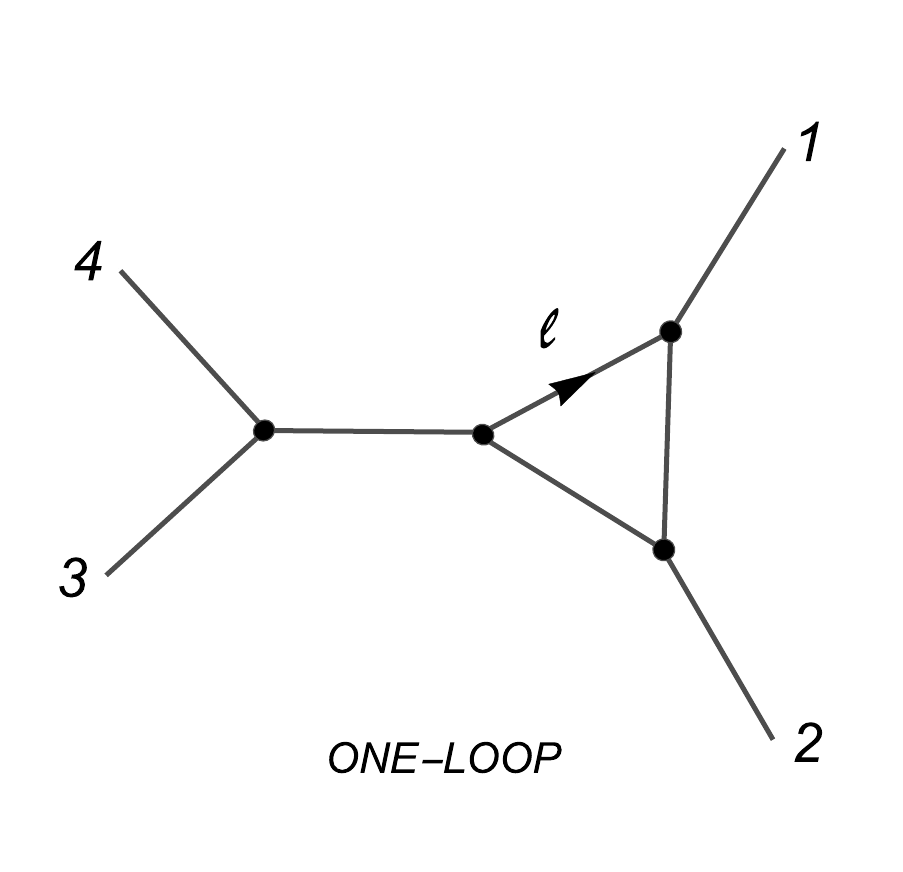}
  \caption{One-loop Feynman diagram with a triangular loop.}\label{FEY_ex2_loop2}
\end{figure}

Let us start considering the Feynman diagram displayed in Figure \ref{FEY_ex2_loop2},

\begin{figure}[h]
  \centering
    \includegraphics[width=1.4in]{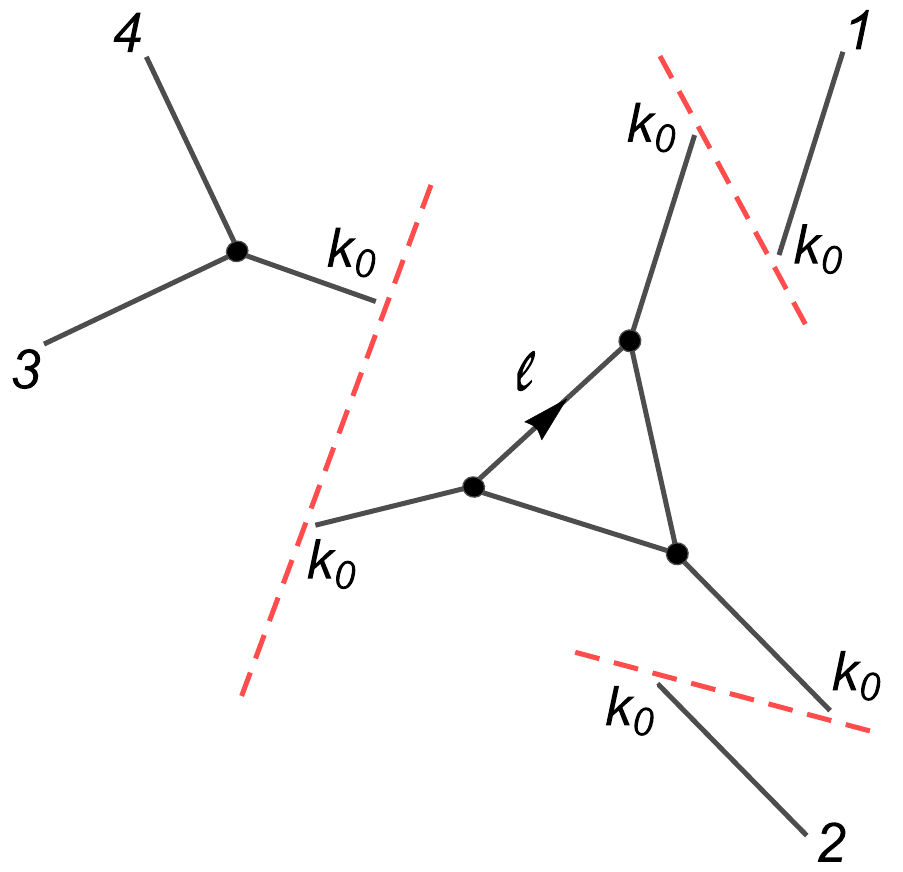}
  \caption{Trimming off the trees from the triangular loop.}\label{FEY_ex2_loop_cut}
\end{figure}

Trimming off the trees attached to the loop and assigning a momentum $k_0$ to every tip of a cut, we get the three tree sub-diagrams represented in Figure \ref{FEY_ex2_loop_cut},

\begin{figure}[h]
  \centering
    \includegraphics[scale=0.5]{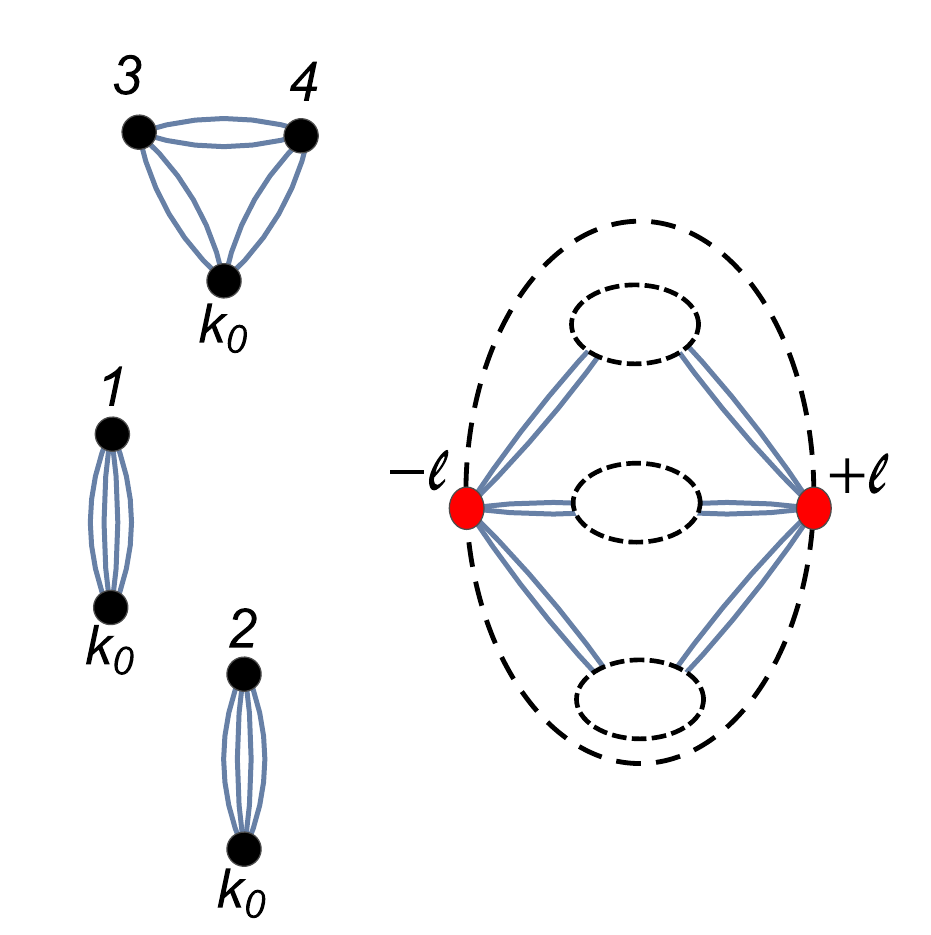}
            \includegraphics[scale=0.45]{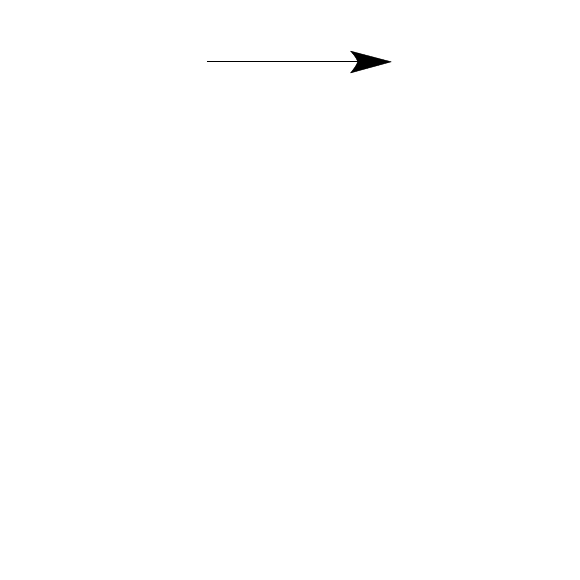}
        \includegraphics[scale=0.48]{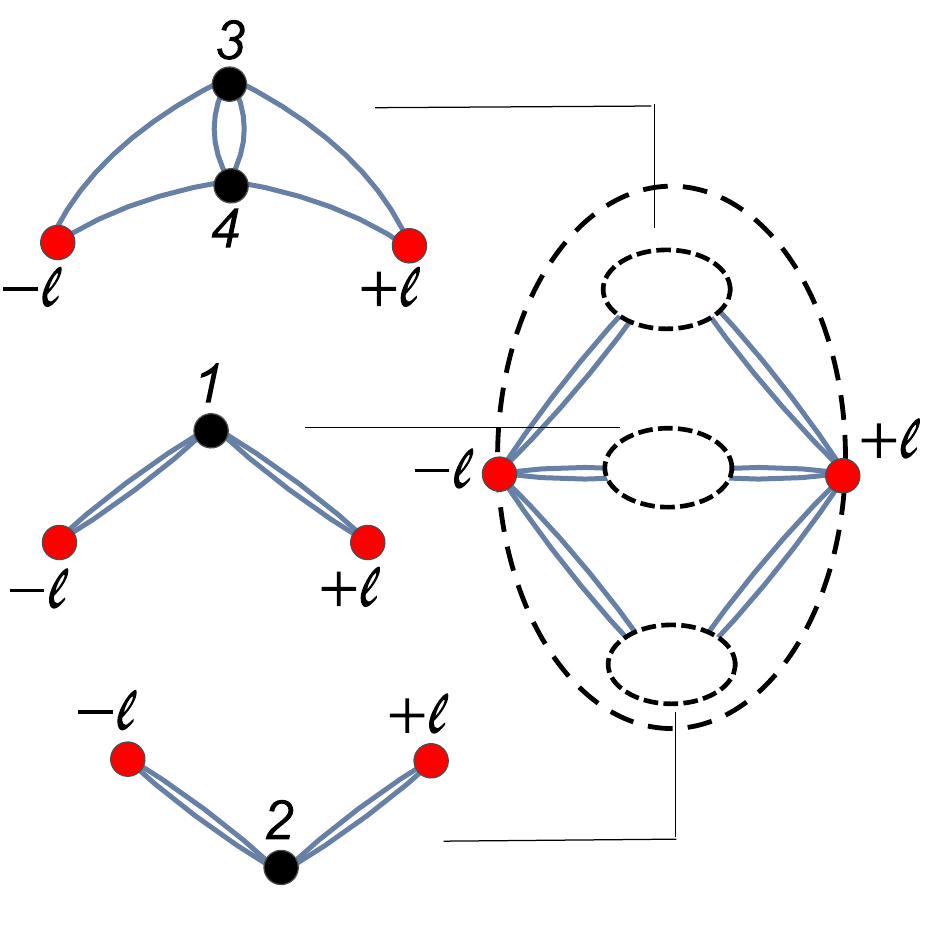}
  \caption{CHY-graph construction for the one-loop diagram at Figure \ref{FEY_ex2_loop2}}\label{FEY_ex2_loop_cut_CHY_1}
\end{figure}

The corresponding CHY-graph for the above Feynman subdiagrams are given in figure \ref{FEY_ex2_loop_cut_CHY_1}.

Splitting the $k_0$ vertex at every tree-CHY sub-graph as two-points with momentum $\ell$ and $-\ell$, we obtain something as the display at the right hand side of figure \ref{FEY_ex2_loop_cut_CHY_1}.

\begin{figure}[h]
  \centering
    \includegraphics[width=1.5in]{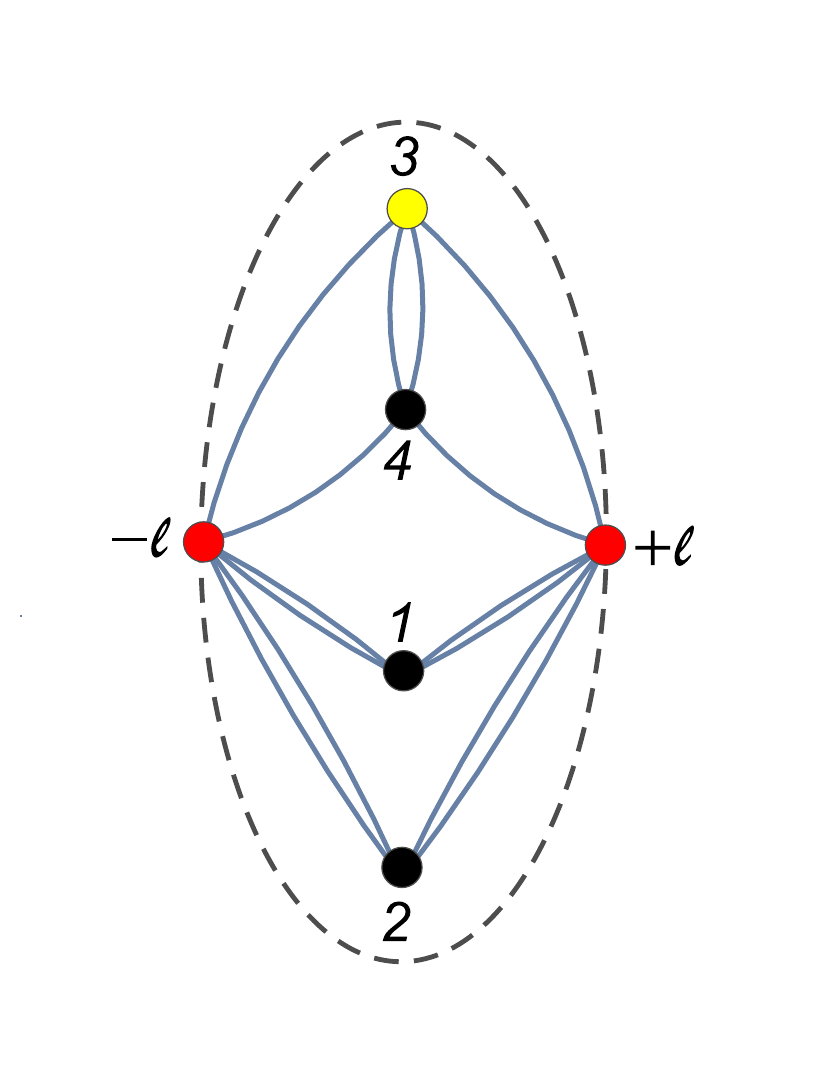}
  \caption{CHY-graph corresponding to the one-loop diagram at Figure \ref{FEY_ex2_loop2}}\label{FEY_ex2_loop_chy}
\end{figure}

Connecting all vertices with momentum $\pm\ell$ to the points $\pm\ell$, we get the corresponding CHY-graph shown in figure \ref{FEY_ex2_loop_chy}, which has been previously obtained in section \ref{Example2}. 

In section \ref{Lalg} this CHY-graph was solved by using the $\Lambda-$algorithm reproducing the result from the Feynman diagram in figure \ref{FEY_ex2_loop2}, which in this section we have used as our starting point instead.

\subsubsection{Six-Point and Box-Loop}

\begin{figure}[h]
  \centering
    \includegraphics[scale=0.5]{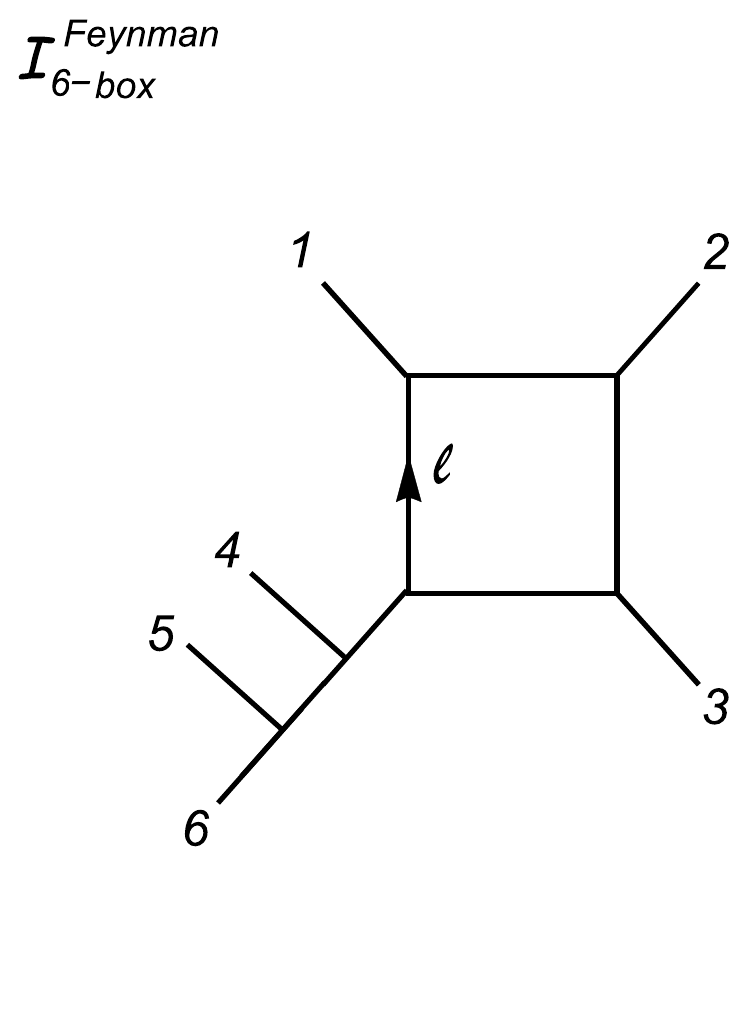}\quad
    \includegraphics[scale=0.5]{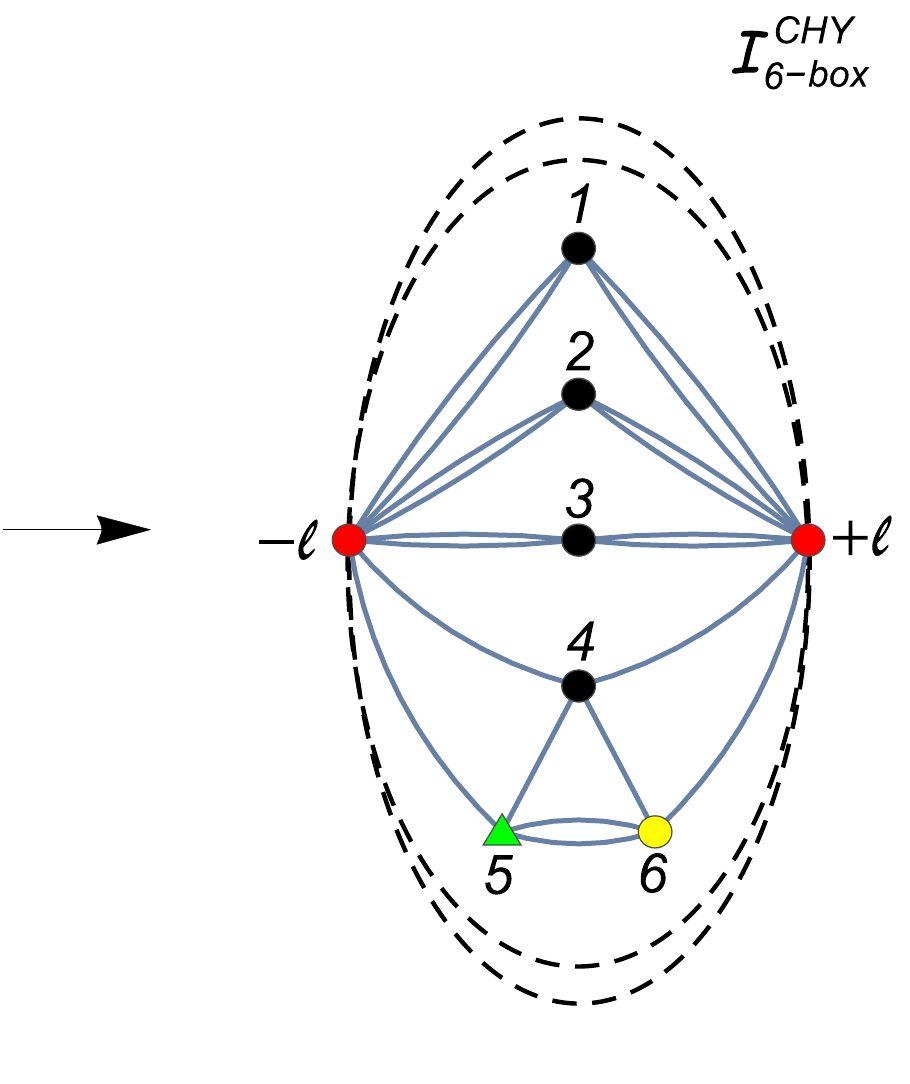}
  \caption{One-loop Feynman diagram with a box and its corresponding CHY-graph.}\label{FeyDog}
\end{figure}

As a slightly more involved example, let us consider the Feynman diagram shown at the left hand side of figure \ref{FeyDog} and containing a box loop.

\begin{figure}[h]
  \centering
   \includegraphics[scale=0.5]{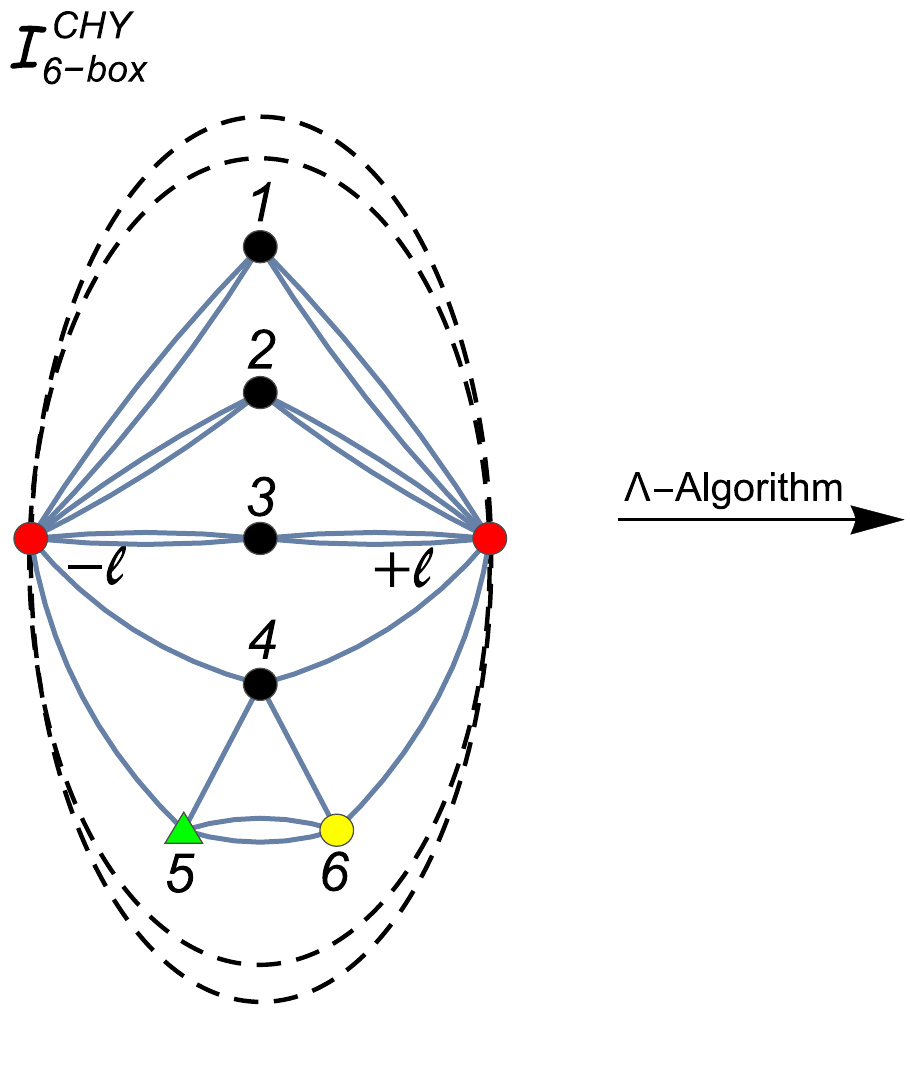}~~~
    \includegraphics[scale=0.5]{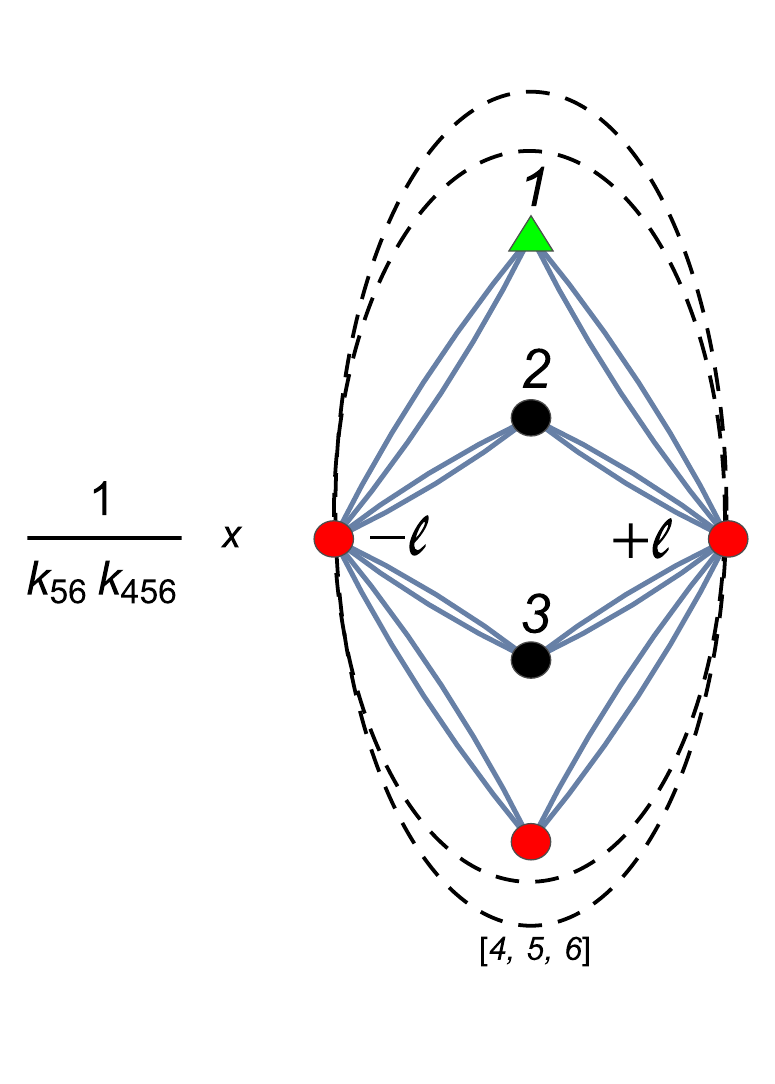}
  \caption{CHY-graph reduction by $\Lambda-$algorithm.}\label{FeyDogCHYGen}
\end{figure}

By using the rules described previously in section \ref{from_FEY_CHY}, we can rapidly read the corresponding CHY-graph, which is given in the right hand side of figure \ref{FeyDog}, where we have already shown explicitly the particular gauge fixing we are going to use to solve it.  By using the $\Lambda-$algorithm with the chosen gauge fixing,  it is straightforward to carry out and the answer can be written as an off-shell ${\rm 4-gon}$, as one given in  figure \ref{FeyDogCHYGen}.

Since the resulting $4-$gon only contains one off-shell vertex, we can still use the $\Lambda-$algorithm to solve it. The explicit result is given by the expression
\begin{equation}\label{RESULT}
{\cal I}_{\rm 6-box}^{\rm CHY}(1,2,3,|4|5,6|\ell,-\ell)=\frac{1}{k_{56}\,k_{456}}  \times {\cal I}_6^{4}(1,2,3,[4,5,6]|\ell,-\ell),
\end{equation}
where ${\cal I}_6^{4}(1,2,3,[4,5,6]|\ell,-\ell)$ is just the $4$-gon (with one off-shell particle) and it was computed in \cite{Cardona:2016bpi}
\begin{align}
{\cal I}_6^{4}(a,b,c,d|\ell,-\ell)&=\frac{{\cal I}_5^3(b,c,d|-\ell, [a,\ell])}{ k_{\ell a}}+ \frac{  {\cal I}_4^2(b,a|[-\ell, c,d  ],\ell) \,\, {\cal I}_4^2(c,d|-\ell, [a,b ,\ell ])   }{k_{\ell ab}}\\
&
+ \frac{ {\cal I}_4^2(c,a|[-\ell, b,d  ],\ell) \,\, {\cal I}_4^2(b,d|-\ell, [a,c ,\ell ])   }{k_{\ell ac}}
+\frac{{\cal I}_5^3(b,c,a|[-\ell,d  ],\ell)   }{k_{\ell abc}}\nonumber\\
&+ (\ell \,\leftrightarrow \, -\ell) \nonumber,
\end{align}
where\footnote{Let us remind, $[b,c]=k_b+k_c$ and  $k_{a [b,c]} = k_{ab}+k_{ac}$.}
\begin{align}\label{triangleOff} 
{\cal I}_5^3(a,b,c|i,j)=&{{\cal I}_4^2 (b,c|i,[a,j])\over k_{ja}}+{{\cal I}_4^2(b,a|[c,i],j)\over k_{jab}}+ (i  \leftrightarrow j),\\
{\cal I}_4^2(a,b|i,j)=&\frac{1}{k_{ai}}+\frac{1}{k_{aj}}.
\end{align}

The result in \eqref{RESULT} has been checked against the corresponding Feynman diagram at figure \ref{FeyDog} after partial fraction decomposition 
\be\label{FeyExpBox}
2^5\,\ell^2\,{\cal I}^{\rm Feynman}_{6-{\rm box}}=
 {1\over k_{45}k_{456}}\sum_{\s\in P_4}{1\over k_{\ell\s_1}k_{\ell\s_1\s_2}k_{\ell\s_1\s_2\s_3}},
\ee
where $P_4$ is defined as 
\begin{equation}
P_4:={\rm permutations} \,\{ a_1,a_2,a_3,a_4 \},\qquad {\rm with}~~a_1=1\,,a_2=2\,,a_3=3\,, a_4=456\,,
\end{equation}
for example, $k_{\ell a_1 a_4} = k_{\ell 1 456}$.


\section{Discussion}\label{Discussion}

In this work we have presented a prescription to build generic CHY-integrands directly over $\M_{1,n}$,  the moduli space of  $n-$puncture Elliptic curves. By generalizing the $\tau_{a:b}$ connectors  on $\M_{0,n}$, the moduli space of  $n-$puncture spheres  \cite{Gomez:2016bmv}, we have proposed  a new  set of connectors,  $\{ H_{a:a}, T_{a:b}, G^{\pm}_{a:b} \}$,  on $\M_{1,n}$. These connectors  implement the different ways to link two punctures lying on different locations on the Torus. Namely, two points connected by circling a $b-$cycle in one direction or in the opposite corresponds to linking them with $G^{\pm}_{a:b}$. Connecting a point to itself by circling a $b-$cycle in any direction corresponds to a link giving by $H_{a:b}$ and finally two points connected without going through a $b-$cycle corresponds to linking them with $T_{a:b}$. 

We have shown through several  examples,  that one way to build physically sensible integrands over $\M_{1,n}$, is by starting with a given known integrand on $\M_{0,n}$ and replacing all $\tau_{a:b}$'s by $T_{a:b}$'s or  $G^{\pm}_{a:b}$'s  in such way that the winding through the $b-$cycle equals zero, or in other words, that the number of $G^{+}$'s equals the number
 of $G^{-}$'s.  This was applied particularly to $\ph^3$ theory.
 
We have also provided a cut and paste graphical process to build $\ph^3$  CHY-integrands on $\M_{1,n}$, by starting from a given Feynman diagram at one loop. Roughtly speaking, by using the rules at $\ph^3$ tree level given in \cite{Baadsgaard:2015ifa}, one can find a CHY-loop graph by gluing CHY-trees in a particular way, as has been schematically shown in figure \ref{CHY_pgon_tree}.

Despite we have applied both constructions to the particular case of bi-adjoint $\Phi^3$ theory, we are 
confident that the rules presented in this paper can be easily extended to any other theory having a CHY representation. 

\begin{figure}
  \centering
   \includegraphics[scale=0.45]{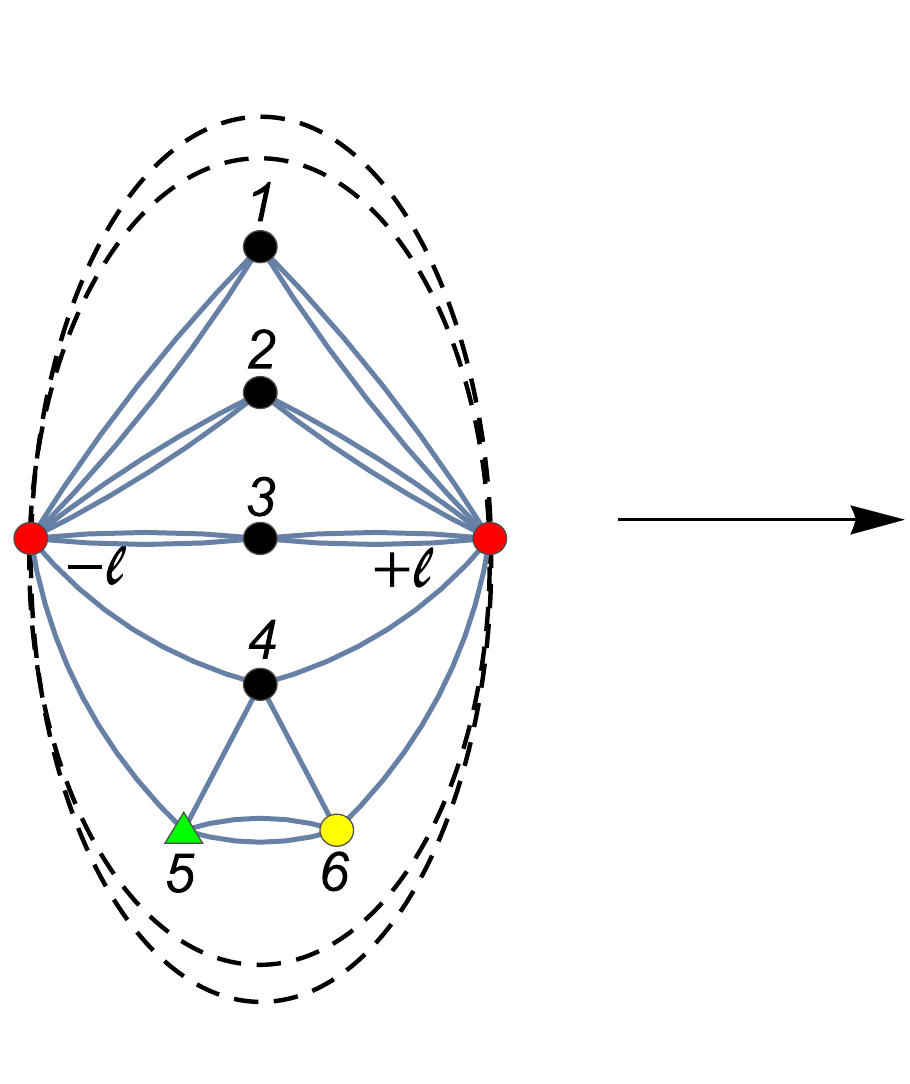}~~~
    \includegraphics[scale=0.45]{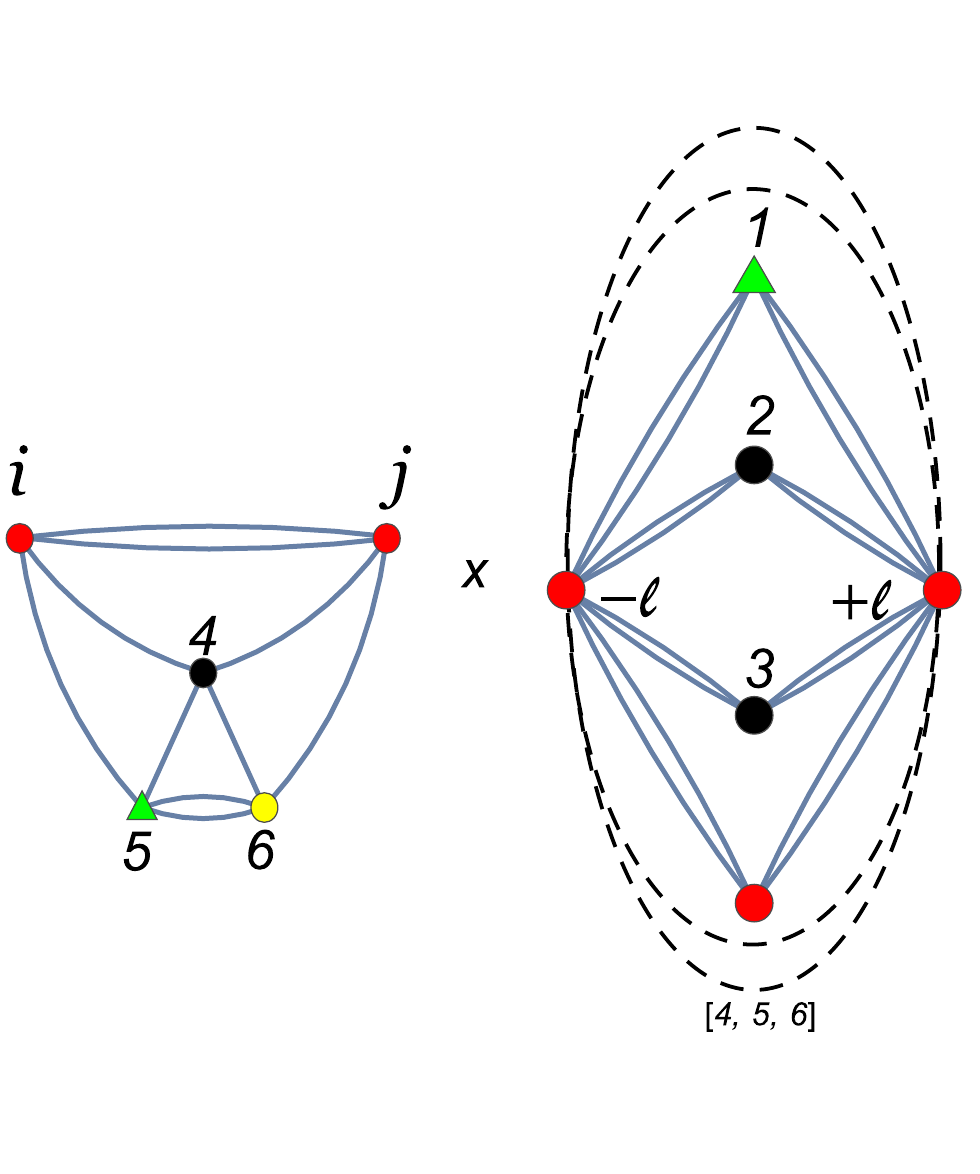}
        \includegraphics[scale=0.55]{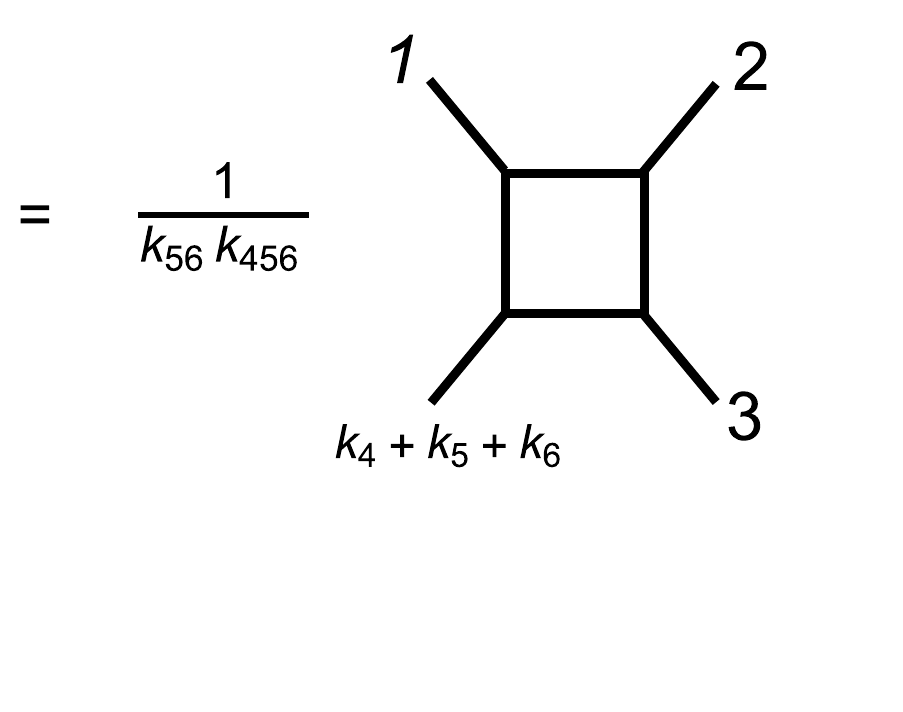}
  \caption{CHY-graph rewrite as the product of an off-shell tree-level CHY-graph  times an off-shell $4-$gon and its Feynman diagram representation.}\label{suposition}
\end{figure}

In section \ref{simple_example} we have noticed an interesting phenomena that happens when the connectors, $\tau_{a:b}$'s, in a given integrand over $\M_{0,n}$ are trade only by $T_{a:b}$'s over $\M_{1,n}$, i.e. when in the CHY-graph on a Torus do not encircle the $b-$cycle at all. It is easy to realize that the resulting CHY-graph on the sphere corresponds to a loop disjoint from the tree(s), which can be interpreted as a  tadpole diagram. It was also shown that this kind of graphs vanish  and hence, we can said that  our approach is free of tadpoles.  
\\
\\
{\it Speculative Perspectives}
\\
We also would like to make an observation induced from the structure of one-loop diagrams in $\Phi^3$ theory. From figure \ref{FeyDogCHYGen},  it is not hard to check that after using the $\Lambda-$algorithm,
the CHY-graph in figure \ref{FeyDog} could be factorized as in figure \ref{suposition}, where $k_i+k_j= k_1+k_2+k_3$.

As we see from this simple example, the one-loop CHY-graph can be rewritten in a factorized form, as a off-shell tree-level graph times a  $4-$gon graph with one off-shell vertex ($k_4+k_5+k_6$), exactly as it happends to the Feynman diagram at the right hand side. We believe that in general, it should be always possible to rewrite a given one-loop CHY graph as a product of the off-shell trees graphs times an off-shell CHY $p-$gon. This looks like a bold statement as it is, but, due that a given one-loop Feynman diagram possess this factorization property it implies that the corresponding CHY-graph should also factorizes in the same manner. Nevertheless, the $\L-$algorithm does not know how to deal with more than 3 off-shell particles, so, one should apply other technique in order to prove it,  perhaps the Feynman rules given in \cite{Huang:2016zzb}.

\begin{figure}
  \centering
      \includegraphics[scale=0.5]{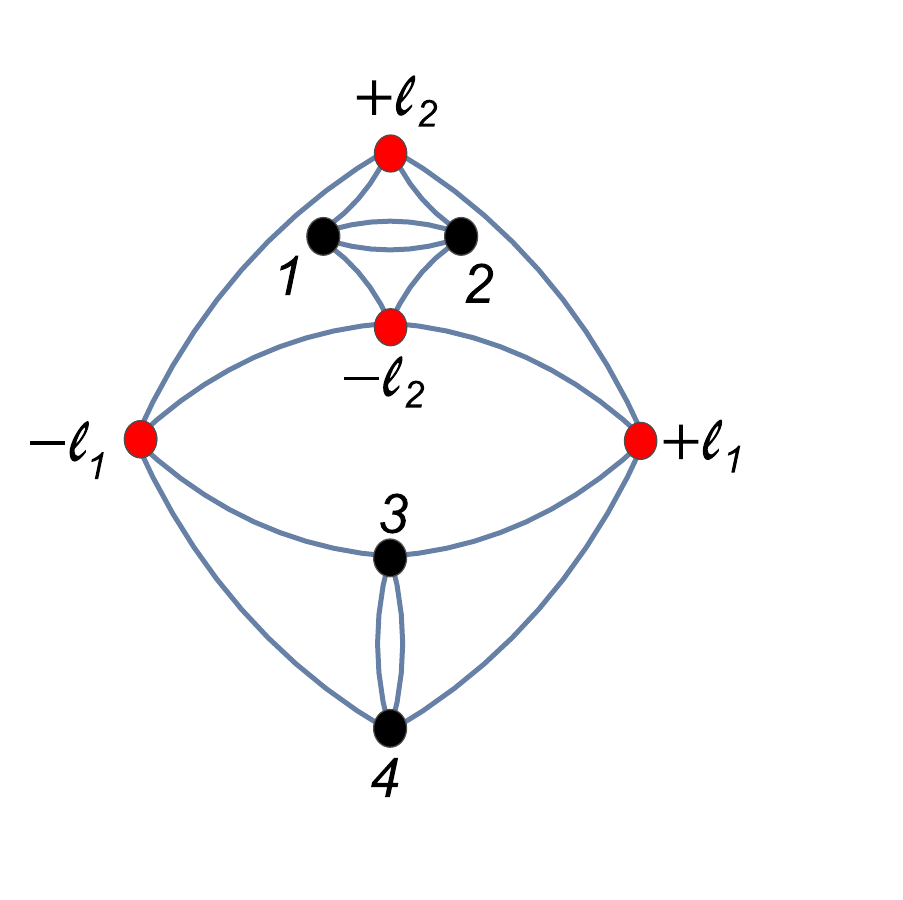}
            \includegraphics[scale=0.45]{arrow-eps-converted-to.pdf}
   \includegraphics[scale=0.5]{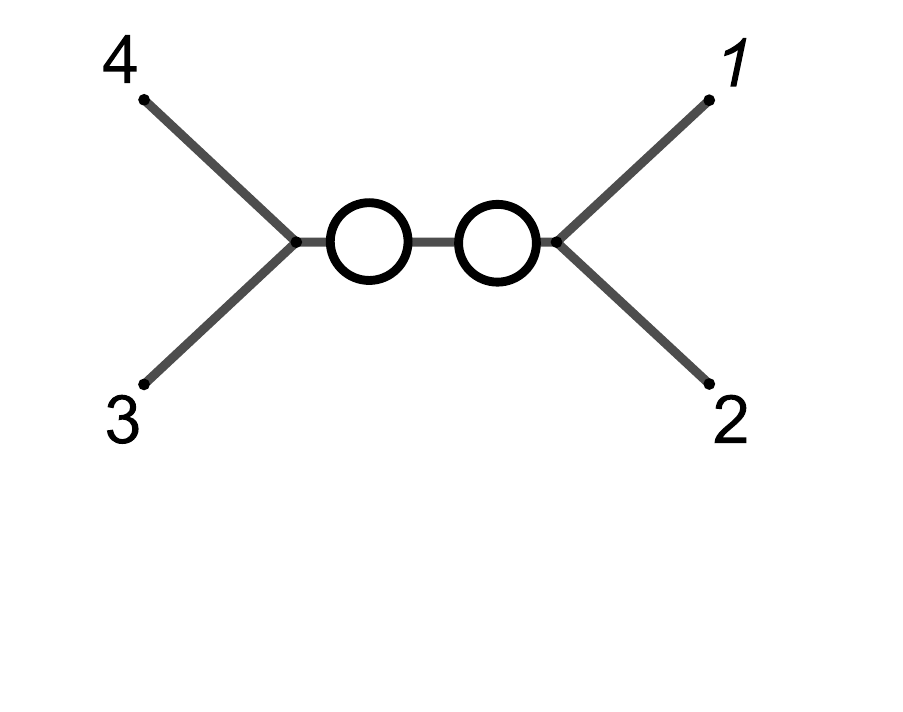}
  \caption{CHY-graph that is presumed to represent the 2-loop Feynman diagram given on the right side, up to $1/(\ell_1^2\, \ell_2^2)$ factor.}\label{2loop}
\end{figure}

Finally, the ideas in this paper can easily be extended to higher loop level  in $\ph^3$ theory. For example, following the rules presented in section \ref{PHI3theory},  we have found the CHY-graph in figure \ref{2loop}, which should represent the two-loop Feyman diagram given over the right side in the same figure.
Nevertheless, although we have not shown this equivalence,  it is a work in progress.

\vspace{5mm}


\acknowledgments

\vspace{3mm}

\noindent
It is our pleasure to thank to F. Cachazo for useful comments and discussions. H.G.  would like to thank the hospitality of  Universidade de S\~ao Paulo (USP)
and Universidad Santigo de Cali,  where this work was developed. The work of C.C. is supported in part by the National Center for Theoretical Science (NCTS), Taiwan, Republic of China. The work of  H.G.  is supported by CNPq  grant 403178/2014-2  and USC grant DGI-COCEIN-No 935-621115-N22.


\bibliographystyle{JHEP}
\bibliography{mybib}
\end{document}